\documentclass[longauth]{aa} 
\usepackage{graphicx}
\usepackage{longtable}
\usepackage{verbatim}
\usepackage{float}
\usepackage{placeins}
\usepackage{array,multirow,makecell}
\usepackage[export]{adjustbox}
\usepackage{lscape}
\newcommand{\pcms}{cm$^{-2}$}
\newcommand{\pcmc}{cm$^{-3}$}
\usepackage{txfonts}
\usepackage{bookmark}
\usepackage[]{hyperref}
\hypersetup{
    colorlinks=True,
    citecolor=blue,
    urlcolor=magenta,      
    linkcolor=blue,
    }
\usepackage{natbib}
\bibpunct{(}{)}{;}{a}{}{,} 
\begin{document}

   \title{An ALCHEMI inspection of sulphur-bearing species towards the central molecular zone of NGC 253 }

   \subtitle{}

   \author{M. Bouvier
          \inst{\ref{Leiden}}
          \and
          S. Viti\inst{\ref{Leiden}}
          \and 
          E. Behrens\inst{\ref{Dept.Astr.Virginia}}
          \and
          J. Butterworth\inst{\ref{Leiden}}
          \and
          K.-Y. Huang\inst{\ref{Leiden}}
          \and 
           J. G. Mangum\inst{\ref{NRAO.Charlottesville}}
           \and
          N. Harada\inst{\ref{NAO.Japan},\ref{Dept.Astr.Japan}}
          \and
          S. Mart\'in\inst{\ref{ESO.Chile},\ref{Alma.obs.Chile}}
          \and
          V. M. Rivilla\inst{\ref{CAB}}
          \and
          S. Muller\inst{\ref{Charlmers}}
          \and
          K. Sakamoto\inst{\ref{Dept.Astr.Japan}}
          \and
          Y. Yoshimura\inst{\ref{Inst.Ast.Japan}} 
          \and
          K. Tanaka\inst{\ref{Dept.Phy.Japan}}
          \and
          K. Nakanishi\inst{\ref{NAO.Japan},\ref{Dept.Astr.Japan}}
          \and
          R. Herrero-Illana\inst{\ref{ESO.Chile}} 
          \and
          L. Colzi\inst{\ref{CAB}}
          \and
          M. D. Gorski\inst{\ref{CIERA}}
          \and
          C. Henkel\inst{\ref{MaxPlanck.Bonn},\ref{Saudi.Arabia},\ref{Xinjiang.China}} 
          \and
          P. K. Humire\inst{\ref{dep.astro.Brazil}}
          \and
          D. S. Meier\inst{\ref{NMInst.Socorro},\ref{NRAO.Socorro}}
          \and 
          P. P. van der Werf\inst{\ref{Leiden}}
          \and 
          Y. T. Yan\inst{\ref{MaxPlanck.Bonn}}
          }

   \institute{\label{Leiden}Leiden Observatory, Leiden University, P.O. Box 9513, 23000 RA Leiden, The Netherlands\\
              \email{bouvier@strw.leidenuniv.nl}
        \and\label{Dept.Astr.Virginia}
            Department of Astronomy, University of Virginia, P.O. Box 400325, 530 McCormick Road, Charlottesville, VA 22904-4325, USA
        \and \label{NRAO.Charlottesville}
             National Radio Astronomy Observatory, 520 Edgemond Road, Charlottesville, VA 22903-2475, USA 
        \and\label{NAO.Japan}     
             National Astronomical Observatory of Japan, 2-21-1 Osawa, Mitaka, Tokyo 181-8588, Japan
        \and\label{Dept.Astr.Japan}
            Department of Astronomy, School of Science, The Graduate University for Advanced Studies (SOKENDAI), 2-21-1 Osawa, Mitaka, Tokyo, 181-1855 Japan
        \and\label{ESO.Chile}     
             European Southern Observatory, Alonso de C\'ordova, 3107, Vitacura, Santiago 763-0355, Chile
        \and\label{Alma.obs.Chile}
             Joint ALMA Observatory, Alonso de C\'ordova, 3107, Vitacura, Santiago 763-0355, Chile
        \and\label{CAB}
            Centro de Astrobiolog\'ia (CAB), INTA-CSIC, Carretera de Ajalvir km 4, Torrej\'{o}n de Ardoz, 28850 Madrid, Spain
        \and\label{Charlmers}
             Department of Space, Earth and Environment, Chalmers University of Technology, Onsala Space Observatory, SE-43992 Onsala, Sweden
        \and\label{Inst.Ast.Japan} 
            Institute of Astronomy, Graduate School of Science, The University of Tokyo, 2-21-1 Osawa, Mitaka, Tokyo 181-0015, Japan
        \and \label{Dept.Phy.Japan}
            Department of Physics, Faculty of Science and Technology, Keio University, 3-14-1 Hiyoshi, Yokohama, Kanagawa 223--8522 Japan
         \and\label{CIERA}
            Center for Interdisciplinary Exploration and Research in Astrophysics (CIERA) and Department of Physics and Astronomy, Northwestern University, Evanston, IL 60208, USA
        \and\label{MaxPlanck.Bonn}
             Max-Planck-Institut f\"ur Radioastronomie, Auf-dem-H\"ugel 69, 53121 Bonn, Germany
        \and\label{Saudi.Arabia}
             Astron. Dept., Faculty of Science, King Abdulaziz University, P.O. Box 80203, Jeddah 21589, Saudi Arabia
        \and\label{Xinjiang.China}
             Xinjiang Astronomical Observatory, Chinese Academy of Sciences, 830011 Urumqi, China
        \and\label{dep.astro.Brazil}
            Departamento de Astronomia, Instituto de Astronomia, Geofísica e Ciências Atmosféricas da USP, Cidade Universitária, 05508-900 São Paulo, SP, Brazil
        \and\label{NMInst.Socorro}
              New Mexico Institute of Mining and Technology, 801 Leroy Place, Socorro, NM 87801, USA
        \and\label{NRAO.Socorro}
            National Radio Astronomy Observatory, PO Box O, 1003 Lopezville Road, Socorro, NM 87801, USA
             }

   \date{Received ; accepted}

 
  \abstract
   {Sulphur-bearing species are detected in various environments within Galactic star-forming regions and are particularly abundant in the gas phase of outflow/shocked regions and photo-dissociation regions. Thanks to the powerful capabilities of millimetre interferometers, studying sulphur-bearing species and their region of emission in various extreme extra-galactic environments (e.g. starburst, active galactic nuclei) and at high angular resolution and sensitivity is now possible.}
   {In this work, we aim to investigate the nature of the emission from the most common sulphur-bearing species observable at millimetre wavelengths towards the nuclear starburst of the nearby galaxy NGC 253. We intend to understand which type of regions are probed by sulphur-bearing species and which process(es) dominate(s) the release of sulphur into the gas phase.  }
   {We used the high-angular resolution (1.6$\arcsec$ or $\sim 27$ pc) observations from the ALCHEMI ALMA Large Program to image several sulphur-bearing species towards the central molecular zone (CMZ) of NGC 253. We performed local thermodynamic equilibrium (LTE) and non-LTE large velocity gradient (LVG) analyses to derive the physical conditions of the gas where the sulphur-bearing species are emitted, and their abundance ratios across the CMZ. Finally, we compared our results with previous ALCHEMI studies and a few selected Galactic environments.}
   {To reproduce the observations, we modelled two gas components for most of the sulphur-bearing species investigated in this work. We found that not all sulphur-bearing species trace the same type of gas: strong evidence indicates that H$_2$S and part of the emission of OCS, H$_2$CS, and SO, are tracing shocks whilst part of SO and CS emission rather trace the dense molecular gas. For some species, such as CCS and SO$_2$, we could not firmly conclude on their origin of emission. }
   {The present analysis indicates that the emission from most sulphur-bearing species throughout the CMZ is likely dominated by shocks associated with ongoing star formation. In the inner part of the CMZ where the presence of super star clusters was previously indicated, we could not distinguish between shocks or thermal evaporation as the main process releasing the S-bearing species.}


   \maketitle
%

\section{Introduction}

Sulphur-bearing species are widely detected in various environments within Galactic star-forming regions such as hot cores/hot corinos \citep[e.g.][]{blake_1996, charnley_sulfuretted_1997, van_der_tak_sulphur_2003, wakelam_resetting_2004, esplugues_line_2013, crockett_herschel_2014a, li_sulfur-bearing_2015, drozdovskaya_alma-pils_2018,luo_sulfur-bearing_2019, codella_svs13-class_2021, bouscasse_sulphur-rich_2022, de_la_villarmois_perseus_2023}, outflows/shocks \citep[e.g.][]{bachiller_chemically_2001, codella_chemical_2005, lefloch_shock-induced_2005, podio_molecular_2014, holdship_h2s_2016, holdship_sulfur_2019, kwon_kinematics_2015, ospina-zamudio_molecules_2019, taquet_seeds_2020, feng_seeds_2020, tychoniec_2021, de_la_villarmois_perseus_2023}, photo-dissociation regions (PDRs; e.g. \citealt{jansen_1995,goicoechea_low_2006, ginard_spectral_2012, goicoechea_bottlenecks_2021, riviere-marichalar_abundances_2019}) and molecular clouds and cores \citep[e.g.][]{wakelam_sulphur-bearing_2004, agundez_chemistry_2013, vastel_sulphur_2018, navarro-almaida2020, hily-blant_sulfur_2022,spezzano_h2cs_2022, fuente_gas_2023,fontani_evolution_2023}. 

In the dense gas phase of star-forming regions, most of the sulphur (S) is locked onto the icy grain mantles and its most efficient release into the gas phase can occur primarily via thermal evaporation in warm regions or non-thermal evaporation through shocks \cite[e.g.][]{minh_1990, charnley_sulfuretted_1997, bachiller_chemically_2001, hatchell_possible_2002,esplugues_modelling_2014,woods_new_2015}. Once in the gas phase, S-bearing species are known to be extremely reactive, meaning their abundance greatly depends on the thermal and kinetic properties of the gas \citep{viti_evaporation_2004}.  S-bearing species are particularly useful for reconstructing the chemical history and dynamics of a variety of objects as they have been used as chemical clocks for the shocked gas and hot cores \citep[e.g.][]{pineaudesforets_1993,charnley_sulfuretted_1997,codella_1999, viti_chemical_2001,wakelam_sulphur-bearing_2004,li_sulfur-bearing_2015,feng_seeds_2020,taquet_seeds_2020,codella_svs13-class_2021,bouscasse_sulphur-rich_2022,esplugues_evolution_2023} .

Thanks to the powerful capabilities of millimetre interferometers such as the Atacama Large Millimeter/(sub-)millimeter Array (ALMA) and the NOrthern Extended Millimetre Array (NOEMA), it is now possible to study many sulphur-bearing species in the extreme environments of external galaxies (e.g. starburst, active galactic nuclei or AGN) compared to our own Galaxy. A variety of sulphur-bearing species have been previously detected in various types of galaxies. For example, SO and SO$_2$ likely trace hot cores in the Large Magellanic Cloud \citep{sewilo_alma_2022} whilst H$_2$S, SO, SO$_2$, and CS seem to probe shocks in AGNs \citep[e.g.][]{scourfield_alma_2020, kameno_probing_2023}.
In starburst galaxies, sulphur-bearing species were found to be associated with PDRs \citep[e.g.][]{meier_2005, bayet_extragalactic_2009}, dense molecular gas \citep[e.g.][]{martin_sulfur_2005, aladro_2011, ginard_chemical_2015}, shocks \citep[e.g.][]{martin_first_2003, martin_sulfur_2005, viti_molecular_2014, meier_alma_2015, sato_apex_2022} or irradiated gas by forming stars (i.e. hot-core-like regions; \citealt{minh_unveiling_2007, sato_apex_2022}). However, these studies were limited by either a low angular resolution \citep{martin_sulfur_2005} or by a low number of available transitions \citep{minh_unveiling_2007} to allow the constrain of the physical conditions of the gas where the S-bearing species are emitted, and hence reveal which environment they trace in starburst galaxies.

The ALMA Comprehensive High-resolution Extragalactic Molecular Inventory (ALCHEMI; \citealt{martin_alchemi_2021}) large programme is the first and broader unbiased imaging molecular survey performed towards the starburst central molecular zone (CMZ) of the nearby galaxy NGC 253 beyond the previous single-dish unbiased surveys by \cite{martin_2006} and \cite{aladro_2015}, giving us access to both high angular resolution ($1.6\arcsec$ or $\sim 27$ pc) and a large number of transitions for many molecular species, including sulphur-bearing species. Several ALCHEMI studies have already been published which derived a high cosmic-ray ionization rate (CRIR) towards the NGC\,253 CMZ using various molecular species and abundance ratios (C$_2$H: \citealt{holdship_distribution_2021}; H$_3$O$^+$/SO: \citealt{holdship_energizing_2022}; HCO$^+$/HOC$^+$: \citealt{harada_starburst_2021}, and HCN/HNC: \citealt{behrens_tracing_2022}), reported the first extragalactic detection of a phosphorus-bearing species, PN \citep{haasler_first_2022}, characterized various types of shocks throughout the CMZ (using HNCO and SiO: \citealt{huang_reconstructing_2023}; methanol masers: \citealt{humire_methanol_2022}; or HOCO$^+$: \citealt{harada_alchemi_2022}), explored the general physical properties of the CMZ of NGC 253 \citep{tanaka_2023}, and performed a full principal component analysis (PCA; \citealt{harada_pca_2024})

In this paper we aim to use the molecular richness and high-angular resolution of ALCHEMI to perform a complete investigation of the most abundant sulphur-bearing species in Galactic star-forming regions (CS, H$_2$S, OCS, SO, H$_2$CS, SO$_2$, CCS; e.g \citealt{hatchell_survey_1998, li_sulfur-bearing_2015,widicus_weaver_deep_2017,humire_2020}) towards the starburst CMZ of NGC 253, where the detection of these species has been previously reported by \cite{martin_alchemi_2021}. We aim to investigate what are the main process(es) dominating the release of sulphur-bearing species into the gas phase towards the CMZ of NGC 253, a prototypical starburst environment.

The paper is structured as follows: An overview of the NGC\,253 CMZ structure is presented in Sec.~\ref{sec:source_background} and the observations in Sec.~\ref{sec:obs}. Sec.~\ref{sec:maps} shows the emission distribution of the targeted S-bearing species and the regions chosen to perform the analyses. The results of Gaussian fit and the LTE and non-LTE analyses are described in Sec.~\ref{sec: results} and then are further discussed in Sec.~\ref{sec:discussion}. Finally, the conclusions are summarised in Sec.~\ref{sec:conclusions}.


\section{The structure of the central molecular zone of NGC 253}\label{sec:source_background}

NGC 253 is a nearby starburst galaxy located at a distance of 3.5 Mpc \citep{rekola_distance_2005} with an inclination of 76$^\circ$ \citep{mccormick_2013}. The inner kpc region exhibits intense star formation with a rate of $\sim 2 $M$_{\odot}$.yr$^{-1}$, which represents about half of the global star formation activity of the galaxy \citep{bendo_2015, leroy_alma_2015}. Within this starburst nucleus, the CMZ is a $\sim300$ pc $\times 100$ pc region containing ten giant molecular clouds (GMCs) of a size of $\sim 30$ pc identified in both molecular and continuum emission \citep{sakamoto_2011, leroy_alma_2015}. The CMZ is particularly chemically rich \cite[e.g.][]{martin_2006,aladro_2015, perez-beaupuits_thorough_2018,mangum_fire_2019,krieger_molecular_2020a, martin_alchemi_2021} and relatively turbulent. Indeed, shocks have been identified in several GMCs lying close to the orbital intersections of the central bar hosted by NGC 253 \citep{harada_alchemi_2022, huang_reconstructing_2023, humire_methanol_2022}, and have been claimed to explain its global chemical abundances \citep{martin_2009a}. Additionally, the presence of the starburst-driven large-scale outflow \citep{bolatto_2013} and superbubbles (due to expanding HII regions of supernovae; \citealt{sakamoto_molecular_2006, gorski_survey_2017, krieger_turbulent_2020b, harada_starburst_2021}) could interact with the CMZ inducing streamers (\citealt{walter_dense_2017, tanaka_2023}, Bao et al. submitted) and cloud-cloud collisions (\citealt{huang_reconstructing_2023}, \citealt{tanaka_2023}).

At a higher angular resolution, the GMCs located in the inner 160 pc of the CMZ can be resolved into several massive star-forming clumps called super proto- super star clusters (proto-SSCs; \citealt{ando_2017, leroy_forming_2018, rico-villas_super_2020, mills_clustered_2021, levy_outflows_2021}). These proto-SSCs are compact ($\leq 1$ pc), massive ($M\geq 10^5$ M$_\odot$) and young ($\leq 3$ Myr). Some of the proto-SSCs host Super Hot Cores (SHCs; \citealt{rico-villas_super_2020}), which are scaled up versions of Galactic Hot Cores with high dust temperatures ($\sim 200-300 $K) and high densities ($\sim 10^6$ \pcmc), and show clear outflow signatures \citep{gorski_2019, levy_outflows_2021}. High kinetic temperatures ($\geq 300$ K) were also measured from H$_2$CO observations at scales $\leq 1\arcsec$ ($\leq 17$ pc), very likely associated with dense gas heating sources \citep{mangum_fire_2019}. Some studies suggested that the proto-SSCs formed following an inside-out formation \citep{rico-villas_super_2020,mills_clustered_2021} but this scenario seems to be contentious \citep{krieger_molecular_2020a, levy_morpho-kinematic_2022}.

\section{Observations, Data Processing and Line Selection}\label{sec:obs}
The observations used in this work are part of the ALCHEMI Large Program (co-PIs.: S. Mart\'in, N. Harada, and J. Mangum). We give here only a summary of the observations.  The full details of the data acquisition, calibration and imaging are described by \cite{martin_alchemi_2021}. The observations were performed during Cycles 5 and 6, under the project codes 2017.1.00161.L and 2018.1.00162.S. The observations covered Bands 3 through 7, with a frequency ranging between 84.2 and 373.2 GHz and were centred at the position $\alpha$(ICRS)$=00^h47^m33.26^s$ and $\delta$(ICRS)$=-25^\circ 17'17.7''$; the common region imaged corresponds to a rectangular area of size $50'' \times 20''$ (850 $\times$ 340 pc) with a position angle of 65$^\circ$, covering the CMZ of NGC 253. All the data cubes have a common beam size of $1.6\arcsec\times 1.6\arcsec$ ($\sim 27$ pc), and a maximum recovered scale of $\geq 15''$ ($\geq 255$ pc). The spectral resolution is $\sim 10$ km.s$^{-1}$. Finally, following \cite{martin_alchemi_2021}, we adopted a flux calibration error of 15\%.

We used the continuum-subtracted FITS image cubes provided by ALCHEMI which contain the sulphur-bearing species and transitions of interest to this work. We processed the FITS image cubes using the packages MAPPING and CLASS from GILDAS\footnote{http://www.iram.fr/IRAMFR/GILDAS} to produce the velocity-integrated maps, extract the spectra and perform a Gaussian fitting of the lines. We extracted the spectra from beam-sized regions (see Sec.~\ref{sec:maps}) and converted them in brightness temperature using the equation:
\begin{equation}
    T(\text{K})=13.6\times \left(\frac{300 \ \text{GHz}}{\nu} \right)^2 \left(\frac{1\arcsec}{\theta_{\text{min}}} \right)\left(\frac{1\arcsec}{\theta_{\text{max}}} \right)S_\nu (\text{Jy})
\end{equation}
where $\nu$, $\theta_{\text{min}}$, $\theta_{\text{max}}$, and $S_\nu$ are the frequency, the minor and major axis of the beam, and the flux density, respectively. For each spectrum, we fit a $0^{\rm{th}}$ order polynomial to the line-free channels from the continuum-subtracted data cubes to retrieve the information about the spectral root-mean-square (rms). 

We targeted the most abundant S-bearing species detected in Galactic star-forming regions, namely CS, H$_2$S, H$_2$CS, OCS, SO, SO$_2$ and CCS \citep[e.g.][]{hatchell_survey_1998,li_sulfur-bearing_2015,widicus_weaver_deep_2017}. All the species used in this work were previously detected in NGC 253 by \cite{martin_alchemi_2021}. We also extracted measurements of the detected isotopologues $^{34}$SO and H$_2^{34}$S. Isotopologues of CS are also detected but they will be the focus of a separate study so we do not consider them in the present paper. For the analysis, the threshold for line detection was set to 3$\sigma$ for the integrated intensity. We did not consider lines that are contaminated by more than 15\% by another transition, or lines that are too blended to allow a multiple Gaussian fit to be produced. To assess whether a line is contaminated, we used the existing line identification (\citealt{martin_alchemi_2021} and priv. comm.), and performed an additional check using the CLASS extension WEEDS \citep{maret_2011}.  WEEDS uses the Jet Propulsion Laboratory (JPL; \citealt{pickett_1998}) and the Cologne Database for Molecular Spectroscopy (CDMS; \citealt{muller_2001, muller_2005}) databases. The list of all transitions used in this work, as well as their spectroscopic parameters, is presented in Tab.~\ref{tab:species_list}.

\begin{figure*}
    \centering
    \includegraphics[width=0.9\linewidth]{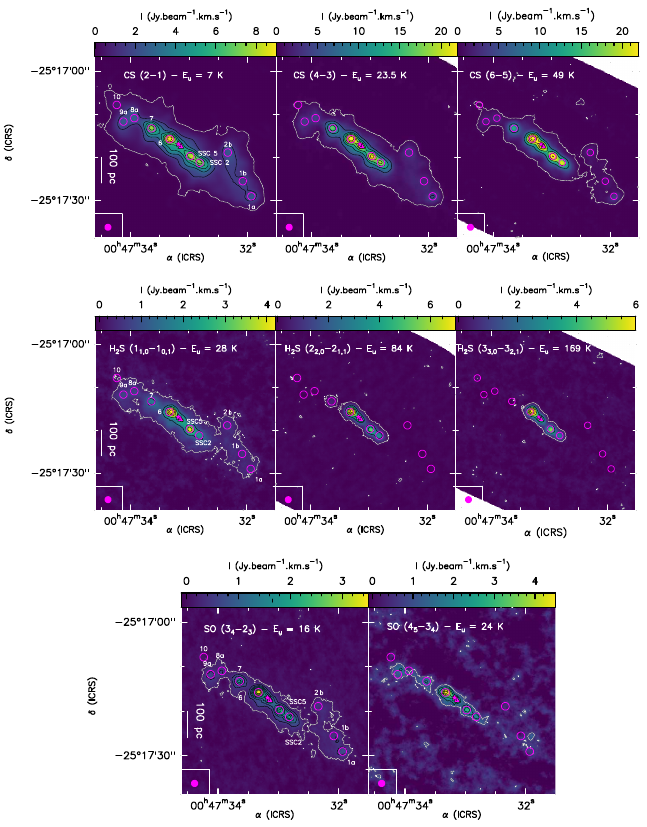}
    \caption{Velocity-integrated maps of CS, H$_2$S, and SO, which show extended emission throughout the CMZ with their peak of intensity towards the inner CMZ. The positions of TH2 \citep{turner_1_1985} and the kinematic centre \citep{muller-sanchez2010} are labeled by a filled magenta star and triangle, respectively. The regions where the spectra were extracted are shown by the magenta circles and are labeled in the left-most panel. The beam is depicted in the lower left corner of each plot. The scale bar of 100 pc corresponds to $\sim 6\arcsec$. \textit{Top:} Velocity-integrated maps of CS (2-1), (4-3) and (6-5). Levels start at 3$\sigma$ (1$\sigma=$46, 150, and 148 mJy.beam$^{-1}$, respectively; white contours) with steps of 30, 30, and 50$\sigma$ (black contours), respectively. \textit{Middle:} Velocity-integrated maps of H$_2$S ($1_{1,0}$-$1_{0,1}$), ($2_{2,0}$-$2_{1,1}$) and ($3_{3,0}$-$3_{2,1}$). Levels start at 3$\sigma$ (1$\sigma=$79, 50, and 81 mJy.beam$^{-1}$, respectively; white contours) with steps of 15, 40, and 30$\sigma$ (black contours), respectively. \textit{Bottom:} Velocity-integrated maps of SO ($3_4$-$2_3$) and ($3_5$-$3_4$). Levels start at 3$\sigma$ (1$\sigma=$50 and 28 mJy.beam$^{-1}$, respectively; white contours) with steps of 7 and 3$\sigma$ (black contours), respectively. }
    \label{fig:maps_extended}
\end{figure*}

\section{Emission Distribution and Targeted Regions }\label{sec:maps}

\begin{figure*}
    \centering
    \includegraphics[width=1\linewidth]{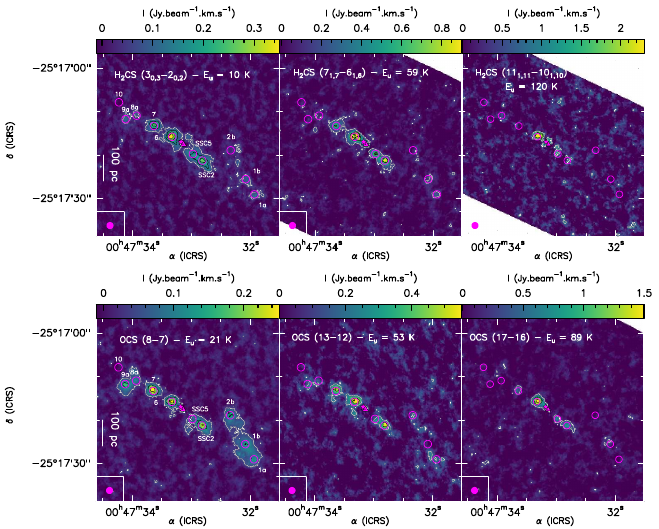}
    \caption{Same as Fig.~\ref{fig:maps_extended} for OCS and H$_2$CS, which show their peak of emission towards each of the GMCs rather than the inner CMZ. \textit{Top:} Velocity-integrated maps of H$_2$CS ($3_{0,3}$-$2_{0,2}$), ($7_{1,7}$-$6_{1,6}$), and ($11_{1,11}$-$10_{1,10}$) . Levels start at 3$\sigma$ (1$\sigma=$ 21, 50, and 27 mJy.beam$^{-1}$, respectively; white contours) with steps of 4, 10, and 4$\sigma$ (black contours), respectively. \textit{Bottom:} Velocity-integrated maps of OCS (8-7), (13-12) and (17-16). Levels start at 3$\sigma$ (1$\sigma=$14, 63.7 and 90 mJy.beam$^{-1}$, respectively; white contours) with steps of 7, 4, and 7$\sigma$ (black contours), respectively.}
    \label{fig:maps_ocs_h2cs}
\end{figure*}

\begin{figure*}
\centering
\includegraphics[width=1\linewidth]{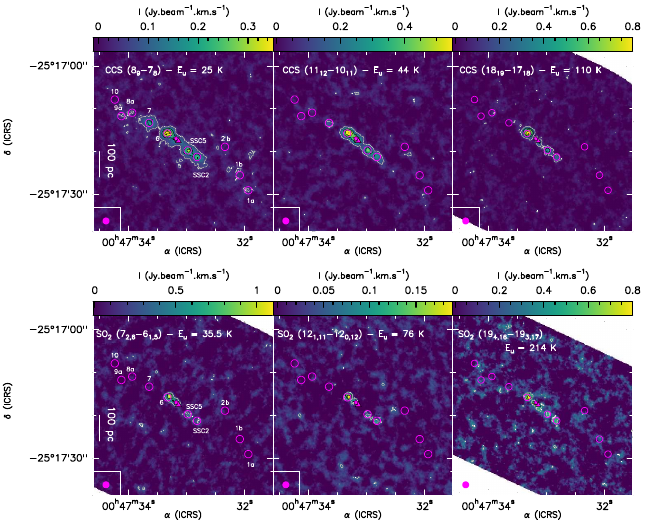}
\caption{Same as Fig.~\ref{fig:maps_extended} for CCS and SO$_2$, which are mostly detected towards the inner part of the CMZ. \textit{Top:} Velocity-integrated maps of CCS ($8_9$-$7_8$), ($11_{12}$-$10_{11}$), and ($18_{19}$-$17_{18}$). Levels start at 3$\sigma$ (1$\sigma=$15.7, 30, and 33 mJy.beam$^{-1}$, respectively; white contours) with steps of 7$\sigma$ (black contours). \textit{Bottom:} Velocity-integrated maps of SO$_2$ ($7_{2,6}$-$6_{1,5}$), ($12_{1,11}$-$12_{0,12}$) and ($19_{4,16}$-$19_{3,17}$), from left to right. Levels start at 3$\sigma$ (1$\sigma=$18.4, 91.5 and 125 mJy.beam$^{-1}$, respectively; white contours) with steps of 5, 3, and 3$\sigma$ (black contours), respectively.}
\label{fig:maps_ccs_so2}
\end{figure*}

The velocity-integrated line intensity maps of the different sulphur-bearing species are shown in Figures \ref{fig:maps_extended}, \ref{fig:maps_ocs_h2cs} and \ref{fig:maps_ccs_so2}. The maps for the isotopologues H$_2^{34}$S and $^{34}$SO, are presented in Fig.~\ref{fig:maps_isotopologues}. For each species, we selected a sample of three transitions which are unblended and isolated or next to a line whose flux does not exceed the level of the flux calibration error of 15\% (see Sec.\ref{sec:obs}) and integrated the emission over the full line profile ranging within [0;400] km.s$^{-1}$. When possible, we selected lines covering different upper-level energies to show the variation of the emission distribution with different excitation levels reflecting the various excitation conditions. Before describing the emission distribution of each species, we give an explanation of how the regions where we extracted the spectra were chosen and how we named them.

We selected the regions based on where the species show the most intense emission. Table \ref{tab:regions} lists the selected regions and their associated coordinates. Each of the selected regions have been named after existing identifications.  Regions GMC10, GMC7, and GMC6 correspond to the Giant Molecular Clouds 10, 7, and 6, respectively, as previously defined by \cite{leroy_alma_2015}. The S-bearing species have a peak intensity that is shifted from the GMC9, GMC8, GMC1 and GMC2 regions defined in \cite{leroy_alma_2015}, but which correspond to the newly defined regions GMC8a, GMC9a, GMC1a, GMC1b and GMC2b in \cite{huang_reconstructing_2023}. We did not notice a particular peak of emission of the S-bearing species towards the position GMC2a and do not consider this position in this work, as we aim to focus on where the emission of the S-bearing species is the strongest. The positions SSC5 and SSC2 have their centre corresponding to the proto-SSCs 5 and 2, respectively, which were defined in \cite{leroy_forming_2018}. The emission of the S-bearing species peaks towards these proto-SSCs rather than GMC3 and GMC4. Hence, we used the coordinates of SSC5 and SSC2 to extract the spectra towards these two positions. It is important to note that as the ALCHEMI beam encompasses several proto-SSCs, the regions named SSC5 and SSC2 include also the SSC4 to SSC7 and the SSC1 to SSC3, respectively. Finally, we did not take into account the GMC5 region, which is close to the position of the strongest radio continuum called TH2 \citep{turner_1_1985} and the kinematic centre of NGC 253 \citep{muller-sanchez2010}. In previous studies, it was found that GMC5 shows particular features, such as absorption, due to the strong radio continuum emission at this location \citep{meier_alma_2015, humire_methanol_2022}. The study of GMC5 is thus impossible to perform with the analysis proposed in this paper. Finally, the S-bearing species studied in this paper are not clearly detected in the streamers associated with the large-scale outflow, except for CS \citep{walter_dense_2017}. We will therefore not investigate vertical structures or gas properties along these features.

Overall, we defined ten regions, shown by magenta circles in Figs.\ref{fig:maps_extended}, \ref{fig:maps_ocs_h2cs} and \ref{fig:maps_ccs_so2}. For the rest of the paper, we will refer to the part of the CMZ containing the regions GMC10, 9a, 8a, 1a, 1b, and 2b as the \textit{outer CMZ} (between $10\arcsec-16\arcsec$ from the kinematic centre) , and we will refer to the part of the CMZ encompassing the regions GMC7 to SSC2 (within $\sim 6\arcsec$ around the kinematic centre) as the \textit{inner CMZ} .

\begin{table}
\centering
\caption{List of selected regions where the S-bearing species emission peaks, their coordinates, and their position within the CMZ. The letter "I" stands for a position in the inner CMZ whilst "O" stands for a position in the outer CMZ (see text). } \label{tab:regions}
\begin{tabular}{lccc}
\hline \hline
\multirow{2}{*}{Region} & R.A. (ICRS) & Dec.(ICRS)& Position\\
& (00$^h$47$^m$) & (-25$^\circ$:17$'$) & (O/I) \\
\hline
GMC10 &34$^s$.2360 &07$^{''}$.836 & O\\
GMC9a &34$^s$.1227  &11$^{''}$.702 & O\\
GMC8a & 33$^s$.9359&10$^{''}$.905& O \\
GMC7 &33$^s$.6432 &13$^{''}$.272& I \\
GMC6 &33$^s$.3312 &15$^{''}$.756& I \\
SSC5 &32$^s$.9811 &19$^{''}$.710& I \\
SSC2 &32$^s$.8199 &21$^{''}$.240 & I \\
GMC2b &32$^s$.3398 &18$^{''}$.869 & O \\
GMC1b &32$^s$.0835 & 25$^{''}$.510 & O\\
GMC1a &31$^s$.9363 &29$^{''}$.018 & O\\
\hline
\end{tabular}
\end{table}

We find that the emission distribution varies depending on the species and the transitions. CS shows the most extended emission among the S-bearing species and is detected throughout the CMZ, with its emission peaking towards the inner CMZ. The emission distribution of H$_2$S and SO are relatively similar to that of CS (Fig.~\ref{fig:maps_extended}). For the isotopologues  H$_2^{34}$S and $^{34}$SO, the former is detected only towards the inner CMZ whilst the latter is detected throughout the CMZ (Fig.~\ref{fig:maps_isotopologues}). However, as H$_2$S is less intense in the outer CMZ, the lack of detection of its isotopologue in these regions could be due to a lack of sensitivity. On the other hand, the emissions of OCS and H$_2$CS peak towards each of the GMCs/proto-SSCs throughout the CMZ rather than towards the GMCs/proto-SSCs of the inner CMZ (Fig.~\ref{fig:maps_ocs_h2cs}). Both species also peak towards SSC2 but not towards SSC5, as found in the case of the other species. Finally, CCS and SO$_2$ are the species which show most of their emission towards the inner CMZ, with SO$_2$, in particular, detected only towards the three innermost regions GMC6, SSC5, and SSC2 (Fig.~\ref{fig:maps_ccs_so2}). For all the species, lower-energy transitions ($E_\text{u} \leq 80-100$ K for H$_2$S and OCS, $E_\text{u} \leq 50-60$ K for H$_2$CS and SO, and $E_\text{u} \leq 25$ K for CCS ) are extended throughout the CMZ whilst the higher-energy transitions are more concentrated towards the inner part of the CMZ. 

\section{Results}\label{sec: results}

\begin{figure*}[ht]
    \centering
    \includegraphics[width=\textwidth]{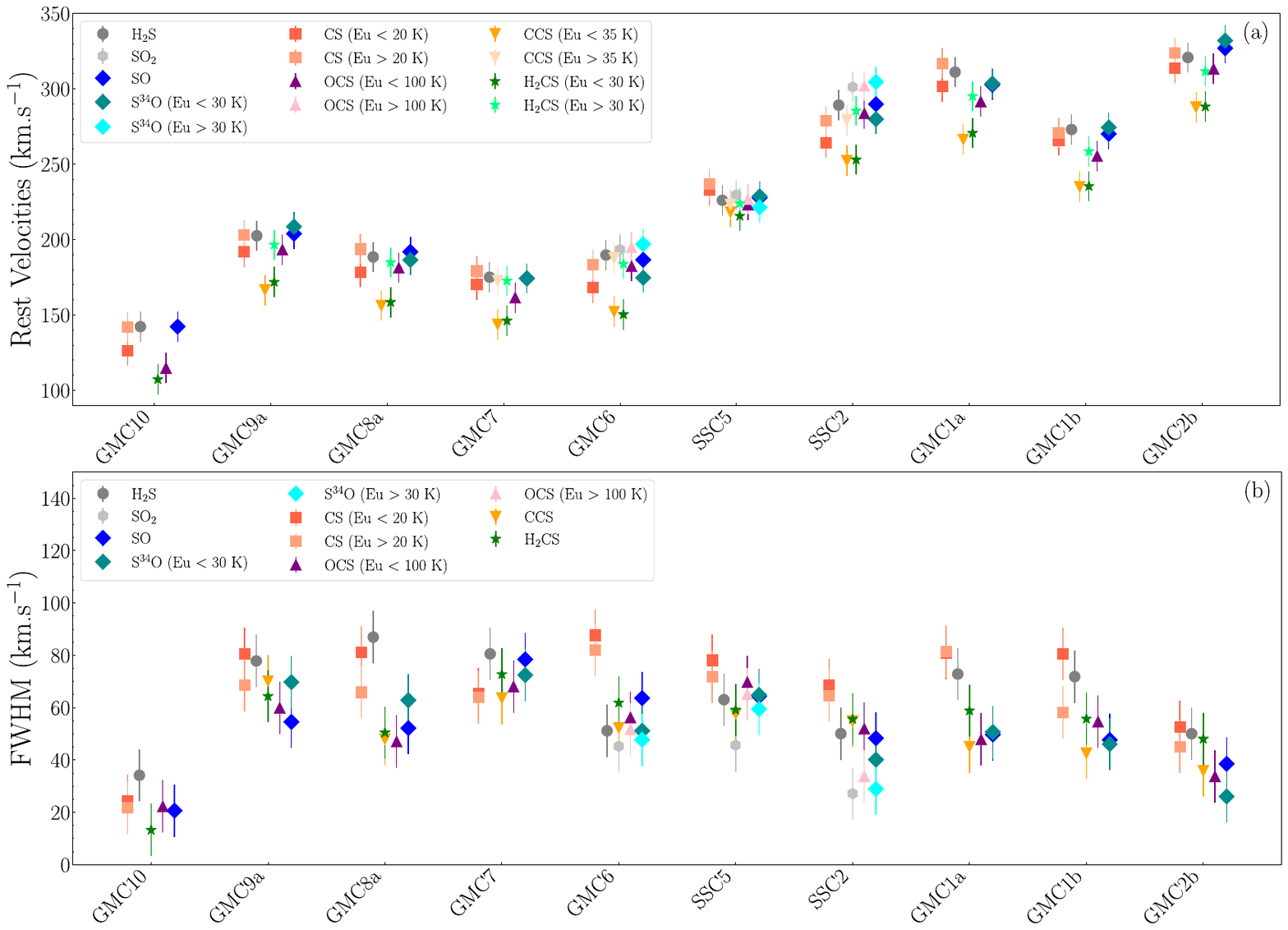}
    \caption{Results of Gaussian fits. The listed sequence of clouds follows the layout of the GMCs of the CMZ as shown in Figs.~\ref{fig:maps_extended} to \ref{fig:maps_ccs_so2}}. (a) Systemic velocities for each species as a function of the regions. If a strong difference is seen in the low- and high upper-level energy transitions of the species, the two velocity components are shown.(b) FWHM for each species as a function of the regions. If, for one species, a difference in line width is seen between the low- and high upper-level energy transitions, the two components are shown. 
    \label{fig:Fit_results}
\end{figure*}

\subsection{Gaussian Fit and Spectra}\label{subsec: gaussian fit}

We extracted each spectrum from a circular region of a beam size (1.6$"$) centred at the coordinates given in Tab.~\ref{tab:regions}. The spectra are shown in Appendix~\ref{app:spectra}. We centred each spectrum at the kinematic local standard of rest (LSRK) frequency of the associated transition before extracting the line parameters. To extract the integrated intensities ($\int T_B.dV$), we used two methods depending on whether the lines are blended or not. In the case of non-blended lines, we measured the direct integration of the channel intensities. This is because most of the lines do not present a clear single-Gaussian profile. We nonetheless performed Gaussian fitting to extract the line widths ($FWHM$) and the peak velocity ($V_{lsr}$) to have a quantitative assessment towards these parameters. For the lines that are blended with another line, we performed a multi-Gaussian fit (with two Gaussian components) and we took the integrated intensity directly from the result of the fit, as well as the $FWHM$ and $V_{lsr}$. If the blending with another line does not allow to perform such a fit, the transition has been dismissed from the analysis. Finally, for the line profiles of CS towards GMC7 and GMC2b, two velocity components could be used for the fitting (Fig.~\ref{fig:spec_part2}). However, for the rest of the analysis, we selected only the component with the highest peak intensity as this component is also detected in the other species (the other component is too weak and/or undetected, i.e. below the 3$\sigma$ threshold). We thus discarded this component as we would not have a complete data set for a comparison with the other component and other regions.  

The line-fitting results and the rms for each species and region, are reported in Table~\ref{tab: fit_results}. The results of the systemic velocities and line widths are discussed in Sec.~\ref{subsub:velocities} and \ref{subsec:linewidths} below.

\subsubsection{Systemic Velocities}\label{subsub:velocities}
Figure ~\ref{fig:Fit_results} (a) shows the averaged systemic velocities derived for each species per region. The uncertainties derived from the Gaussian fit being usually smaller than the spectral resolution of 10 km.s$^{-1}$ (see Table~\ref{tab: fit_results}), we thus adopted this value for the error bars.
First, it appears that not all transitions of the same species peak at the same systemic velocity: the low upper-level energy transitions of CS ($E_\text{u} < 20$ K), H$_2$CS ($E_\text{u} < 30$ K), OCS ($E_{\text{u}} < 100$ K), $^{34}$SO ($E_{\text{u}} < 30$ K) and CCS ($E_{\text{u}} < 35$ K) show a different peak compared to the higher upper-level transitions. However, depending on the species, this difference is not necessarily significant. Indeed, for CS the difference varies in the range $\sim 14-16$ km.s$^{-1}$ in most regions. The line widths vary between 65 and 81 km.s$^{-1}$. The differences in peak velocity are thus less than 1/3 of the line widths and cannot be considered as significant. Similarly, the difference in peak velocities for OCS is at most 1/3 of the line widths.Additionally, and this is particularly visible with the CS lines (see Figs.~\ref{fig:spec_part1} and \ref{fig:spec_part2}), non-Gaussian line profiles could also lead to a shift in the derived systemic velocity. Therefore, for CS and OCS the difference in peak velocities between the low- and high-upper-energy level transitions is not significant. On the other hand, for H$_2$CS, $^{34}$SO, and CCS, the difference in peak velocity between the low- and high upper-level energy transitions varies between 1/3 (CCS) and more than half (H$_2$CS and $^{34}$SO) the line widths of these species and, hence, can be considered as significant.

 A second result is that, within each region, not all species peak at the same systemic velocity. Throughout the CMZ, a clear distinction is seen between the velocity peak of the low upper-energy levels of H$_2$CS and CCS, with that of the other species and transitions. On average, the difference in peak velocity is 10--20 km.s$^{-1}$ compared to the low upper-level energy transitions of CS and OCS, and the high upper-level energy transitions of CCS. Here again, non-Gaussian line profiles could lead to a difference in the derived systemic velocities. Such small shifts in velocity might thus not be significant. A larger difference is seen compared to H$_2$S, SO, the low upper-level energy transitions of $^{34}$SO, and the high upper-level energy transitions of CS and H$_2$CS with a shift of 30--40 km.s$^{-1}$. Finally, the most important shift in peak velocity (40--50 km.s$^{-1}$) is seen for SO$_2$ and the high upper-level energy transitions of $^{34}$SO and OCS, when comparing with the systemic velocities of the other species and transitions. If we consider the line widths derived, all these peak differences are likely significant towards GMC10 where the peak difference is similar to or greater than the line width of the various transitions. Towards the regions SSC2 and GMC2b, where the line widths range between 20 and 60 km.s$^{-1}$ (see Fig.~\ref{fig:Fit_results}b), peak velocity shifts higher than 30 km.s$^{-1}$ can be considered as significant. Finally, in the other regions where most line widths are within the 40--80 km.s$^{-1}$ range, shifts of at least 40 km.s$^{-1}$ can be considered as significant. SSC5 is the only region where all species and transitions peak at almost the same velocity of $\sim 220-230$ km.s$^{-1}$, without any clear difference between transitions or species.

\subsubsection{FWHMs}\label{subsec:linewidths}
Figure~\ref{fig:Fit_results} (b) shows the FWHM analysis results. Towards GMC10, GMC2b, and SSC2 line widths are narrower compared to the other regions. The narrowest FWHMs are found towards GMC10, with all the species having FWHM$< 40$ km.s$^{-1}$. For SSC2, the smaller line width ($\sim30$ km.s$^{-1}$) compared to most of the other regions is only seen for some species and transitions: $^{34}$SO, SO$_2$ and OCS ($E_{\text{u}} > 100$ K). The rest of the species and transitions have line widths varying between 40 and 90 km.s$^{-1}$. Then, within each region, there is some scatter in the FWHMs: in the outer CMZ, H$_2$S and CS have the largest FHWM, except towards GMC10 where only H$_2$S shows the larger FWHM. In the inner CMZ, CS has the largest line widths compared to the other species whilst SO$_2$ and $^{34}$SO ($E_{\text{u}} > 30$ K) show the narrowest line widths. No general trend is found for the other species and transitions. Finally, for species where the low- ($E_\text{u}< 20-30$ K for CS, H$_2$CS, and $^{34}$SO; $E_\text{u}< 50$ K for CCS; $E_\text{u}< 100$ K for OCS) and high upper-level energy transitions are peaking at different systemic velocities (see Sec.~\ref{subsub:velocities}), there is no clear difference in the line widths, except for GMC1b where the line width of CS ($E_{\text{u}} < 20$ K) is larger than that of the other transitions by $\sim 22$ km.s$^{-1}$. Although some trends can be seen, it is important to keep in mind that optically thick lines and presence of multiple components (which cannot be disentangled at our current spectral resolution) can induce larger line widths or non-Gaussian line profiles (see Figs.~\ref{fig:spec_part1} and \ref{fig:spec_part2}). The optical depths are derived as part of the non-LTE LVG analysis (see Sec.~\ref{subsec:results_LVG}) and the results of both line optical depths and whether the derived trends in line widths can be trusted will be discussed in Sec.~\ref{subsec:results_LVG}.

\subsection{Derivation of physical parameters} \label{sec:phys_params}

To derive the physical parameters of each species and for each region, we used two different methods. As a first step, we performed rotation diagrams (RDs) without beam correction. As we cover a large range of $E_{\text{u}}$ for most of the species, it can allow us to identify whether different components with different excitation conditions (temperature and density) are present within each region studied. In this case, we expect to clearly see different slopes (hence different excitation temperatures) in the RDs. The main assumptions of the RD method are local thermodynamic equilibrium (LTE) conditions and optically thin emission of the lines. Although we do not expect these two assumptions to hold for all transitions and species, the results we obtain can be used as a first indication of the rotation temperature and beam-averaged column densities of each species. As examples of the results we derived from the LTE analysis, we will present below only those for the species for which we could not perform a non-LTE analysis as a second step.

We could not perform the non-LTE analysis for CCS, SO$_2$, and OCS. In the case of CCS, no collisional excitation rates are available in the online databases. For SO$_2$, the results of the RD predicted temperatures in the range $55-79$ K for the first component. The LAMDA database \citep{schoier_lamda_2005} provides two sets of collisional excitation rates; one for temperatures lower than 50\,K and the other one for temperatures larger than 100\,K. None of the two sets were thus usable to model the physical conditions of the first component of SO$_2$. For the second component, although we could use the set for which the rates were calculated for temperatures above 100 K, the non-LTE code we use cannot converge towards a solution if the first 200 levels of a molecule are populated, which is the case here. Finally, in the case of OCS, the collisional coefficient rates were initially calculated by \cite{green_1978} for the first thirteen levels and for a temperature up to 100 K. An extrapolation of these rates is available on the LAMDA database for transition levels up to J=99 and for a temperature up to 500 K \citep{schoier_lamda_2005}. However, no matter which range of physical conditions ($T,n$) we run, the code does not converge and is unable to find a solution\footnote{The collisional rate for a same temperature and transition differs by one order of magnitude between the LAMDA and BASECOL files, even though both databases are using the \cite{green_1978} paper. We notified the LAMDA database team about this issue}. We could find a solution by using the rates available in the BASECOL database \citep{dubernet_basecol_2013} but the available rates only concern the 13 first levels which corresponds to the first five transitions that are detected in this work. Consequently, we could  not model the second component on the regions GMC6, SSC5, and SSC2. In this section, we will thus present the LTE results for OCS, SO$_2$, and CCS. The LTE results for the other species are presented in Appx.~\ref{app:RDs}.

\subsubsection{LTE analysis: OCS, CCS, and SO$_2$}\label{subsec:LTE_main}

Thanks to a large enough ($\geq 10$ for most species and regions) number of transitions detected, as well as the wide range of upper-level energy transitions available per species, we could perform rotation diagram (RD) analyses. The beam filling factor was set to 1 as (1) we do not know the size of emission of the species and (2) the emission is more extended than the beam size in most cases. We included the calibration error of 15\% and the spectral RMS in the integrated flux (see Sec.~\ref{sec:obs}). The best-fit was calculated by minimizing the reduced chi-square, $\chi^2_r$. Depending on the regions and species, we could fit more than one rotation temperature. This behaviour can be due to the presence of multiple components with different physical conditions, or to non-LTE and/or opacity effects \citep{goldsmith_1999}. For CCS and OCS, the break in the temperature found at $\sim 30$ K and $\sim 100$ K for CCS and OCS, respectively, is consistent with the difference in systemic velocity seen in Sec.~\ref{subsub:velocities}, which could favour the multiple component scenario. For SO$_2$, no clear shift in systemic velocity was found. The various possibilities which could cause the presence of more than one rotation temperature cannot be distinguished with rotation diagrams only but this will be investigated during the non-LTE analysis, except for CCS, SO$_2$.

\begin{figure*}
    \centering
    \includegraphics[width=\textwidth]{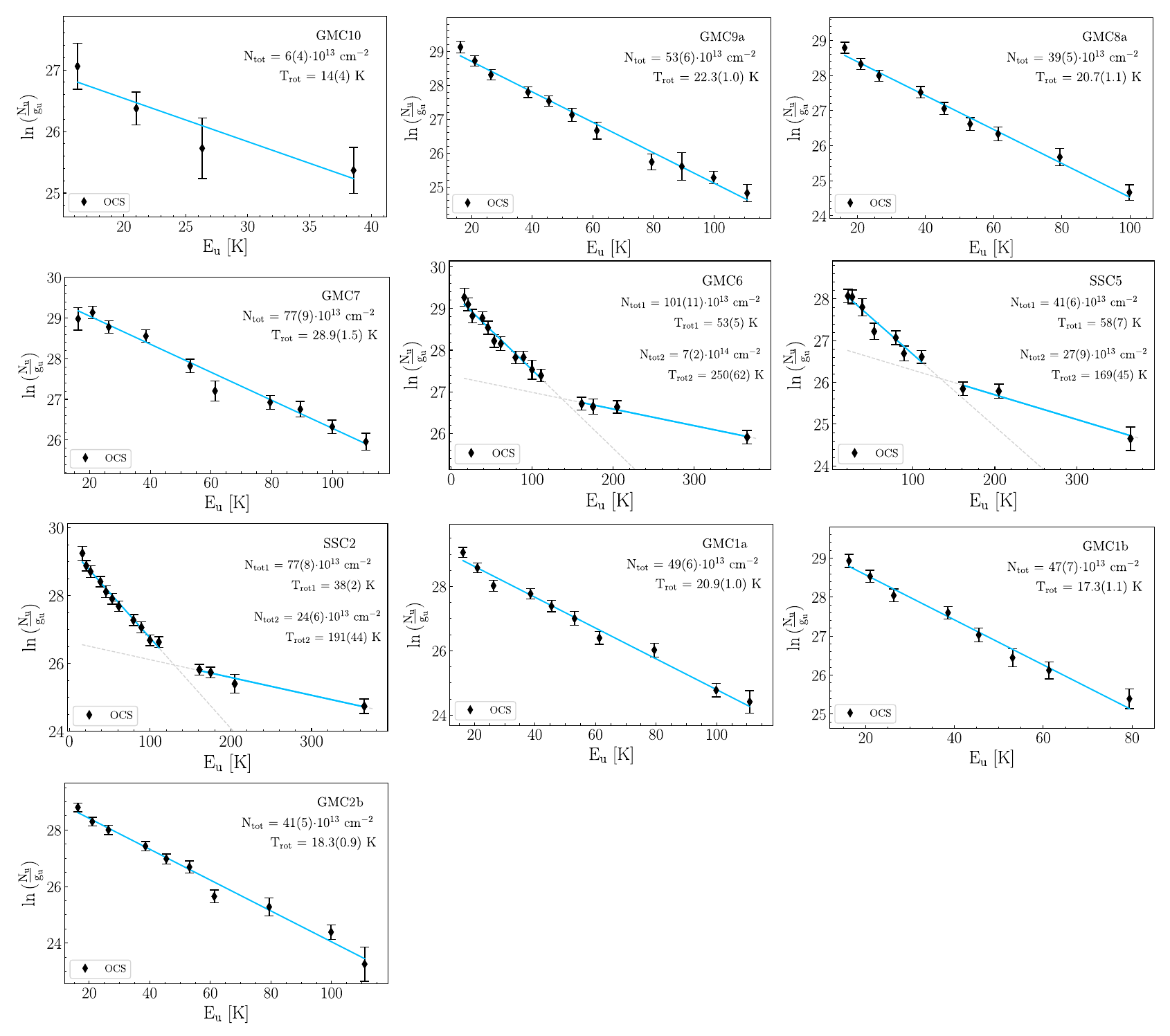}
    \caption{Rotation diagrams of OCS. The parameters $N_{\text{u}}, g_{\text{u}},$ and $E_{\text{u}}$ are the column density, degeneracy and level energy (with respect to the ground state level) of the upper level, respectively. The error bars on ln($N_{\text{u}}$/$g_{\text{u}}$) include the 15\% calibration error. The blue line represent the best fits and the dashed grey lines are the extrapolations of the fit for the full range of $E_{\text{u}}$. }
    \label{fig:RD_OCS}
\end{figure*}

\begin{figure*}
    \centering
    \includegraphics[width=\textwidth]{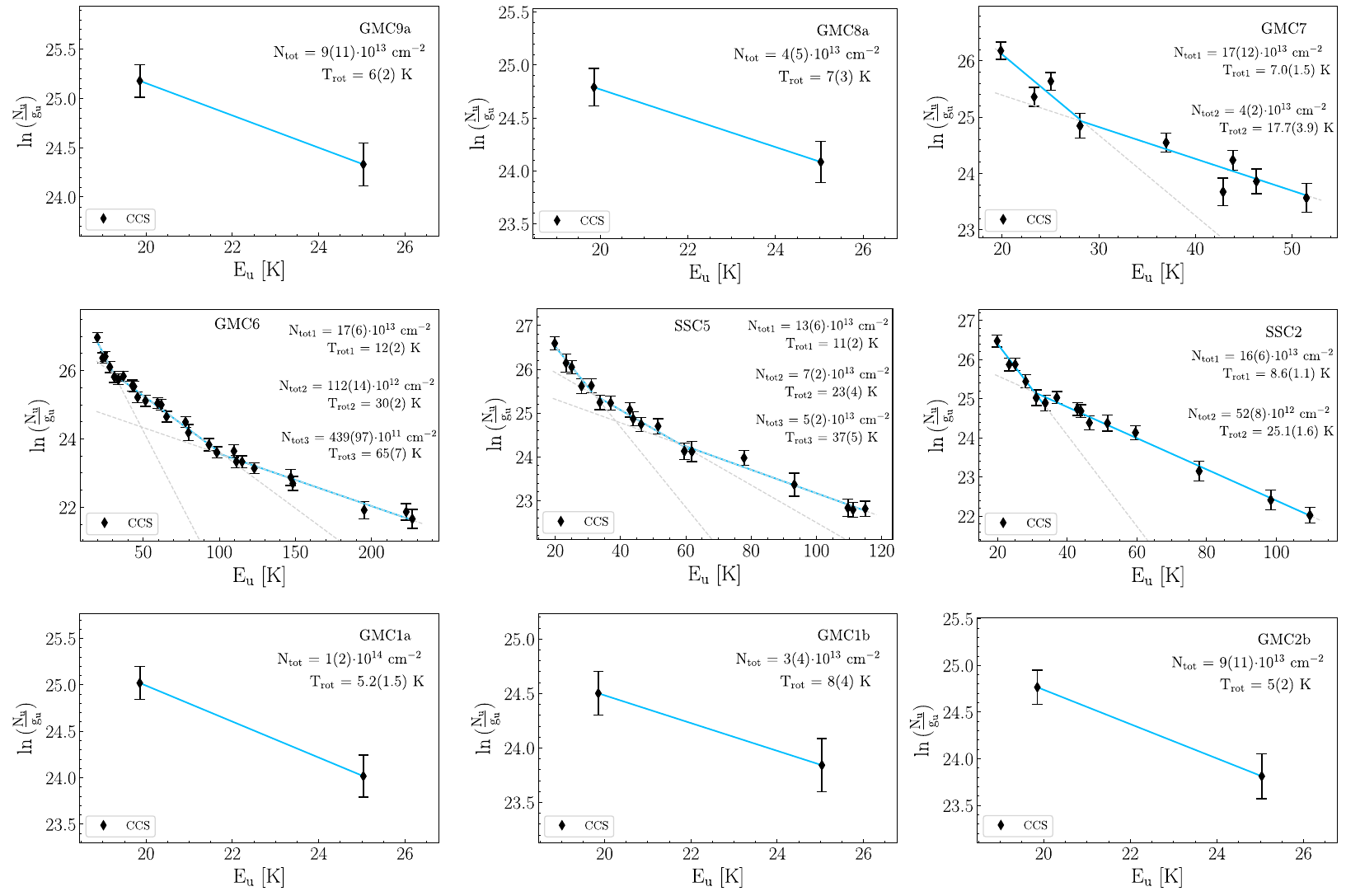}
    \caption{Rotation diagrams of CCS. The parameters $N_{\text{u}}, g_{\text{u}},$ and $E_{\text{u}}$ are the column density, degeneracy and level energy (with respect to the ground state level) of the upper level, respectively. The error bars on ln($N_{\text{u}}$/$g_{\text{u}}$) include the 15\% calibration error. The blue line represent the best fits and the dashed grey lines are the extrapolations of the fit for the full range of $E_{\text{u}}$.}
    \label{fig:RD_CCS}
\end{figure*}

\begin{figure*}
    \centering
    \includegraphics[width=\textwidth]{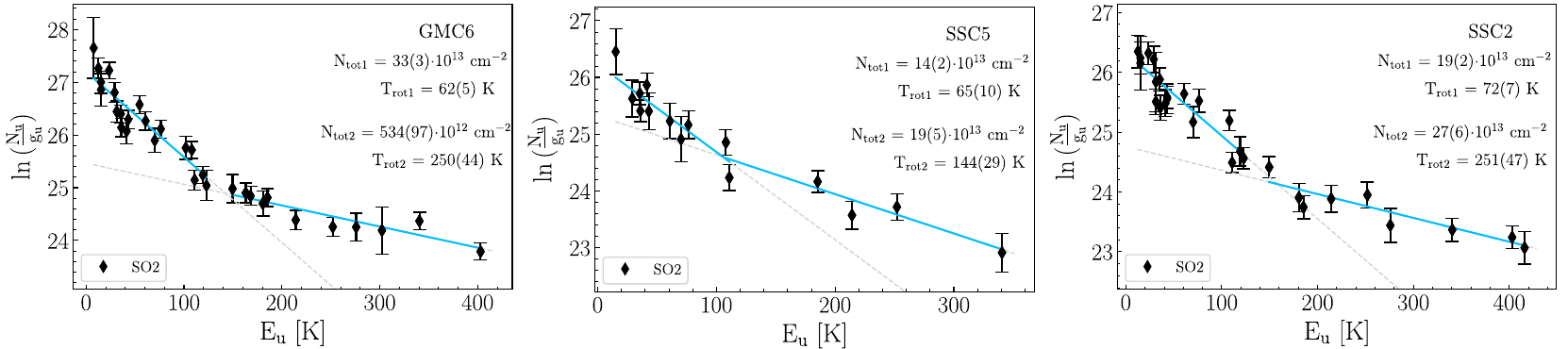}
    \caption{Rotation diagrams of SO$_2$. The parameters $N_{\text{u}}, g_{\text{u}},$ and $E_{\text{u}}$ are the column density, degeneracy and level energy (with respect to the ground state level) of the upper level, respectively. The error bars on ln($N_{\text{u}}$/$g_{\text{u}}$) include the 15\% calibration error. The blue line represent the best fits and the dashed grey lines are the extrapolations of the fit for the full range of $E_{\text{u}}$.}
    \label{fig:RD_SO2}
\end{figure*}

Figures \ref{fig:RD_OCS}, \ref{fig:RD_CCS}, and \ref{fig:RD_SO2} show the rotation diagrams OCS, CCS, and SO$_2$, respectively. The rotation temperatures ($T_{\text{rot}}$) and beam-averaged column densities ($N_{\text{beam}}$) derived are reported in Table~\ref{tab: LTE_results}.

For OCS, two rotation temperature components were needed in the three innermost regions, GMC6, SSC5, and SSC2. The first temperature component includes transitions up to $E_{\text{u}}=111$ K and ranges from $T_{\text{rot}}=10$ K in GMC10 to  $T_{\text{rot}}=65$ K in SSC5. The range of the second temperature component is $T_{\text{rot}}=124-312$ K. The beam-averaged column densities range between $(0.2-11)\times 10^{14}$ \pcms.  

For CCS, several transitions with upper-level energy ranging between 20 and 230 K were detected. For the regions GMC6, SSC5, and SSC2, we found the fit was better if we considered three rotation temperature components ($\chi^2_r= 1.5 \ \text{vs} \ \chi^2_r =0.8$). As mentioned above, the first break in rotation temperature corresponds to the shift seen in the systemic velocity for transitions at $E_{\text{u}}\sim30$ K (see Sec.~\ref{subsub:velocities}). For GMC6 and SSC5, the second break in temperature corresponds to $E_{\text{u}}\sim 50$ K. From the Gaussian fit, another shift in systemic velocity at this upper-level energy seem to be occur but the difference being of the order of the spectral resolution, we could not conclude whether this shift is significant, and we thus did not report and discuss it in Sec.~\ref{subsub:velocities}.  In the outer CMZ, only two transitions were detected. This resulted in relatively large errors, in particular for the beam-averaged column densities. Overall, the rotation temperatures range from $\leq$ 10 K in the outer CMZ to 72 K in the inner CMZ (towards GMC6) and the beam-averaged column densities are $\leq 3\times10^{14}$ \pcms. 

Finally, SO$_2$ was detected towards only the three innermost regions of the CMZ. Two rotation temperatures components were needed to fit the observations. The first and second components have $T_{\text{rot}}= (55-79)$ K and $(115-298)$ K, respectively. The beam-averaged column densities are $(1.4-6)\times10^{14}$ \pcms. 

\subsubsection{Methodology of the LVG analysis}\label{subsec:method_LVG}

As a second step, we performed a non-LTE analysis using the large velocity gradient (LVG) code \textit{grelvg}\footnote{Initially developed by \cite{ceccarelli_2003}}. The collisional excitation rates used for the non-LTE analysis, as well as the corresponding references and temperature ranges for which the coefficients are calculated, can be found in Table \ref{tab:coeffs}. The collision excitation rates were taken from the BASECOL database, except for H$_2$CS and SO$_2$ for which the rates were taken from the LAMDA database. For each region and species, we ran a large grid of models ($\geq 10,000$) varying the column density of the species, the gas density and the gas temperature. The range of values chosen are shown in Table \ref{tab:coeffs} and are large enough to cover the physical conditions in molecular clouds, outflows/shocks and hot cores, as we expect the emission from S-bearing species to be associated with either of these types of environments. For the gas temperature, we used the range of temperature for which the collisional rates were calculated for each species.

As a result of the LTE analysis (see Sec.~\ref{subsec:LTE_main} and Appx.~\ref{app:RDs}), we identified between one and three temperature ($T_{\text{rot}}$) components, depending on the species and the region. For each species, we tried to fit all the transitions together to understand whether all the transitions could probe one same gas component (and thus the multiple rotation temperature we identified in Sec.~\ref{subsec:LTE_main} are due to the presence of non-LTE and/or opacity effects) but we could not find models reproducing simultaneously the low- and high-upper-level energy transitions. We, therefore, performed the non-LTE analysis for each of these components individually. We left as free parameters the column density, $N_{\text{tot}}$, the gas temperature, $T_{\text{gas}}$, and the gas density, $n_{\text{gas}}$. For components with at least 4 lines available, we left the source size, $\theta_s$, as a free parameter. In \textit{grelvg} the sizes can vary between 0.1$\arcsec$ and 150$\arcsec$. Given that the beam size of the observations ($\theta_b$) is $1.6\arcsec$, leaving the source size as a free parameter is equivalent of varying the beam filling factor defined as $\theta_s^2/(\theta_s^2+\theta_b^2)$ from 0.004 to 1. We could leave the size as a free parameter in most cases, except when three or less transitions were available. In these cases, we fixed the source size as a first step before varying it around the fixed value to see the range of sizes for
which the solution stays the same (same temperature and density range, and same $\chi^2$). The choice of the initial fixed size is indicated in Appx.\ref{app:LVG} for the corresponding species and components. In some cases, the size could be left as a free parameter, but it did not result in constrained column densities and the best-fitted column density provided is in fact the product $\theta_s\times N_{\text{tot}}$. This can happen if multiple gas components are still mixed together and, thus, our two component assumption was not suitable. Another possibility is if all the lines are optically thin ($\tau<1$). In these cases, we ran again the best-fitting procedure, this time by fixing the source size to the best-fitted value obtained initially (i.e. the size associated with the best $\chi^2$ value obtained when leaving the size as a free parameter). We then varied the source size around its best-fit value until the degeneracy disappears, hence when the $\theta_s \times N_{\text{tot}}$ product does not give the same chi square. We had to use this method for H$_2$CS, and SO. 

Finally, we assumed the geometry to be a semi-infinite slab \citep{scoville_1974, dejong_1975}. Choosing this geometry seemed relevant in case S-bearing species are tracing mostly shocks, as in our Galaxy. We nonetheless performed some modelling using the geometry of a uniform sphere which led to similar results (not shown). The model takes into account the cosmic background radiation (CBR) field in the level population excitation, with a temperature of 2.7 K. The ortho-to-para ratio (OPR) for H$_2$ is assumed to be set to the statistical value of 3. The assumed line widths are the average ones measured from the spectral lines for each species (and component in the case of multiple component identified) in each region. We included a calibration error of 15\% in the observed intensities (see \ref{sec:obs}). The lines optical depths $\tau_L$, are also an output of the LVG analysis. Further information concerning the LVG modeling of each species (e.g. $^{32}$S/$^{34}$S ratio, OPR of H$_2$S and H$_2$CS) are presented in Appx.~\ref{app:LVG}.

\begin{table*}[ht]
    \centering
        \caption{References for the collision rates (and associated reference for the calculated potential energy surfaces on which the rates are based) used for the LVG analysis and the corresponding available range for the gas temperature, column density and gas density ranges chosen for the LVG analysis.}
    \label{tab:coeffs}
    \begin{tabular}{cccc}
    \hline \hline
        Species & $T_{\text{gas}}$ (K) & References  \\
        \hline
         CS& $5-305.5$ & \cite{denisAlpizar_2012, denisAlpizar_2018} \\
         H$_2$S(+H$_2^{34}$S)& $5-500$ & \cite{dagdigian_2020a, dagdigian_2020b} \\
         OCS & $10-100$ & \cite{green_1978} \\
         H$_2$CS &$10-300$ & \cite{troscompt_2009,wiesenfield_2013}\\
         SO (+$^{34}$SO) & $10-300$ & \cite{yang_2020,price_2021} \\
         \hline
         $N_{\text{Species}}$ (\pcms) & \multicolumn{2}{c}{$5\times 10^{12} - 10^{18}$} \\
         $n_{\text{gas}}$ (\pcmc) &\multicolumn{2}{c}{$10^{3} - 10^{8}$} \\
         \hline
    \end{tabular}
\end{table*}

\subsubsection{Results of the LVG analysis}\label{subsec:results_LVG}

\begin{figure*}
    \centering
    \includegraphics[width=\textwidth]{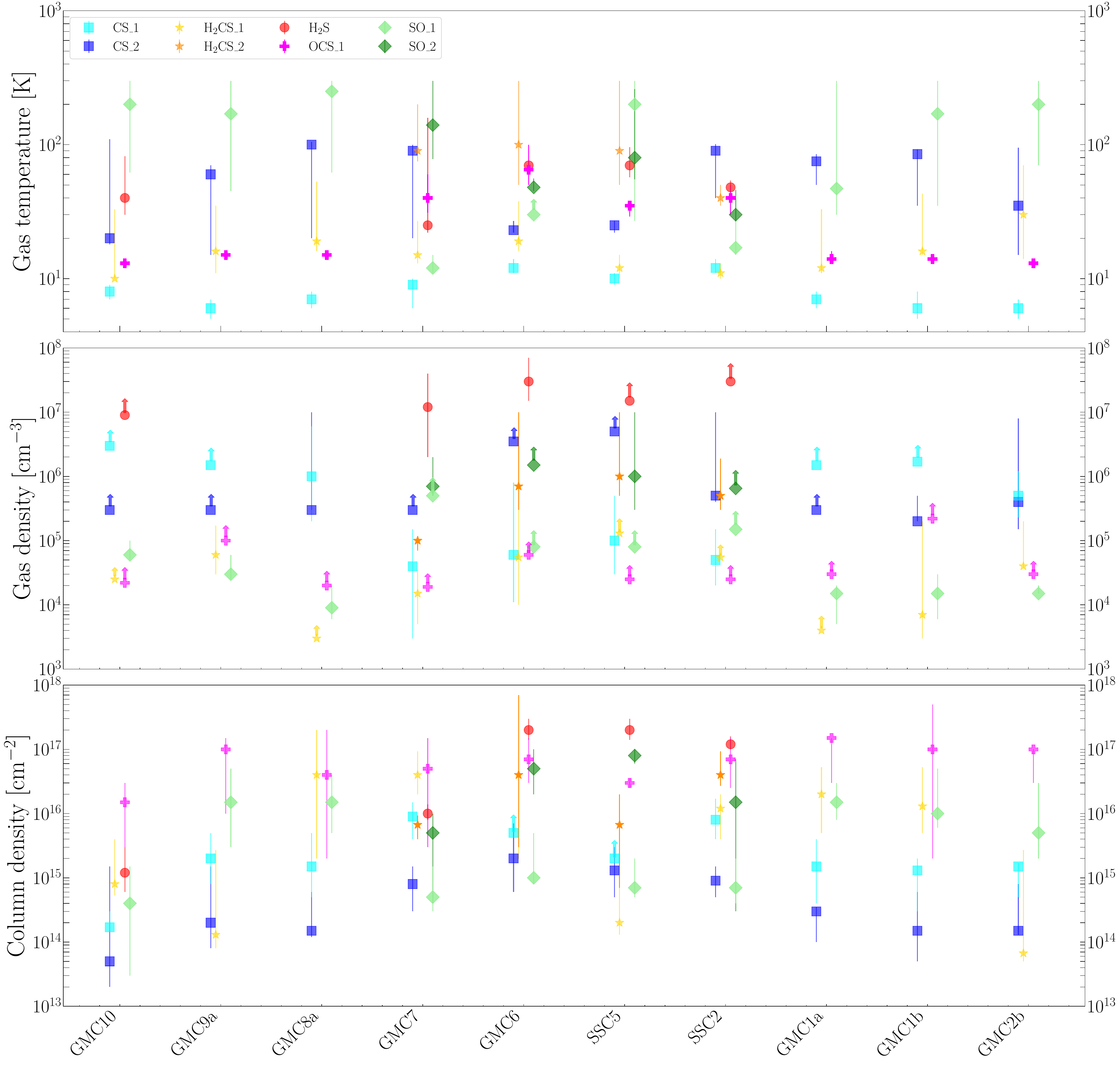}
    \caption{Results of the LVG analysis for CS (square markers), H$_2$CS (starred markers), H$_2$S (circle marker), OCS (crossed marker), and SO (diamond marker) as a function of the regions. The first and second components for CS, H$_2$CS, OCS, and SO are labeled "\_1" and "\_2", respectively. Lower limits are indicated by upward arrows. The listed sequence of clouds follows the layout of the GMCs of the CMZ as shown in Figs~\ref{fig:maps_extended} to \ref{fig:maps_ccs_so2}. From top to bottom, the results for the gas temperature, gas density, and column densities, are shown.}
    \label{fig:res_LVG}
\end{figure*}

\begin{figure*}
    \centering
    \includegraphics[width=0.85\textwidth]{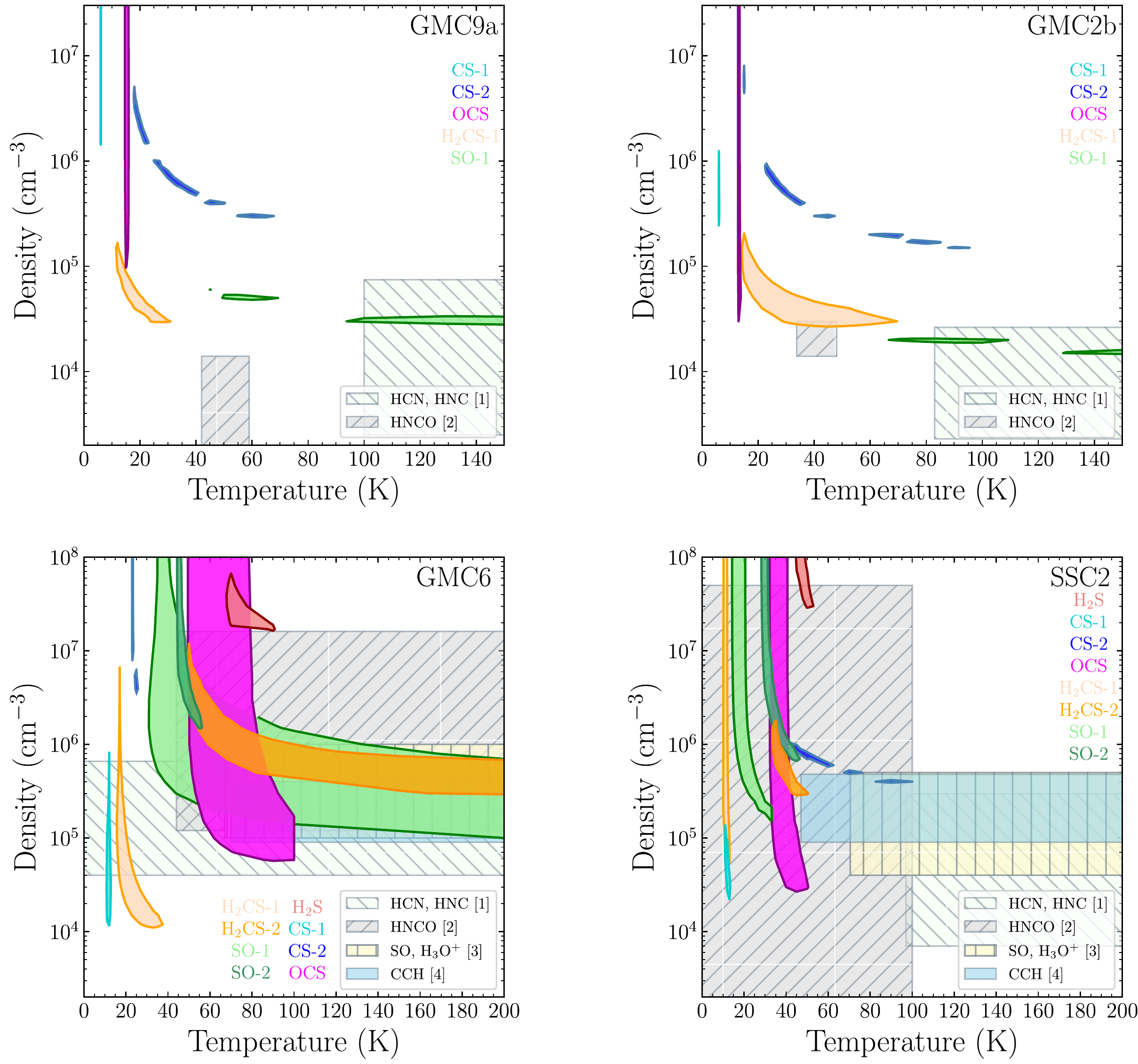}
    \caption{Density and temperature contour plots derived from the LVG analysis for representative regions in the outer CMZ (GMC9a and GMC2b) and for the inner CMZ ( GMC6, and SSC2). The contours show the $1\sigma$ solutions obtained for the minimum value of the $\chi^2$ in the column density parameter derived for each species (and components) and region. Derived gas density and parameters for other species (HCN, HNC; HNCO; SO, H$_3$O$^+$; CCH) studied in previous ALCHEMI studies were plotted for comparison: [1] \cite{behrens_tracing_2022}, [2] \cite{huang_reconstructing_2023}, [3] \cite{holdship_energizing_2022}, [4] \cite{holdship_distribution_2021}}
    \label{fig:LVG_main}
\end{figure*}

We present the results for individual species, followed by a comparison between the species across regions. Figure~\ref{fig:res_LVG} shows an overview of the derived gas temperature, gas density, and column densities for each species analysed in this Section, and for each region.  Some regions were found to be relatively similar in terms of physical conditions derived for each species. In the outer CMZ, GMC2b shows different results from the other outer regions. In the inner CMZ, SSC2 shows fairly different results from the other inner regions. Therefore, Figure \ref{fig:LVG_main} and Table \ref{Tab:LVG_results_main} summarise the results for GMC9a and GMC2b as representative regions for the outer CMZ, and GMC6 and SSC2, as representative regions for the inner CMZ. The results for the other regions are shown in Appendix~\ref{app:LVG}. 

\paragraph{\textbf{CS}}
Based on the RD analysis, we modelled two different components for CS with the LVG code, the first one comprising the N=2-1 to N=4-3 levels, and the second one comprising the levels N=5-4 to N=7-6. The column densities range between $2\times10^{13}$ and $1.7\times10^{16}$ \pcms. The lowest values are found towards GMC10, whilst the highest are found towards GMC6 and SSC5. For these regions, only a lower limit for the column density of the first component could be derived. This is because all the observed lines are optically thick, and the emission becomes that of a black body. In this case, the range of optical depths derived ($\tau \sim 4-7$) are likely lower limits to the true optical depths. The first component is very likely optically thick with high $\tau_L$ values derived for all the regions whilst the second component is likely not ($\tau_L\leq 0.8$ in all the regions). Therefore, the possible trend seen in FWHM in Sec.~\ref{subsec:linewidths} with the low upper-level energy transitions of CS having larger line widths than the other species is likely due to the line's optical depth and could thus not be related to the kinematics of the gas.The high upper-level energy transitions seem to have larger line widths compared to other species in some of the regions. The output of the LVG analysis indicates that these transitions are optically thin. However, if the emission size were to be smaller, the optical depths could be underestimated. Therefore, we cannot conclude firmly whether the higher line width for these transitions is due to the kinematics of the gas or to the line's optical depth. 
The gas temperature derived for both components is relatively constant throughout the CMZ, with values between $T_{\text{gas}}=5-14$ K, and between $T_{\text{gas}}=15-110$ K for the first and second components, respectively. The highest values are derived towards GMC6 and SSC2 for the first component. For the second component, the lower values are derived towards GMC6 and SSC5, where the temperature does not exceed 30 K. The gas density traced by CS varies depending on the component and the region. Indeed, for the first component, gas densities traced are  $n_{\text{gas}}>10^5-10^6$ \pcmc \ in the outer CMZ, and  $n_{\text{gas}}<10^6$ \pcmc \ in the inner CMZ. For the second component, the gas densities traced are  $n_{\text{gas}}>10^5$ \pcmc \ for all the regions. We could not leave the size as a free parameter (see Sec.~\ref{app:LVG}), which means that other solutions might be possible. Here, we consider the very likely extension of CS to arise on scales of GMCs ($\sim 30$ pc). For the first component, the lower limit found for the size is $\sim 1-1.5\arcsec$ for the entire CMZ whilst for the second component, a lower limit can be as low as $\sim 0.3-0.6\arcsec$ ($\sim 5-10$ pc) indicating that this component could arise from a more compact region. 
Overall, due to the low number of transitions available for each of the component, the results for CS are among the less constrained and should be taken with caution. In particular, it is not really possible to distinguish between low temperature/high density and high temperature/low density solutions for the high upper-level energy transitions which is a classical problem occurring for linear molecules.

\paragraph{\textbf{H$_2$S}}
We could derive the physical conditions of H$_2$S only in the inner CMZ and GMC10. The derived total column densities range between $8\times10^{15}-3\times10^{17}$\pcms \ in the inner CMZ, and between $(0.6-3)\times10^{15}$\pcms \ for GMC10. The gas temperatures and densities derived are constant in all the regions with $T_{\text{gas}}=30-159$ K and $n_{\text{gas}}\geq 10^7$\pcmc \ ($\geq 2\times10^6$ for GMC7), respectively. Concerning the emission size, the LVG model predicts relatively compact emissions with values ranging between 0.17 and 0.5$\arcsec$ ($\sim 3-8$ pc). Finally, the lines are optically thin in GMC10 and GMC7 ($\tau_L=0.01-0.4$), whilst they are optically thick in GMC6, SSC5, and SSC2. In these regions, the H$_2^{34}$S ($1_{1,0}-1_{0,1}$) transition is, however, optically thin. As we could not run the LVG code for the regions GMC8a, GMC1a, and GMC1b, we do not have the information on the line optical depths. Hence, we cannot conclude whether in these regions, the fact that H$_2$S seems to have a larger line width compared to the other species is a significant feature or not.

\paragraph{\textbf{H$_2$CS}}
Based on the rotation diagrams, we performed the LVG analysis for two components of H$_2$CS in the inner CMZ. The two components are comprised of transitions with $E_\text{u}<50$ and $>50$\,K, respectively. The derived column densities range between $5\times10^{13}$ and 7$\times10^{17}$ \pcms. The highest values are found towards GMC6 and SSC2 for both components. For the gas temperature, the derived temperature lies in the range 10--70 K for the first component and $\geq 35$ K for the second component. The gas densities are $\geq 3\times10^3$ \pcmc \ and between $7\times10^4$ and $10^7$ \pcmc, for the first and second components, respectively. We could leave the size as a free parameter most of the time and derived sizes of emission of $\leq 0.70\arcsec$ ($\sim 12$ pc) for most regions. In the inner CMZ, where a second component of H$_2$CS has been identified, the sizes derived are more compact than for the first component, with an average best-fit value of 0.20$\arcsec$ ($\sim 3$ pc). Finally, for the first component, the lines are usually optically (moderately) thick for all the regions but GMC2b and GMC9a, for which the lines were found to be optically thin within the range of size of emission derived. For the second component the lines are optically thin towards SSC5 and GMC7 and likely optically thick towards GMC6 and SSC2. 

\paragraph{\textbf{OCS}}
As explained in Sec.~\ref{subsec:method_LVG}, we could only use the first five transitions of OCS to derive the physical conditions of the gas emitting the first component of OCS. We assume that the results associated with these first five transitions are representative enough for the entire first component of OCS, which includes transitions up to the level N=19--18 ($E_{\text{u}}=111$ K). As we derive  gas temperature around $T_{\text{gas}}\sim 14$ K in the outer CMZ, and around $30-50$ K in the inner CMZ, which correspond to the excitation temperatures found in the rotational diagrams, our assumption holds and we are likely not underestimating the temperature for this component of OCS. The total column densities derived range between $2\times10^{15}$ and $5\times10^{17}$ \pcms. The highest values are found towards the regions of the outer CMZ. For the gas density, only a lower limit of $\sim 2\times10^4$ \pcmc \ for all regions except for GMC9a (lower limit of $10^5$ \pcmc) could be derived. The size was left as a free parameter, and the best-fit value of $\sim 0.17\arcsec$ ($\sim 3$ pc) on average is the same for the entire CMZ. Across the range of sizes derived, the lines could be optically thick as the range of optical depths lies between $\tau_L=0.1-3.0$, except for SSC5 for which the lines are always optically thin ($\tau_L=0.1-0.2$) over the range of size derived.

\paragraph{\textbf{SO}}
Both the RDs of SO and $^{34}$SO show two components in the inner CMZ. We thus ran separately the LVG code for the two components, this first one including transitions with $E_{\text{u}}<40$ K. For the first component, the derived total column densities in the range $(0.1-5)\times10^{16}$ \pcms \ are highest towards the outer CMZ (except GMC10) compared to the inner CMZ, where derived values lie in the range $(0.3-5)\times10^{15}$ \pcms. The gas temperatures are also poorly constrained with lower limits of $\sim30$
K for most of the regions. SSC2 and GMC7 have the lower range of derived gas temperature with $T_{\text{gas}}=15-40$ K and $T_{\text{gas}}=11-15$ K, respectively. For the second component, the gas temperature of $\leq 60$ K is found towards GMC6 and SSC2, whilst for GMC7 and SSC5, $T_{\text{gas}}\geq 50$ K are found. In the outer CMZ, the gas densities all lie around $n_{\text{gas}}=10^4-10^5$ \pcmc \ whilst for the inner CMZ, a lower limit for the two components of $n_{\text{gas}}=10^5$ \pcmc \ is found. For the emission size, two cases were identified between the outer and inner CMZ for the first component. In the outer CMZ, relatively compact sizes of emission were found, with $\theta_s \leq 0.50\arcsec$ ($\sim$ 8.5 pc), whilst in the inner CMZ, the emission is more extended with $\theta_s \geq 0.8\arcsec$ ($\sim 13.5$ pc). The second component of the inner CMZ is more compact with best-fit values in the range $\theta_s=0.20\arcsec$ ($\sim 3$ pc) on average. Finally, for the first component, the emission is optically thin towards the inner CMZ and GMC 10 ($\tau_L \leq 0.1$) whilst potentially optically thick in the outer CMZ where $\tau_L$ ranges between $0.1-1.6$. For the second component from the inner CMZ, the emission could be (moderately) optically thick with $\tau_L =0.1 1.4$ except for GMC7, for which the emission is optically thin ($\tau_L \leq 0.2$). The $^{34}$SO transitions are always optically thin. \\

\begin{table*}[ht]
\tiny
    \centering
    \caption{Best-fit results and $1\sigma$ confidence level range from the non-LTE LVG analysis for GMC9a, GMC6, SSC2, and GMC2b, respectively (from top to bottom).}
    \label{Tab:LVG_results_main}
    \begin{tabular}{c|c|cc|cc|cc|cc|c}
    \hline \hline
         \multirow{2}{*}{Species}& \multirow{2}{*}{Component} & $N_{\mathrm{tot}}$ & Range $N_{\mathrm{tot}}$ & $T_{\mathrm{kin}}$ & Range $T_{\mathrm{kin}}$ &$n_{\mathrm{H2}}$ & Range $n_{\mathrm{H2}}$ & Size ($\theta_s)$ &Range $\theta_s$&\multirow{2}{*}{Range $\tau_L$\tablefootmark{b}}\\
         & & (\pcms) &(\pcms) &(K) & (K)&(\pcmc) &(\pcmc) &($\arcsec$) &($\arcsec$)& \\
\hline
\multicolumn{11}{c}{GMC9a}\\
\hline
\multirow{2}{*}{CS} & 1 &$2\times 10^{15}$ &$(0.8-5)\times10^{15}$ &6 &$5-7$ &... &$\geq 1.5\times10^6$ &$1.60\tablefootmark{a}$ &$\geq 1.20$&$0.6-1.6$ \\
 & 2 &$2\times 10^{14}$ &$(0.8-15)\times10^{14}$ &60 &$15-70$ & ...&$\geq 3\times10^5$ &$1.60\tablefootmark{a}$&$\geq 0.44$ &$<0.5$\\
\hline
H$_2$CS & 1 &$1.3\times10^{14}$ &$(0.8-27)\times10^{14}$ &16 &$11-35$ &$6\times10^4$ &$(3-17)\times10^4$ &...&$\geq 0.35$ &$\leq 0.3$  \\
\hline
OCS & 1 &$1\times10^{17}$ & $(0.1-1.5)\times10^{17}$&15 &$14-16$ &... &$\geq 10^5$ &0.18 &$0.15-0.40$ &$0.1-2.0$\\
\hline
SO & 1 & $1.5\times10^{16}$&$(0.3-5)\times10^{16}$ &170 &$45-300\tablefootmark{c}$ &$3\times10^4$ &$(2.5-6)\times10^4$ &0.17 &$0.13-0.34$&$0.3-1.0$\\
\hline
\multicolumn{11}{c}{GMC6}\\
\hline
\multirow{2}{*}{CS} & 1 &$1.5\times10^{16}$ &$\geq 5\times10^{15}$ &12 &$11-14$ &$6\times10^4$ &$(1.1-80)\times10^4$ &1.60\tablefootmark{a} &$\geq 1.60$ &$\leq 6.0$ \\
 & 2 &$2\times10^{15}$ &$(0.6-7)\times10^{15}$ &23 &$22-27$ &... &$\geq 3.5\times10^6$ &1.60\tablefootmark{a}&$\geq 1.00$ &$\leq 0.8$\\
\hline
H$_2$S  & 1 &$2\times10^{17}$ &$(1.4-3)\times10^{17}$ &70 &$68-91$ &$3\times10^7$ &$(1.5-7)\times10^7$ &0.26&$0.24-0.32$&$0.2-1.9$ \\
\hline
\multirow{2}{*}{H$_2$CS} & 1 &$4\times10^{16}$ &$(0.2-20)\times10^{16}$ &19 &$16-38$ &$5.5\times10^4$ &$(0.1-70)\times10^5$ &0.22&$0.21-0.70$ &$0.1-2.8$ \\
& 2 &$4\times10^{16}$ &$(0.3-$70$)\times10^{16}$ & 100& $50-300\tablefootmark{c}$&$7\times10^5$ &$(0.3-10)\times10^6$ &0.11& $0.10-0.30$& $0.01-1.5$\\
\hline
OCS & 1 &$7\times10^{16}$ &$(3-15)\times10^{16}$ &65 &$50-100\tablefootmark{c}$ &... &$\geq 6\times10^4$ &0.17 &$0.13-0.25$ &$0.2-1.0$ \\
\hline
\multirow{2}{*}{SO} & 1 &$1\times10^{15}$ &$(0.8-5)\times10^{15}$ &... &$\geq 30$ &... &$\geq8\times10^4$ &... &$\geq 0.80$ &$\leq 0.1$ \\
 & 2 &$5\times10^{16}$ &$(2-10)\times10^{16}$ &48 & $45-56$&... &$\geq 1.5\times10^6$ &0.26& $0.17-0.42$& $0.1-0.8$\\
\hline
\multicolumn{11}{c}{SSC2}\\
\hline
\multirow{2}{*}{CS} & 1 &$8\times10^{15}$ &$(4-17)\times10^{15}$ &12 & $11-14$&$5\times10^4$ &$(2-15)\times10^4$ &1.60\tablefootmark{a} &$1.50-2.00$  & $3.0-4.0$\\
 & 2 &$9\times10^{14}$ &$(5-15)\times10^{14}$ &90 & $40-101$&$5\times10^5$ & $(0.4-10)\times10^6$&1.60\tablefootmark{a}&$\geq 1.50$ &$\leq 0.3$ \\
\hline
H$_2$S  & 1 &$1.2\times10^{17}$ &$(1-1.6)\times10^{17}$ &48 &$45-54$ &... &$\geq 3\times10^7$ &0.20&$0.18-0.22$ &$0.2-2.5$  \\
\hline
\multirow{2}{*}{H$_2$CS} & 1 &$1.2\times10^{16}$ &$(0.4-2)\times10^{16}$ & 11& $10-13$&... &$\geq5.5\times10^4$ &0.30&$0.25-0.45$ &$0.5-1.3$  \\
& 2 &$4\times10^{16}$ &$(2.7-9.3)\times10^{16}$ &40 &$35-50$ &$5\times10^5$ &$(3-19)\times10^5$ &0.10&$0.10-0.12$ &$0.2-1.5$  \\
\hline
OCS & 1 &$7\times10^{16}$ &$(2.5-15)\times10^{16}$ &40 & $30-50$&...&$\geq2.5\times10^4$ &0.17\tablefootmark{a} &$0.12-0.25$ &$0.2-3.0$ \\
\hline
\multirow{2}{*}{SO} & 1 &$7\times10^{14}$ &$(4-20)\times10^{14}$ &17 &$15-40$ & ...&$\geq1.5\times10^5$ &... &$\geq 1.00$ &$\leq0.05$ \\
& 2 &$1.5\times10^{16}$ &$(0.3-70)\times10^{15}$ &30 &$27-46$ &... &$\geq6.5\times10^5$ &...& $\geq 0.20$& $\leq1.4$ \\
\hline
\multicolumn{11}{c}{GMC2b}\\
\hline
\multirow{2}{*}{CS} & 1 &$1.5\times10^{15}$ &$(0.5-1.7)\times10^{15}$ &6 &$5-7$ &$5\times10^5$ &$(2.3-12)\times10^5$&1.60\tablefootmark{a}&$\geq1.50$ &$\leq 1.7$  \\
 & 2 &$1.5\times10^{14}$ &$(1.3-8)\times10^{14}$ &35 &$15-95$ &$4\times10^5$ &$(1.5-80)\times10^5$ &1.60\tablefootmark{a} &$0.60-2.00$ &$0.01-0.4$ \\
\hline
H$_2$CS & 1 &$6.7\times10^{13}$ &$(0.5-27)\times10^{14}$ &30 & $14-70$&$4\times10^4$ &$(3.8-20)\times10^4$ & ...&$\geq 0.3$ &$\leq 0.3$  \\
\hline
OCS & 1 &$1\times10^{17}$ &$(0.3-1)\times10^{17}$ &13 &$12-14$ &...&$\geq3\times10^4$ &0.18& $0.15-0.25$&$0.3-1.8$ \\
\hline
SO & 1 &$5\times10^{15}$&$(2-30)\times10^{15}$  &200 &$70-300\tablefootmark{c}$ & $1.5\times10^4$& $(1.5-2)\times10^4$& 0.30& $0.26-0.46$&$0.1-0.8$ \\
\hline
    \end{tabular}
    \tablefoot{
\tablefoottext{a}{Initial fixed size assumed before exploring the range of similar solution (see text Sec.\ref{subsec:method_LVG})}
\tablefoottext{b}{Average value for the optical depths of all the transitions over the range of size $\theta_s$.}
\tablefoottext{c}{Maximum limit due to the maximum temperature for which the collisional rates are available.}
}
\end{table*}

Different physical conditions were found for the gas emitting the various S-bearing species. In the outer CMZ, the first component of CS shows the coldest temperature with $T_{\text{gas}}< 10$ K, followed by OCS with $T_{\text{gas}}< 20$ K, and H$_2$CS with $T_{\text{gas}}< 50$ K, except in GMC2b where the gas temperature goes up to 70 K. Such cold temperatures for CS were already derived in a previous study performing an LVG analysis but with lower angular resolutions  \citep{bayet_extragalactic_2009}.  The second component of CS and the SO show the warmest gas temperature with $T_{\text{gas}}\geq 30$ K. In terms of gas densities, within each region, the lower ones are usually derived for H$_2$CS with $n_{\text{gas}}>10^3-10^4$\pcmc \ except for GMC2b, for which the lower values are found for SO ($n_{\text{gas}}=[1.5-2]\times10^4$ \pcmc). Then, OCS has larger density values derived with lower limits around $2\times10^4$ \pcmc, followed by CS with $n_{\text{gas}}\geq 10^5-10^6$ \pcmc). 

In the inner CMZ, the situation is different. First, comparing the gas densities, the lowest values are found for OCS and the first component of CS and of H$_2$CS with $n_{\text{gas}}\geq10^3-10^4$ \pcmc \ on average.  H$_2$S shows the largest gas density derived with $n_{\text{gas}}\geq 10^7$ \pcmc  ($n_{\text{gas}}\geq 10^6$ \pcmc \ in GMC7). Within the inner CMZ, smaller densities are found towards GMC7 for H$_2$S and the first component of CS and H$_2$CS. A striking result concerns the gas temperatures derived with values below $\sim 50$ K ($\leq 100$ K for the second component of CS) towards the region SSC2. In the remaining inner regions, the first components of CS and H$_2$CS have the lowest gas temperatures ($T_{\text{gas}} \leq 40 K$) whilst H$_2$S and the second components of SO and H$_2$CS have the highest ones ($T_{\text{gas}} > 45 K$).

Comparing with the outer CMZ, OCS, H$_2$S, H$_2$CS, and the second component of CS show relatively constant gas density throughout the CMZ, whilst the first component of CS is emitted from a less dense gas in the inner CMZ compared to the outer CMZ but this result could be a result of fixing the emission size and, hence, possibly underestimating the density of CS in the regions. Concerning the gas temperature, the first components of CS and H$_2$CS, and SO in the case of GMC7, show the coldest gas temperatures within each region. Finally, when comparing the temperatures derived in the inner CMZ to those from the outer CMZ (see Tables ~\ref{Tab:LVG_results_main} and ~\ref{Tab:LVG_results_others}), the derived gas temperatures are higher in the inner CMZ for CS (first component) and OCS, the same for CS (second component) and H$_2$CS, and lower for SO in the case of SSC2 and GMC7.


\section{Discussion} \label{sec:discussion}

For the purpose of the following discussion and to ease the reading, we will refer to the \textit{low-$E_\text{u}$} component when we discuss the component corresponding to the low upper-level energy transitions of the species, and the \textit{high-$E_\text{u}$} component when we discuss the component associated with the high upper-level energy transitions of the species. For CCS where we identified 3 components, we will refer as the \textit{low-}, \textit{mid-}, and \textit{high-$E_\text{u}$} component, respectively. Table~\ref{tab:summ_comp} summarises the range of upper-level energies associated to each component of each species, as well as the derived physical conditions ($T_{\text{gas}},n_{\text{gas}}$) across the CMZ.

\begin{table}[]
    \centering
        \caption{Description of each component for each species in terms of range of upper-level energies and physical conditions derived towards the CMZ of NGC 253. The ranges for the temperature and density cover all the positions. }
    \label{tab:summ_comp}
    \begin{tabular}{cccc}
    \hline \hline
    Species & $E_{\text{u}}$ [K]& $T_{\text{gas}}$ [K] & $n_{\text{gas}}$  [\pcmc]\\
    \hline
      H$_2$S & $28-169$ &$30-159$  &$\geq 2\times10^6$  \\
      CS low-$E_\text{u}$ &$7-23.5$ &$5-14$ &$\geq 3\times10^3$ \\
      CS high-$E_\text{u}$ &$35-66$ &$15-110$ &$\geq 1.5\times10^5$ \\
      SO low-$E_\text{u}$ &$9-34$ &$15-300$ &$\geq 5\times10^3$ \\
      SO high-$E_\text{u}$ &$35-86$ &$27-260$ & $\geq 3\times10^5$\\
      OCS low-$E_\text{u}$ &$16-111$ &$12-100$&$\geq 2\times10^4$ \\
      OCS high-$E_\text{u}$ &$161-254$ &$124-312$\tablefootmark{a}&... \\
      H$_2$CS low-$E_\text{u}$ &$10-38$ &$10-70$ &$\geq 10^3$ \\
      H$_2$CS high-$E_\text{u}$ &$59-120$ &$50-300$ &$7\times10^4-10^7$ \\
      CCS low-$E_\text{u}$ &$20-37$ & $3-14$\tablefootmark{a}& ...\\
      CCS mid-$E_\text{u}$ &$37-65$ &$19-32$\tablefootmark{a}&... \\
      CCS high-$E_\text{u}$ &$65-227$ &$33-72$\tablefootmark{a} &... \\
      SO$_2$ low $E_\text{u}$ &$8-123$&$55-79$\tablefootmark{a} &... \\
      SO$_2$ high $E_\text{u}$ &$123-416$ &$115-298$\tablefootmark{a}& ...\\
         \hline
    \end{tabular}
    \tablefoottext{a}{The temperature is the rotation temperature $T_{\text{rot}}$.}
\end{table}

\subsection{Comparison with previous ALCHEMI studies}\label{subsec: alchemi_comp}

A first step towards understanding what the S-bearing species trace is to compare the physical conditions of the gas from which they emit with the conditions derived from other species. Previous ALCHEMI studies derived physical conditions of the gas emitting C$_2$H \citep{holdship_distribution_2021}, H$_3$O$^+$ and SO \citep{holdship_energizing_2022}, HCN and HNC \citep{behrens_tracing_2022}, and SiO and HNCO \citep{huang_reconstructing_2023}. From these studies, it was found that C$_2$H, SO, H$_3$O$^+$ and the HCN/HNC abundance ratio were all sensitive to cosmic rays (CRs), C$_2$H, SO and H$_3$O$^+$ being enhanced with the cosmic-ray ionization rate (CRIR), and the HCN/HNC abundance ratio being anti-correlated with the CRIR. On the other hand, \cite{huang_reconstructing_2023} studied the low- and high-velocity shock tracers, HNCO and SiO, respectively, and found that shocks were present in most of the GMCs. To compare this with the S-bearing species, we reported the derived gas density and temperature of C$_2$H, H$_3$O$^+$ and SO, HCN and HNC, and HNCO in Figures \ref{fig:LVG_main} and \ref{fig:LVG_others}. The physical conditions of the gas from which SiO is emitted could not be compared with our results, as the gas density is usually lower ($\sim 10^2-10^4$ \pcmc) and the gas temperature much higher ($\sim 500-600$ K) than what we derived here for the S-bearing species.

In the outer CMZ, the derived gas density and temperature for SO usually correlates well with that derived for HCN and HNC. For GMC 10, the gas density traced by SO is reaching that of HCN and HNC at the very high range of the gas temperature ($T_{\text{gas}}>200$ K). No clear trend is seen between the other S-bearing species and HNCO. For C$_2$H and H$_3$O$^+$/SO,, studies were performed only for GMC3 to GMC7. 

In the inner CMZ, the gas properties traced by the S-bearing species are quite different between regions. First, OCS and H$_2$S trace the same gas conditions than HNCO, except for SSC5 in the case of OCS. However, the result towards SSC5 could have been impacted by the fact that we had to fix the emission size due to the low number of OCS lines available (because of blending) in this region. The low-$E_\text{u}$ component of CS and H$_2$CS also show the same gas physical conditions across the inner CMZ and do not seem to consistently trace the same type of gas of any of the tracers from the previous ALCHEMI studies. Then, towards GMC6 and SSC5, the high-$E_\text{u}$ component of CS does not trace the same gas as any other tracer whilst same physical conditions with multiple tracers are found towards SSC2 and GMC7 (e.g. C$_2$H, HNCO). The high-$E_\text{u}$ component of H$_2$CS and SO also shows different behaviour as they probe the same gas conditions as HNCO in SSC2, SSC5, and GMC6 but not towards GMC7.  
Finally all the S-bearing species and components in SSC2 show similar physical conditions to those of HNCO, which could indicate that this region is largely governed by slow shocks. The region seems to be indeed located at the position of the south-west (SW) streamer associated with the large-scale outflow of the galaxy \citep[e.g.][]{walter_dense_2017, zschaechner_2018, krieger_2019}. Enhancement of shocks at the position of SSC2 are likely occurring coherently with the location of the base of the SW streamer (e.g. Bao et al. submitted).

We note the discrepancy between the physical parameters we derive for SO towards GMC7 and SSC2 and what was previously found in \cite{holdship_energizing_2022}. This could result from the fact that we performed different LVG analyses (SO versus H$_3$O$^+$/SO) and that we did not take into account the SO transitions with $A_{\text{ij}}\sim 10^{-6}$ s$^{-1}$ as they could not be well reproduced by the LVG models (see Sec. \ref{app:LVG_SO}). \\

Overall, from comparing the physical conditions of the gas traced by the various S-bearing species with that of previous ALCHEMI studies, we can get a first hint of the type of gas traced by some of the S-bearing species. Indeed, in the outer CMZ, SO traces the same type of gas (same gas density and temperature) as HCN and HNC which are well probing the GMCs of NGC 253 \citep{leroy_alma_2015}. In the inner CMZ, SSC2 could be a region where shocks are relatively prominent compared to the other regions, leading to the release/formation of the sulphur-bearing species  mostly via non-thermal desorption due to the shocks. Throughout the inner CMZ, H$_2$S and OCS show similar gas density and temperature as the low-velocity shock tracer HNCO. This is also the case for the high-$E_\text{u}$ component of H$_2$CS and SO (also the low-$E_\text{u}$ component of SO but the solutions are unconstrained in GMC6 and SSC5), except in GMC7. As none of the S-bearing species trace the same gas as SiO, it is likely that strong and high-velocity shocks are not at the origin of the release of S-bearing species but rather low-velocity and milder shocks (traced by HNCO; \citealt{huang_reconstructing_2023}).  None of the S-bearing species seems to trace consistently the same gas as C$_2$H.

\subsection{Comparison with Galactic environments}\label{subsec:Galactic_comp}

A second step towards understanding the origin of emission of the S-bearing species in the CMZ of NGC 253 is to compare our results with what is found in different Galactic environments. In this work, we focused on the most abundant S-bearing species typically found in Galactic star-forming regions \citep[e.g.][]{hatchell_survey_1998, li_sulfur-bearing_2015, widicus_weaver_deep_2017, humire_2020}. As other S-bearing species than those investigated in this work are detected towards NGC 253 (e.g. NS, SO$^+$; \citealt{martin_alchemi_2021}), we cannot determine the total S-budget of NGC 253. Therefore, we cannot accurately determine the level of S depletion in NGC 253 and assess how it compares to what is found in various Galactic environments. In order to remove this uncertainty, we compare various S-bearing abundance ratios. 

We selected the abundance ratios depending on the region of emission of the S-bearing species constrained by the LVG analysis in Sec.~\ref{subsec:results_LVG}. For H$_2$S, OCS (low-$E_\text{u}$ component), H$_2$CS (both low- and high-$E_\text{u}$ components), and SO (both low- and high-$E_\text{u}$ components, in the outer and inner CMZ, respectively), we derived a range for the emission size relatively similar, in the range ($\sim 0.1\arcsec-0.5\arcsec$). To compare with the Galactic environments, we investigated all the possible abundance ratios involving these six species or their low-/high-$E_\text{u}$ components and selected those which showed the most interesting trends helping us understand their region of emission. The selected ratios are: [H$_2$S]/[OCS], [H$_2$S]/[SO], [H$_2$S]/[H$_2$CS], [OCS]/[H$_2$CS], [OCS]/[SO], and [H$_2$CS]/[SO]. For the other species, CS (both low- and high-$E_\text{u}$ components), SO (low-$E_\text{u}$ component in the inner CMZ), CCS, SO$_2$, and OCS (high-$E_\text{u}$ component), the emission size were either unconstrained, or assumed to be extended (for the RDs). For these species, after investigating the various abundance ratios available, we selected the following ones: [CS]/[CCS], [SO]/[SO$_2$], [CS]/[SO], [OCS]/[SO$_2$], and [OCS/CS]. The range of values for each abundance ratios and for each region of NGC 253 are shown in Table~\ref{tab:abundance_ratios}.

For the choice of Galactic regions, we selected those most relevant to compare with the starburst CMZ of NGC 253. This includes regions dominated by shocks and/or outflows, Hot cores, PDRs, and molecular clouds. We would like to raise a word of caution as we are comparing GMC-size scales in NGC 253 with parsec and sub-parsec scales in the various Galactic environments. Nonetheless, by performing this comparison, we aim to perform a first order comparison of the abundance ratios found towards the CMZ of NGC 253 and those derived in different Galactic environments. We selected the regions for which the column densities or abundances of several S-bearing species relevant to this work were available. 

For the shocks/outflows dominated regions, we selected the prototypical L1157-B1 bow shock, a well-known shocked region within the protostellar outflow of L1157-mm, and the Orion-KL (Kleinmann-Low) Plateau regions located in the well-studied Orion high-mass star-forming region, which is a mixture of shocks, outflows, and interactions with the ambient cloud \citep{blake_1987, wright_1996, lerate_2008, esplugues_line_2013}. These two different shocked/outflow regions are representative of two different shock scenarios: the L1157-B1 bow shock could be representative of sporadic shocks due to star formation, whilst the Orion-KL plateau could be representative of the interaction between outflows and an ambient cloud, hence faster (up to 100 km.s$^{-1}$ for Orion Plateau; e.g. \citealt{odell_2001, goddi_2009} vs 20--40 km.s$^{-1}$ for L1157-B1; e.g. \citealt{lefloch_2010,viti_2011,benedettini_2013, benedettini_2021}) and more turbulent shocks. These two scenarios are part of the three shock scenarios \cite{huang_reconstructing_2023} proposed to explain the SiO and HNCO emission in the CMZ of NGC 253. If the abundance of S-bearing species varies between these two types of shocked regions, we could put constraints on which of the shocked scenarios proposed by \cite{huang_reconstructing_2023} is the more likely to happen in the CMZ of NGC 253.

Concerning the Hot Cores (HCs), we selected the Orion KL \citep{feng_2015, luo_sulfur-bearing_2019}, Sagittarius B2 N core \citep{neill_herschel_2014}, and Cyg X-N12 \citep{el_akel_unlocking_2022}. For Cyg X-N12, we used the column densities derived for the warm chemistry  (Table 5 in \citealt{el_akel_unlocking_2022}). Cyg X-N12 is close to another hot core, Cyg X-N30, for which abundances of sulphured species were also derived. However, the authors classified Cyg-X N30 as a "non-traditional" hot core as the molecular richness towards this source are likely the result of the combination of processes rather than only the hot core chemistry \citep{vanderwalt_2021}. We thus did not include this source in our comparison.  

For PDRs, we included only highly UV-irradiated PDRs. These types of PDRs are particularly relevant for our study as we expect high-UV irradiation to occur towards the CMZ of NGC 253 due to the intense star formation. The presence of strong PDRs was deduced by \cite{harada_starburst_2021} using the HCO$^+$/HOC$^+$ abundance ratios. We thus selected the Orion Bar PDR \citep{jansen_1995, leurini_apex_2006} and Monoceros R2 (Mon R2; \citealt{ginard_spectral_2012}) PDRs.

Finally, for the molecular clouds, we selected two different types of clouds. On the one hand, we selected prototypical dark clouds such as L1544 \citep{vastel_sulphur_2018}, L483 \citep{lattanzi_molecular_2020}, and B1-b \citep{fuente_ionization_2016}. On the other hand, we compare with Galactic Center clouds (towards Sgr B2 and A molecular complexes; \citealt{nummelin_2000}, \citealt{martin_sulfur_2005} and references therein, \citealt{armijos-abendano_3_2015}). Choosing these two types of clouds allows us to compare our results with cold ($T_{\text{kin}} \sim 10$ K) and dense ($n_{\text{gas}} \geq 10^5$\pcmc) quiescent clouds (Dark clouds) and with Galactic Center clouds which are much warmer ($T=50-200$ K) with an average density of $n_{\text{gas}} \sim 10^4$\pcmc \ \citep{guesten_2004}. 

\begin{figure*}
    \centering
    \includegraphics[width=\textwidth]{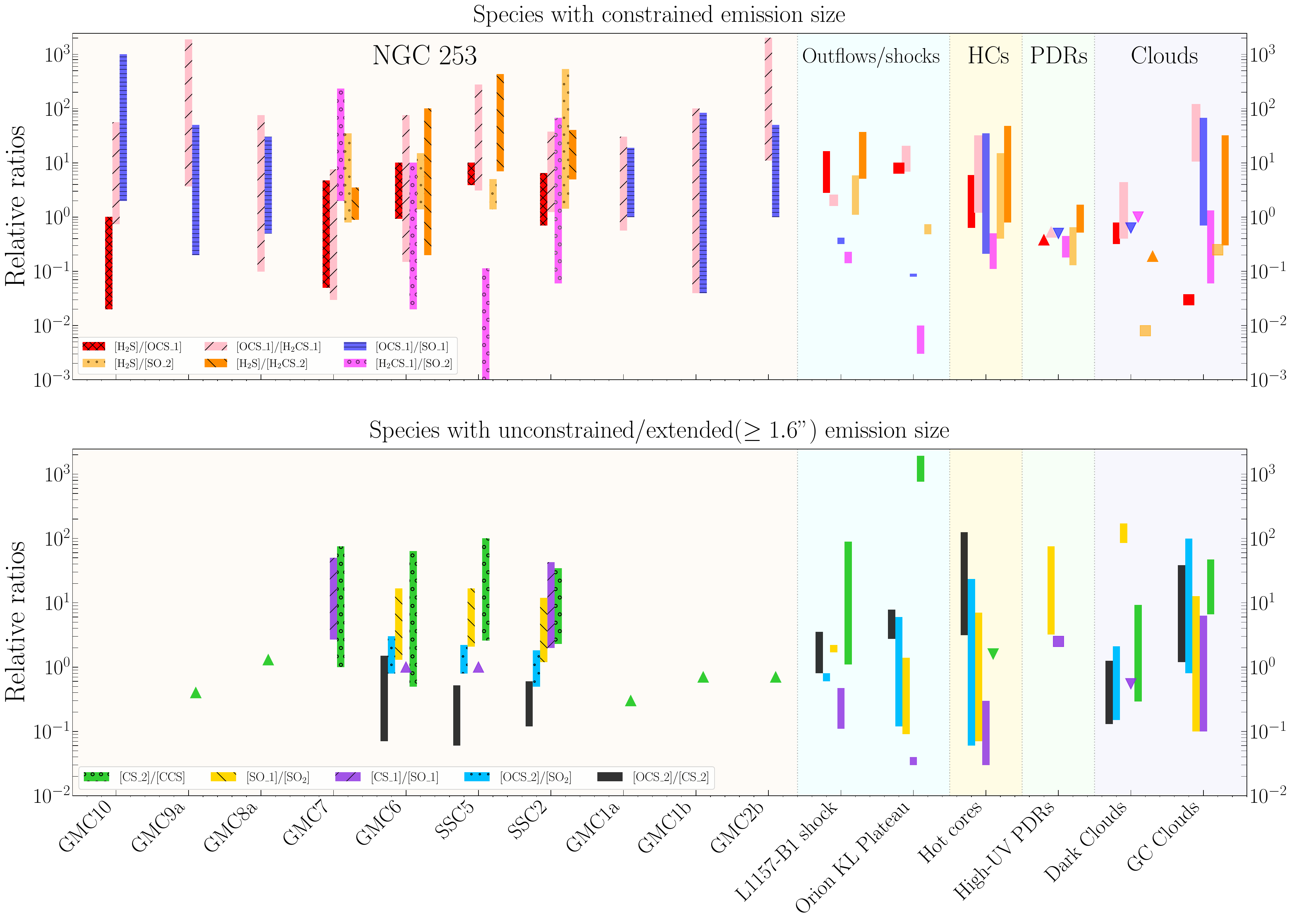}
    \caption{Abundance ratio of the S-bearing species in each region of NGC 253 (red shaded area), and in various  Galactic environments. For NGC 253, the listed sequence of clouds follows the layout of the GMCs of the CMZ as shown in Figs.~\ref{fig:maps_extended} to \ref{fig:maps_ccs_so2}. The ratios are between species for which the emission size is similar, and is constrained (\textit{top panel}) or unconstrained/extended ($\geq 1.6\arcsec$; \textit{bottom panel}). The underscore numbers 1 and 2 refer to the low- and high-$E_\text{u}$ component, respectively. For the outflows and shocks (blue shaded area), we compare with the prototypical shock L1157-B1 \citep[][and references therein]{ holdship_sulfur_2019} and the Orion KL Plateau \citep{persson_spectral_2007, tercero_line_2010, esplugues_line_2013,crockett_herschel_2014a, crockett_herschel_2014b,gong_2015}. The hot cores (HCs; blue shaded area) are the Orion KL hot core \citep{feng_2015, luo_sulfur-bearing_2019}, Sgr B2 (N) \citep{neill_herschel_2014}, and Cyg X-N12 \citep{el_akel_unlocking_2022}. The PDRs (yellow shaded region) are the Orion Bar \citep{jansen_1995,leurini_apex_2006} and Mon R2 \citep{ginard_spectral_2012}, two high-UV irradiated PDRs. Finally, the molecular clouds category (purple shaded area) comprises dark clouds on the one hand (L1544, L483, and B1-b;  \citealt{fuente_ionization_2016,vastel_sulphur_2018,lattanzi_molecular_2020}), and Galactic Centre (GC) Clouds on the other hand (\citealt{nummelin_2000}, \citealt{martin_sulfur_2005} and references therein, \citealt{armijos-abendano_3_2015}). Lower and upper limits are marked with filled triangles. The legend on the bottom part of the plot is associated to NGC 253. For Galactic environments, the same color code applies.}
    \label{fig:Comp_abundances}
\end{figure*}

Figure \ref{fig:Comp_abundances} shows the various abundance ratios derived for the S-bearing species in NGC 253 as a function of the selected regions and for various Galactic environments described above. For CCS and SO$_2$, as we could not confirm whether the multiple temperature component is correct are not due to non-LTE or optically thick effects, we used the lowest and highest value for the beam-averaged column density derived across the components to calculate the abundance ratios. We listed below the main results from this comparison:

\begin{itemize}
    \item The [H$_2$S]/[OCS\_1] abundance ratios of the inner regions are similar to those found in  HCs, PDRs, and the two outflows/shocks regions. The ratio seems to increase towards SSC5.
    \item The [H$_2$S]/[SO\_2] and [H$_2$S]/[H$_2$CS\_2] abundance ratios are similar to what is found in the L1157-B1 shock, HCs, and clouds (only the case for [H$_2$S]/[H$_2$CS\_2]). Whilst [H$_2$S]/[SO\_2] remains constant across the inner regions, [H$_2$S]/[H$_2$CS\_2] is the highest towards SSC5 and the lowest towards GMC7.
    \item The [OCS\_1]/[H$_2$CS\_1] ratio seems to increase towards the inner CMZ. In the outer CMZ, the ratios derived agree with what is found in most Galactic regions. In the inner CMZ,  the ratio is different to what is found in GC clouds in GMC7, whilst it is different to what is found in the L1157-B1 shock towards SSC5. 
    \item The [OCS\_1]/[SO\_1] ratio mostly agrees with what is found in HCs and GC clouds.
    \item The [H$_2$CS\_1]/[SO\_2] ratio decreases towards SSC5, where the ratio is similar to what is found in the Orion KL Plateau region. In GMC6 and SSC2, the ratio is similar to those found in the Galactic regions, except the Orion KL Plateau region. In GMC7, the ratio is higher than what is found in the Galactic environments.  
    \item The [CS\_2]/[CCS] ratio is unconstrained in the outer regions due to the lower limit in beam-average column density derived for CCS (see Sec.~\ref{subsec:LTE_main}). However, in the inner regions, this ratio is constant and agrees well with what is derived in the L1157-B1 shock and the molecular cloud regions.
    \item The [SO\_1]/[SO$_2$] ratio in NGC 253 is similar to most of the Galactic regions, except for the dark clouds and the Orion KL Plateau region.
    \item The[CS\_1]/[SO\_1] abundance ratio agrees well with what is found in PDRs and GC clouds.
    \item The [OCS\_2]/[SO$_2$] and [OCS\_2]/[CS\_2] ratios are both constant in the three innermost regions of NGC 253. The [OCS\_2]/[SO$_2$] ratio agrees with all Galactic regions, whilst the [OCS\_2]/[CS\_2] ratio mostly agrees with dark clouds and the L1157-B1 shock (only towards GMC6).
\end{itemize}

\subsection{Assembling the puzzle: what do S-bearing species trace?}\label{sec:puzzle}

After comparing the physical conditions of the gas emitting the S-bearing species in Sec.~\ref{subsec: alchemi_comp} and comparing their abundance ratios with prototypical Galactic regions in Sec.~\ref{subsec:Galactic_comp}, we can attempt to understand what S-bearing species trace in the CMZ of NGC 253. 

\subsubsection{Primary shock tracers}\label{subsec:shock_tracers}

First, from Sec.~\ref{subsec:Galactic_comp}, we found that the [H$_2$S]/[OCS\_1] and [H$_2$S]/[SO\_2] abundance ratios all agree with the abundance ratios found in the L1157-B1 shock and HCs. This suggests that H$_2$S, the low-$E_\text{u}$ component of OCS, and the high-$E_\text{u}$ component of SO emit from either a shocked gas due to star formation or a warm gas due to the presence of proto-SSCs in the inner CMZ.

In hot cores, we expect temperatures of a few hundred Kelvins, much higher than the temperatures derived in this study for H$_2$S and the low-$E_\text{u}$ component of OCS. Additionally, the gas temperature derived for these species is lower than their desorption temperature ($T_{\text{des}}$) ($T_{\text{des}} > 100$ K for H$_2$S and $T_{\text{des}}\sim 150$ K for OCS; e.g. \citealt{collings_2004, jimenez-escobar_sulfur_2011, jimenez-escobar_2014, cazaux_photoprocessing_2022}) in most regions. If we assume that H$_2$S and the low-$E_\text{u}$ component of OCS trace the same gas component in all the regions of the CMZ, then we can conclude that thermal desorption is not the main process explaining the H$_2$S emission in the CMZ of NGC 253. On the other hand, the physical conditions of H$_2$S and the low-$E_\text{u}$ component of OCS are similar to those of HNCO (only in the inner CMZ for OCS) and the derived emission size of these two species, which is a few parsec (3-8 pc) scale, are consistent with a shock origin. This scenario can be further strengthened for OCS: the derived gas temperatures are lower in the outer CMZ ($T_{\text{gas}}<20$ K) compared to the inner CMZ ($T_{\text{gas}}=30-50$ K), indicating that the shocks traced by OCS could be older in the outer CMZ as the gas needs time to cool down after the shock passage, consistent with the shock timescale derived by \cite{huang_reconstructing_2023}, where older shocks are present towards the outer CMZ and younger shocks towards the inner CMZ. Finally, although the physical conditions of the gas emitting OCS and HNCO are different in the outer CMZ, this discrepancy could be explained by a difference in the cooling timescale of the two species. Indeed, since the gas densities derived for HNCO are, on average, lower than those derived for OCS ($n_{\text{gas}}\leq 2\times10^4$ \pcmc \ vs $\geq 2\times 10^4$\pcmc \ for OCS), the gas traced by HNCO would cool slower than the gas traced by OCS. Hence, HNCO would thus trace a warmer gas than OCS, as in the case of the outer CMZ. Another scenario to explain the difference in gas physical conditions of OCS and HNCO in the outer CMZ could be that the two species probe different types of shocks (e.g. sporadic shocks, outflow-induced shocks, cloud-cloud collision), as proposed by \cite{huang_reconstructing_2023}. 

Then, we can also conclude a shock origin for the high-$E_\text{u}$ component of SO. The LVG analysis showed that this SO component has a high density ($10^5-10^6$ \pcmc). The gas temperature is either lower than $\sim 60$ K (GMC 6 and SSC2) or higher than 60 K (GMC7 and SSC5). Finally, the species arise from a compact region ($\sim 3$ pc on average) and the physical conditions are usually similar to those derived for HNCO in most regions. SO is known to be a post-shocked gas tracer as it is a product of the subsequent reaction of H$_2$S with OH in the gas phase \citep[e.g.][]{pineaudesforets_1993, charnley_sulfuretted_1997, viti_chemical_2001}. More recently, \cite{taquet_seeds_2020} suggested that SO could be abundant both in the front shock and behind the shock. The absence of this component in the outer CMZ would indicate that SO is present in the front shock and has not yet been fully converted to SO$_2$. Additionally, SO is thought to co-desorb with H$_2$O from the ice grain mantles (e.g. \citealt{viti_evaporation_2004}) so thermal desorption of SO is unlikely to occur in GMC6 and SSC2 where the gas temperature is lower than 100 K. If we assume SO is tracing the same feature throughout the inner CMZ, then this SO component is likely tracing shocks.

Finally, we also propose that the low-$E_\text{u}$ component of H$_2$CS traces shocks. As OCS, H$_2$S, and the high-$E_\text{u}$ component of SO, the low-$E_\text{u}$ component of H$_2$CS emits on $\sim 3$ pc scales. In the outer CMZ, the physical conditions traced by H$_2$CS are close to that of OCS, although H$_2$CS shows a lower density most of the time ($\geq 10^4$\pcmc \ for OCS and $\sim 10^3-10^5$\pcmc \ for H$_2$CS) and a larger range of gas temperature ($T_{\text{gas}}=10-70$ K for H$_2$CS versus $T_{\text{gas}}=12-16$ K for OCS).
However, whilst OCS and H$_2$S are known to trace mainly front shock events as they would directly come off the ice mantles of grains (e.g. \citealt{hatchell_survey_1998, podio_molecular_2014, taquet_seeds_2020}), H$_2$CS rather traces post-shocked gas as it would be formed in the gas phase after the release of H$_2$S and OCS \citep[e.g.][]{millar_1990, charnley_sulfuretted_1997, hatchell_survey_1998, codella_chemical_2005}. The abundance ratios [OCS\_1]/[H$_2$CS\_1] and [H$_2$CS\_1]/[SO\_2] do not give much constrain as the ratios are similar to those found in most Galactic regions, but they indicate that the amount of low-$E_\text{u}$ component of H$_2$CS decreases towards the inner CMZ, where shocks are younger \citep{huang_reconstructing_2023}, which is consistent with H$_2$CS tracing the post-shocked gas rather than the front shocked gas. The difference seen in the systemic velocities of the low-$E_\text{u}$ component of H$_2$CS compared to the rest of the species (see Sec.~\ref{subsub:velocities}), also strengthens this scenario. 

In summary, H$_2$S, the low-$E_\text{u}$ component of OCS and H$_2$CS and the high-$E_\text{u}$ component of SO likely trace shocks, although they do not seem to trace the same shock component or event based on the information we could gather. However, none of these S-bearing species are probing fast and, thus, strong shocks since they do not trace the same gas as SiO, a well-known fast shock tracer \citep[e.g.][]{schilke_1997, gusdorf_2008a, kelly_2017}, and as most of the abundance ratios investigated here agree on average with what is found in the L1157-B1 shock rather than in the Orion KL Plateau. In the region SSC5, however, some of the ratios agree better with what is found towards the Orion KL Plateau which could indicate the presence of stronger shocks (likely due to a more intense star formation activity) compared to the rest of the regions. Overall, our results seem to agree with a scenario where H$_2$S, the low-$E_\text{u}$ components of OCS, H$_2$CS, and the high-$E_\text{u}$ component of SO probe shocks which are likely associated with star formation in both the outer and inner CMZ, although star-formation is known to be particularly intense in the inner CMZ where most of radio continuum sources are located \citep[][]{ulvestad_1997}. Our results are consistent with previous low-angular resolution studies of S-bearing species which concluded that S-bearing species are associated with low-velocity shocks similar to those found in the Sgr B2 molecular cloud complex, where star-formation is ongoing \citep{martin_first_2003, martin_sulfur_2005}. Nonetheless, performing chemical modelling will be crucial to confirm these results and assess more accurately the nature of the shocks traced by these species. 

\subsubsection{Primary molecular cloud tracers}

The [OCS\_1]/[SO\_1] abundance ratio derived in the outer CMZ is constant and agrees consistently with what is found in hot cores and GC clouds. However, we just determined that the low-$E_\text{u}$ component of OCS arises likely from shocks, indicating that the low-$E_\text{u}$ components of OCS and SO do not trace the same gas. The physical conditions of the gas emitting the low-$E_\text{u}$ component of SO correlate well with that traced by HCN and HNC \citep{behrens_tracing_2022}, which are well-known dense gas tracers of the GMCs of NGC 253 \citep{leroy_alma_2015}. From the LVG analysis, this component of SO emits from a warm ($\geq 40$ K) and moderately dense ($n_{\text{gas}}\sim 10^4-10^5$\pcmc) gas in most regions. The physical conditions derived are similar to that of the high-density component of the NGC 253 GMCs ($T_{\text{gas}}\sim 85$ K and $n_{\text{gas}}\sim 10^4$ \pcmc) derived by \cite{tanaka_2023} suggesting that SO traces the dense clouds, in particular the compact regions ($\leq 0.3-0.4\arcsec$ or $\leq 5-7$pc) as derived with the LVG analysis in the outer CMZ. This conclusion agrees with a previous study of SO towards NGC 253 by \cite{holdship_energizing_2022}, who concluded that the SO abundance was unlikely driven by PDRs or shocks. Finally, the physical conditions of the low-$E_\text{u}$ component of SO in the inner CMZ being different than those in the outer CMZ (cooler gas temperature and larger emission region) could either indicate a different origin of emission of SO in the inner CMZ or could result from the influence of the higher concentration of CRs in the inner towards the inner CMZ \citep{mangum_fire_2019, behrens_tracing_2022}, which are known to destroy SO \citep{bayet_2011, holdship_energizing_2022}.

The low-$E_\text{u}$ component of CS also very likely traces the dense gas from molecular clouds. The [CS\_1]/[SO\_1] abundance ratio is similar to that found in the PDRs and GC clouds. The low-$E_\text{u}$ component of CS emits from a cold ($< 20$ K) and dense ($\geq 10^4$\pcmc) gas on GMC scales ($\theta_s\geq 1\arcsec$ or $\sim 17$ pc) which would favor the molecular cloud origin rather than the PDR one. These results needs to be taken with caution as they originate from the fact that we initially fixed the emission size to 1.6$\arcsec$ (see Sec.~\ref{subsec:results_LVG}). If the emission region were smaller, the physical conditions derived could differ. Nonetheless, our results suggest that the low-$E_\text{u}$ component of CS traces the dense GMCs of NGC 253, thus agreeing with previous observations \citep[e.g.][]{bayet_extragalactic_2009, martin_photodissociation_2009, aladro_2011, leroy_alma_2015, holdship_energizing_2022}. 

\begin{figure*}
    \centering
    \includegraphics[width=\linewidth]{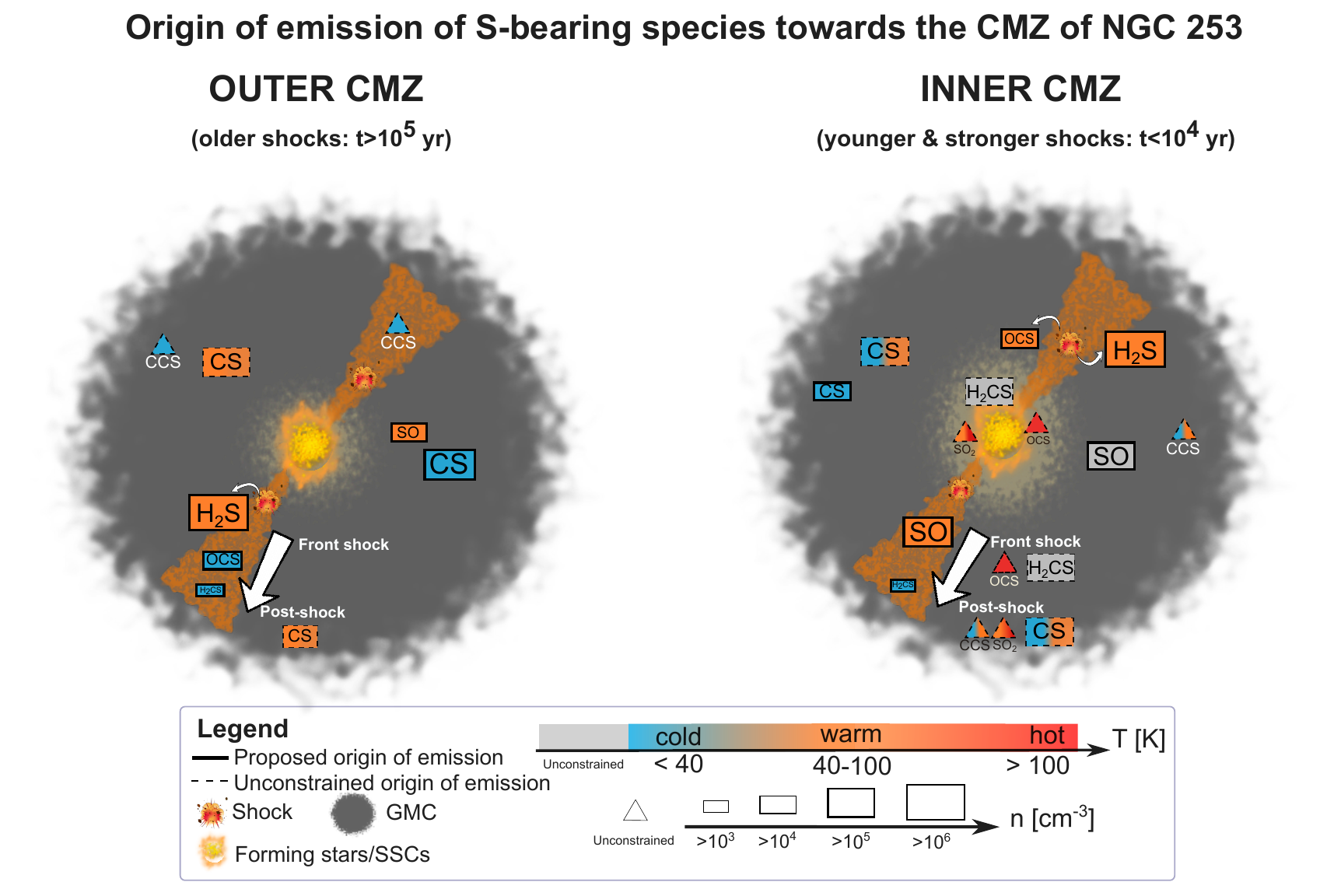}
    \caption{Schematic (not to scale) showing the proposed origin of emission of the S-bearing species towards the CMZ of NGC 253 (Sec.~\ref{sec:puzzle}) and summarising the main results for the physical properties ($T_{\text{gas}},n_{\text{gas}}$) of the gas emitting the S-bearing species (Sec~\ref{sec:phys_params}). We illustrate two scenarios, for a giant molecular cloud in the outer (left) CMZ and inner (right) CMZ. The timescale for the shocks induced by star-formation are from \cite{huang_reconstructing_2023}. The temperatures and densities reported are averaged over the regions and variations could be present in some of the regions.}
    \label{fig:scheme}
\end{figure*}

\subsubsection{Shocks, proto-SSCs or molecular clouds?}\label{subsec:diff_origins}

For some of the species or components, we cannot constrain the region of emission. However, it is still possible to propose some hypotheses based on our results.

For the high-$E_\text{u}$ component of H$_2$CS and OCS, and SO$_2$, we derived temperatures high enough to be consistent with thermal desorption ($T_{\text{des}}\geq 100$ K\footnote{New calculations of the binding energies of SO$_2$ \citep{perrero_binding_2022} indicate that SO$_2$ can sublimate at a lower temperature \citep{bianchi_2023}}; e.g. \citealt{collings_2004, viti_evaporation_2004,kanuchova_2017, vidal_new_2018}). These hot components are only present towards GMC6, SSC2, and SSC5 (also GMC7 in the case of H$_2$CS) and could thus arise from the proto-SSCs. This hypothesis could be strengthened by the fact that such high temperatures for H$_2$CS and SO$_2$ were found in higher angular resolution observations towards the proto-SSCs of NGC 253 by \cite{krieger_molecular_2020a}.  However, we cannot rule out a possible shock origin. Young shocks create dense and warm environments by compressing and heating the surrounding gas before cooling (e.g. \citealt{frank_2014}). In the example of the L1157 outflow, the younger shock position called B0 is much warmer (T$>120$ K) than the older shock positions B1 and B2 (T$< 80$K) \citep[e.g.][]{lefloch_2012, feng_2022}. If both the low-$E_\text{u}$ and high-$E_\text{u}$ components of OCS trace shocks in the inner CMZ, then the high-$E_\text{u}$ component would trace hot young shocks in the inner CMZ whilst the low-$E_\text{u}$ component traces older cooled shocks. In the case of H$_2$CS, this species is efficiently formed on the grains at high densities \citep[e.g.][]{jimenez-escobar_sulfur_2011, drozdovskaya_alma-pils_2018, laas_modeling_2019} and strong enough shocks could directly sputter H$_2$CS into the gas phase. The higher concentration of the fast shock tracer SiO towards the inner CMZ \citep{huang_reconstructing_2023} could indicate the presence of stronger shocks in these regions. In sec.~\ref{subsub:velocities}, we found a significant difference between the systemic velocity of the low- and high upper-level energy transitions of H$_2$CS, suggesting the two components being emitted from two different gas or from different shock events. Finally, SO$_2$ is a well-known post-shocked gas tracer due to the reaction between SO and OH in the gas phase, which takes place within $10^4$ yr \citep[e.g.][]{pineaudesforets_1993}. The sole detection of SO$_2$ in the inner CMZ could be explained by the fact that for timescales $\geq 3\times10^4$ yr, SO$_2$ is converted to atomic sulphur \citep[e.g.][]{pineaudesforets_1993, hatchell_survey_1998, wakelam_sulphur-bearing_2004,feng_2015}. Since \cite{huang_reconstructing_2023} derived that the shocks in the outer CMZ have $\sim 10^5$ yr time scales. SO$_2$ could have been already converted to atomic sulphur. The [H$_2$S]/[H$_2$CS\_2] and [OCS\_2]/[SO$_2$] abundance ratios do not give further constrain on the possible region of emission of these species as the values are similar to that of most Galactic environments.

Finally, for the high-$E_\text{u}$ component of CS and CCS, we cannot distinguish between a molecular cloud origin or a shock/outflow origin. The [OCS\_2]/[CS\_2] abundance ratio is consistently similar to those found in dark clouds throughout the three innermost regions of the CMZ. We mentioned that the high-$E_\text{u}$ component of OCS could either arise from shocked regions or proto-SSCs. The abundance ratio indicates that the high-$E_\text{u}$ component of CS and OCS are not arising from the same gas or the same shock event, if they both trace shocks. If CS traces shocks, it would likely trace a post-shocked gas following the destruction of OCS and H$_2$S \citep[e.g.][]{charnley_sulfuretted_1997,wakelam_sulphur-bearing_2004,wakelam_sulphur_2005,codella_chemical_2005, viti_molecular_2014, gomez-ruiz_density_2015,taquet_seeds_2020}. On the other hand, the physical conditions of the high-$E_\text{u}$ CS component ($n\geq 10^5$ \pcmc \ and $T_{\text{gas}}\sim 20-100$ K) could indicate that it traces more compact and dense regions of the GMCs than the low-$E_\text{u}$ transitions. As mentioned already, the physical conditions of CS need to be taken with caution as we had to fix the source emission size for the LVG analysis.

The [CS$\_2$]/[CCS] abundance ratio agrees well with the ratios found in the L1157-B1 shock and the molecular cloud regions. Since we mentioned the high-$E_\text{u}$ component of CS traces either dense molecular gas or post-shock gas, CCS could also trace these types of regions: On the one hand, CCS is widely detected in the galactic young and cold dark clouds \citep[e.g.][]{suzuki_1992, rathborne_2008, hirota_2009, roy_2011, marka_2012, tatematsu_2014, seo_2019, koley_2022} and the excitation temperatures derived for the various components ($T_{\text{ex}}\sim 10-72$ K) could correspond to cold dark clouds but also to GMCs (for which the temperature can be several tens of Kelvins). On the other hand, CCS could originate from shocked gas. In the L1157-B1 shock, \cite{holdship_sulfur_2019} also derived two different excitation temperatures ($\sim 6$ and $50$ K) which are within the range of what we derived in this work. Additionally, the low-$E_\text{u}$ component of CCS seem to be well correlated with the systemic velocity and line width of that of the low-$E_\text{u}$ component of H$_2$CS (see Sec.~\ref{subsec: gaussian fit}) which could strengthen the shock origin of the emission of this species. Additionally, if CCS mainly traces shocks, its concentration towards the inner CMZ could be explained by the fact that its lifetime is thought to be around $10^4$ yr \citep{de_gregorio-monsalvo_ccs_2006} which is lower than the age of the shock induced star formation derived by \cite{huang_reconstructing_2023} towards the outer CMZ ($\sim 10^5$ yr). It is also possible that the various components of CCS are emitted from different gas. This would agree with the difference in systemic velocity seen between the low- and high- upper-level energy transitions in Sec.~\ref{subsub:velocities}. \\

In summary, we found that most of the S-bearing species could be associated with shocks due to star formation activity within the CMZ of NGC 253. Figure~\ref{fig:scheme} shows a schematic of the proposed scenario for the outer and inner CMZ, as well as a summary of the average physical conditions traced by each species. In the inner CMZ, the presence of stronger and younger shocks due to an intense star forming region would lead to the emission of most S-bearing species, although we cannot exclude that the presence of proto-SSCs contributes to thermally desorb some of the S-bearing species. In the outer CMZ, shocks could also account for the emission of most of S-bearing species, although they are likely to be be less strong or older (due to less intense/older star formation) than those from the inner CMZ. Performing chemical modelling will be the next crucial step to disentangle the various possible scenarios for these species (Bouvier et al. in prep). 

\subsection{Other considerations from the general morphology of the CMZ}

We mainly discussed the emission and origin of sulphur-bearing species in the context of ongoing star formation, since these species are known to be enhanced in these regions in the Milky Way. However, the CMZ of NGC 253 is relatively turbulent and other physical effects could be involved in the enhancement of S-bearing species.

First, as mentioned in Sec.~\ref{sec:source_background}, shocks could be caused by large-scale cloud-cloud collisions, induced by orbital intersections between the central bar and the spiral arms \citep{harada_alchemi_2022, humire_methanol_2022,huang_reconstructing_2023}, the streamers from the large-scale outflow (\citealt{walter_dense_2017, tanaka_2023}, Bao et al. submitted), or the superbubles from supernovae remnants \citep[e.g.][]{sakamoto_molecular_2006, gorski_survey_2017, krieger_turbulent_2020b}. Orbital intersections are thought to happen at the northeast and southwest part of the CMZ, i.e. close to the positions we labelled as GMC 1a, 1b, 8a, 9a, and 10 (see e.g. \citealt{sorai2000, das2001, levy_morpho-kinematic_2022} for more details of the various bar orbits). Previous ALCHEMI studies explained the enhancement of HOCO$^{+}$ and HNCO \citep{harada_alchemi_2022, huang_reconstructing_2023}, and the presence of CH$_3$OH masers \citep{humire_methanol_2022} with potential cloud-cloud collisions happening in these regions. Among the S-bearing species analysed in this study, OCS is the species for which the emission distribution is the most resembling that of the above-mentioned species. However, our analysis showed that in the outer CMZ, hence where these cloud-cloud collisions occur, OCS and HNCO do not trace the same gas (as we also mentioned in Section ~\ref{subsec:shock_tracers}), which could indicate that these two species are likely not tracing the same shock event or type. Further investigations about the type of shock using chemical modelling (Bouvier et al. in prep) are needed to conclude whether cloud-cloud collisions are responsible for the OCS emission in the outer CMZ.

Then, studies performed on the large-scale outflow of NGC253 and its link with the CMZ showed the presence of molecular streamers emerging from the centre of the CMZ \citep[][]{bolatto_2013, walter_dense_2017, krieger_2019}, one of them, in the south-west (SW) being particularly prominent. These streamers are powered by the ongoing star formation occurring inside the inner GMCs and the presence of strong shocks is detected in these GMCs, as well as at the base of the streamers, near the GMCs 7, 6, and 3 (corresponding to our region SSC2) (e.g. \citealt{walter_dense_2017, krieger_2019, huang_reconstructing_2023}, Bao et al. submitted). The present analysis of the S-bearing also indicates the possibility of stronger shocks happening in the inner CMZ (see Sec.~\ref{subsec:shock_tracers} and \ref{subsec:diff_origins}, and Fig.~\ref{fig:scheme}.), due to more enhanced ongoing star formation in the inner CMZ. 

Finally, in the inner CMZ, other types of sources associated with star formation are present, due to the inflow of gas along the bar which triggers star formation inside the GMCs. Several supernovae explosions, supernovae remnants (SNRs), and HII regions were identified in the inner CMZ \citep{ulvestad_1997}, which could contribute to shocking and heating the molecular gas. Whilst GMC6 is bright in all the S-bearing species, differences can be seen between the positions SSC5 and SSC2. Indeed, the emission of CS, H$_2$S, and CCS seems to be slightly more intense towards SSC5 compared to SSC2 (see Fig.~\ref{fig:maps_extended} and \ref{fig:maps_ccs_so2}). \cite{behrens_tracing_2022} showed that a higher number of HII regions was located towards SSC5 (GMC 4 in their study) compared to the other inner GMCs. Enhancement of CS, H$_2$S, and CCS due to the presence of these HII regions is however unlikely, as we showed in Sec.~\ref{sec:puzzle} that the emission of S-bearing species is not associated with PDRs, powered by HII regions. On the other hand, \cite{behrens_tracing_2022} showed that SNRs are more numerous towards GMC 6 compared to the other GMCs studied in this paper, but there is no distinct patterns seen in the gas properties or column densities of the S-bearing species that could indicate a specific enhancement in GMC 6 due to the presence of SNRs.
 

\section{Conclusions}\label{sec:conclusions}

We investigated the origin of emission of the most abundant sulphur-bearing species (CS, H$_2$S, SO, SO$_2$, H$_2$CS, OCS, CCS) detected in Galactic star-forming regions towards the central molecular zone of the nearby starburst galaxy NGC 253. Thanks to the high sensitivity and broad frequency range of the ALMA Large Program ALCHEMI, we had access to a variety of transitions for each species, allowing us to perform rotational diagram and LVG analyses. We then first compared our results with other various species from previous ALCHEMI studies, and we then compared several molecular abundance ratios found in NGC 253 with those found in various Galactic regions. We summarise the main results here below. 

\begin{itemize}
    \item CS, H$_2$S, SO, SO$_2$, H$_2$CS, OCS, CCS show different emission distributions and peaks. The low upper-level energy transitions of CS and H$_2$S show the most extended emission throughout the CMZ. On the other hand, SO$_2$ is concentrated within the 6$\arcsec$ region around the kinematic centre of the CMZ. OCS and H$_2$CS emissions peak towards each of the GMCs rather than towards the centre of the CMZ as the other species.
    \item We performed Gaussian fits to derive the line widths and systemic velocities. We found significant differences in the systemic velocity peak of the low- and high upper-level energy transitions of H$_2$CS, $^{34}$SO, and CCS indicating differences in the emission likely due to excitation conditions, and between the low upper level energy transitions of H$_2$CS and CCS with the other species and their higher upper-level energy transitions. The most narrow line widths for each species were found towards GMC 10.
    \item We performed both LTE and non-LTE LVG analyses to derive the physical conditions of the gas emitting the sulphur-bearing species. We found that towards the inner CMZ, we need at least two gas phase components (associated to low- and high upper-level energy transitions) to reproduce the observed line intensities indicating the presence of different excitation conditions between the inner and the outer CMZ
    \item Across the CMZ, the low upper-level energy transitions of CS and H$_2$CS trace the coldest gas ($T_{\text{gas}}< 50$ K). In the outer CMZ, the highest gas temperatures are derived for SO and the high J-transitions of CS whilst in the inner CMZ  H$_2$S and the high upper-level energy transitions of SO, H$_2$CS show the highest gas temperature. Towards SSC2, the overall gas temperature is colder ($T_{\text{gas}}< 50$ K) than in the other inner regions. The highest density is traced by H$_2$S ($n_{\text{gas}}\geq 10^6-10^7$ \pcmc) throughout the CMZ. SO and the low upper-energy level transitions of CS trace higher densities in the inner CMZ whilst for the other species, the density stays relatively constant.
    \item We found that most of the sulphur-bearing species investigated in this work are likely tracing shocks throughout the CMZ. We also found that they are not likely probing strong shocks as SiO would do. Only the low upper-level energy transitions of SO and CS are tracing dense molecular gas. In the inner CMZ, where the proto-SSCs are located, we could not always disentangle whether the S-bearing species are released through shocks or thermal evaporation. Chemical modelling will be a crucial next step to distinguish between the two possibilities (Bouvier et al. in prep).
\end{itemize}

We investigated the region of emission of various S-bearing species assuming that they are emitting from the same gas throughout the CMZ, or distinguishing between the inner and outer part of the CMZ. However, as seen with the LVG analysis (Sec.~\ref{subsec:results_LVG}), the physical condition of a species could differ in some regions compared to the others, like in the case of SSC2 (see Sec.\ref{subsec: alchemi_comp}). \cite{harada_pca_2024} also concluded that the various GMCs of the CMZ were distinct following their PCA analysis. Therefore, our conclusions could be not entirely valid and further investigation would be needed for these specific regions.

\begin{acknowledgements}
We thank the anonymous referee for their comments that helped improve the paper.
This work is founded by the European Research Council (ERC) Advance Grant MOPPEX 833460. L.C. and V.M.R. acknowledge support from the grants No. PID2019-105552RB-C41 and PID2022-136814NB-I00 by the Spanish Ministry of Science, Innovation and Universities/State Agency of Research MICIU/AEI/10.13039/501100011033 and by "ERDF A way of making Europe".  VMR also acknowledges support from the grant number RYC2020-029387-I funded by MICIU/AEI/10.13039/501100011033 and by "ESF, Investing in your future", and from the Consejo Superior de Investigaciones Cient{\'i}ficas (CSIC) and the Centro de Astrobiolog{\'i}a (CAB) through the project 20225AT015 (Proyectos intramurales especiales del CSIC). N.H. acknowledges support from JSPS KAKENHI grant No. JP21K03634. This paper makes use of the following ALMA data: ADS/JAO.ALMA\#2017.1.00161.L and ADS/JAO.ALMA\#2018.1.00162.S. ALMA is a partnership of ESO (representing its member states), NSF (USA) and NINS (Japan), together with NRC (Canada), MOST and ASIAA (Taiwan),
and KASI (Republic of Korea), in cooperation with the Republic of Chile. The Joint ALMA Observatory is operated by ESO, AUI/NRAO and NAOJ.
\end{acknowledgements}

\bibliographystyle{aa} 
\bibliography{refs}

\onecolumn
\appendix

\section{Velocity-integrated maps}
In this section, the velocity integrated maps of H$_2^{34}$S and $^{34}$SO are shown.

\begin{figure*}
    \centering
    \includegraphics[width=0.8\linewidth]{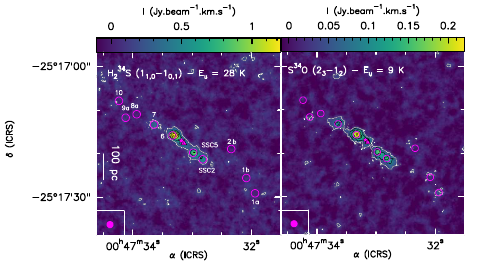}
    \caption{Same as Fig.~\ref{fig:maps_extended} for H$_2^{34}$S and $^{34}$SO. \textit{Top:} Velocity-integrated maps of $^{34}$SO ($2_3$-$1_2$). Levels start at 3$\sigma$ (1$\sigma=$50, 28 and 10.4 mJy.beam$^{-1}$, respectively; white contours) with steps of 7, 3, and 5$\sigma$ (black contours), respectively. \textit{Bottom:} Velocity-integrated maps of H$_2^{34}$S ($1_{1,0}$-$1_{0,1}$). Level starts at 3 $\sigma$ (1$\sigma=56$ mJy.beam$^{-1}$; white contours) with steps of 7$\sigma$ (black contours). }
    \label{fig:maps_isotopologues}
\end{figure*}

\section{Targeted species and transitions}

\begin{longtable}{lccccc}
   \caption{Species and transitions detected (above a $3\sigma$ level on the integrated intensity, see Section \ref{sec:obs}) and used in this work, as well as their spectroscopic parameters taken from CDMS, except for H$_2$CS and CCS from which the parameters are from JPL.}
    \label{tab:species_list} \\
    \hline\hline 
    Species & Transition & Rest Frequency & $E_{\text{u}}$ & $A_{\text{ij}}$ & $g_{\text{u}}$ \\
     && (MHz) & (K) & (s$^{-1})$ &\\
    \hline
    o-H$_2$S\tablefootmark{a}  & $1_{1,0} - 1_{0,1}$ & 168762 & 27.9 & 2.65$\times 10^{-5}$ & 9 \\
    & $3_{3,0} - 3_{2,1}$ & 300505 & 168.9 & 1.01$\times 10^{-4}$ & 21 \\
     p-H$_2$S\tablefootmark{a} & $2_{2,0} - 2_{1,1}$ & 216710 & 84.0 & 4.83$\times 10^{-5}$ & 5\\
    \hline
    o-H$_2^{34}$S\tablefootmark{a}  &$1_{1,0} - 1_{0,1}$ &167910 &28.0 &2.64$\times 10^{-5}$ &9 \\
    \hline
     CS\tablefootmark{b} & 2 $-$ 1 & 97980 & 4.9 & 1.68$\times 10^{-5}$ & 5\\
      & 3 $-$ 2& 146969 & 14.1 & 6.07$\times 10^{-5}$ & 7 \\
      &4 $-$ 3& 195954 & 23.5 & 1.49$\times 10^{-4}$ & 9\\
      &5 $-$ 4& 244935 & 35.3 & 2.98$\times 10^{-4}$ & 11\\
      &6 $-$ 5& 293912 & 49.4 & 5.23$\times 10^{-4}$ & 13\\
      &7 $-$ 6& 342883 & 65.8 & 8.39$\times 10^{-4}$ & 15\\
     \hline
     OCS\tablefootmark{b} & 7 $-$ 6 & 85139 & 16.3 & 1.71$\times 10^{-6}$ & 15 \\
     & 8 $-$ 7 & 97301 & 21.0 & 2.58$\times 10^{-6}$ & 17 \\
     & 9 $-$ 8 & 109463 & 26.3 & 3.70$\times 10^{-6}$ & 19\\
     & 11 $-$ 10 & 133785 & 38.5 & 6.82$\times 10^{-6}$ & 23 \\
     & 12 $-$ 11 & 145946 & 45.5  & 8.88$\times 10^{-6}$ & 25\\
     & 13 $-$ 12 & 158107 & 53.1  & 1.13$\times 10^{-5}$ & 27\\
     & 14 $-$ 13 & 170267 & 61.3 & 1.42$\times 10^{-5}$ & 29\\
     & 16 $-$ 15 & 194586 & 79.4  & 2.13$\times 10^{-5}$& 33\\
     & 17 $-$ 16 & 206745 & 89.3 & 2.56$\times 10^{-5}$ & 35\\
     & 18 $-$ 17 & 218903 & 99.8 & 3.04$\times 10^{-5}$& 37\\
     & 19 $-$ 18 & 231060 & 110.9 & 3.58$\times 10^{-5}$ & 39 \\
     & 23 $-$ 22 & 279685 & 161.1 & 6.37$\times 10^{-5}$ & 47\\
     & 24 $-$ 23 & 291839 & 175.1 & 7.25$\times 10^{-5}$ & 49\\
     & 26 $-$ 25 & 316146 & 204.9 & 9.23$\times 10^{-5}$ & 53\\
     & 29 $-$ 28 & 352599 & 253.9 & 1.28$\times 10^{-4}$ & 59\\
     \hline
     o-H$_2$CS\tablefootmark{a}  & $3_{1,3} - 2_{1,2}$  & 101477 & 22.9 & 1.26$\times 10^{-5}$ & 21 \\
     & $3_{1,2} - 2_{1,1}$ & 104617 & 23.2 & 1.38$\times 10^{-5}$ & 21 \\
     & $4_{1,3} - 3_{1,2}$ & 139483 & 29.9 & 3.57$\times 10^{-5}$& 27 \\
      & $5_{1,5} - 4_{1,4}$ & 169113 & 37.5 & 6.66$\times 10^{-5}$& 33\\
      & $7_{1,7} - 6_{1,6}$ & 236727 & 58.6& 1.91$\times 10^{-4}$& 45\\
      & $8_{1,7} - 7_{1,6}$ & 278886 & 73.4& 3.17$\times 10^{-4}$& 51\\
      & $9_{1,8} - 8_{1,7}$ & 313716 & 88.5& 4.56$\times 10^{-4}$& 57 \\
      & $10_{1,9} - 9_{1,8}$ & 348534 & 105.2& 6.30$\times 10^{-4}$ &63 \\
      & $11_{1,11} - 10_{1,10}$ & 371847 & 120.3 & 7.70$\times 10^{-4}$ & 69 \\
      p-H$_2$CS\tablefootmark{a} & $3_{0,3} - 2_{0,2}$ & 103040 & 9.9& 1.48$\times 10^{-5}$& 7\\
      & $4_{0,4} - 3_{0,3}$ & 137371 & 16.5 & 3.64$\times 10^{-5}$& 9 \\
      & $6_{0,6} - 5_{0,5}$ & 205987 & 34.6& 1.27$\times 10^{-4}$ & 13 \\
      & $7_{2,5} - 6_{2,4}$ & 240548 & 98.9& 1.88$\times 10^{-4}$& 15 \\
    \hline
    CCS\tablefootmark{c} & $7_6 - 6_5$& 86181 &23.3 & 2.82$\times 10^{-5}$ &13\\
    & $7_8 - 6_7$ & 93870 &19.9 & 3.80$\times 10^{-5}$ &17\\
    & $8_7 - 7_6$& 99866 & 28.1& 4.46$\times 10^{-5}$ &15\\
    & $8_8 - 7_7$& 103640 &31.1 & 5.05$\times 10^{-5}$ &17\\
    & $8_9 - 7_8$& 106347& 25& 5.56$\times 10^{-5}$ &19\\
    & $9_8 - 8_7$& 113410 &33.6 & 6.63$\times 10^{-5}$ &17\\
    & $10_{10} - 9_9$& 129548 &42.9 &1.00$\times 10^{-4}$ &21\\
    & $10_{11} - 9_{10}$& 131551 &37.0 &1.06$\times 10^{-4}$ &23\\
    & $11_{10} - 10_9$& 140180& 46.4&1.27$\times 10^{-4}$ &21\\
    & $11_{12} - 10_{11}$& 144244 &43.9 &1.41$\times 10^{-4}$ &25\\
    &$12_{13} - 11_{12}$ & 156981 &51.5 &1.82$\times 10^{-4}$ &27\\
    \endfirsthead
    \caption{continued.}\\
    \hline \hline
    Species & Transition & Rest Frequency & $E_{\text{u}}$ & $A_{\text{ij}}$ & $g_{\text{u}}$ \\
     && (MHz) & (K) & (s$^{-1})$ &\\
    \hline
    \endhead
    \hline
    \endfoot
    & $13_{12} - 12_{11}$&166662 & 61.8 &2.16$\times 10^{-4}$ &25\\
    & $13_{13} - 12_{12}$&168406 & 65.3&2.24$\times 10^{-4}$ &27\\
    & $13_{14} - 12_{13}$&169753 & 59.6 &2.31$\times 10^{-4}$ &29\\
    CCS\tablefootmark{c}& $15_{14} - 14_{13}$&192961 & 79.7& 3.38$\times 10^{-4}$ &29\\
    &$15_{16} - 14_{15}$ &195961 & 77.8&3.53$\times 10^{-4}$ &33\\
    & $16_{16} - 15_{15}$&207260 & 93.3& 4.21$\times 10^{-4}$ &33\\
    &$17_{18} - 16_{17}$ &221071 & 98.4& 5.14$\times 10^{-4}$ &37\\
    & $18_{17} - 17_{16}$&232201 & 111.2& 5.94$\times 10^{-4}$ &35\\
    & $18_{18} - 17_{17}$&233159 & 115.0& 6.02$\times 10^{-4}$ &37\\
    &$18_{19} - 17_{18}$ &233938 & 109.6& 6.10$\times 10^{-4}$ &39\\
    & $19_{18} - 18_{17}$&245244 & 123.0& 7.01$\times 10^{-4}$ &37\\
    & $21_{20} - 20_{19}$&271292 & 148.4&9.52$\times 10^{-4}$ &41\\
    & $21_{22} - 20_{21}$&272592 & 147.0&9.69$\times 10^{-4}$ &45\\
    & $24_{24} - 23_{23}$&310837 & 195.2&1.44$\times 10^{-3}$ &49\\
    & $26_{25} - 25_{24}$&336258 & 222.8&1.82$\times 10^{-3}$ &51\\
    & $26_{25} - 25_{24}$&336722 & 226.9&§.83$\times 10^{-3}$ &53\\
    \hline
     SO\tablefootmark{d} &$2_2 - 1_1$ & 86093 & 19.3& 5.25$\times 10^{-6}$ & 5\\
     &$5_4 - 4_4$ & 100029 & 38.6& 1.08$\times 10^{-6}$ & 9\\
     &$3_2 - 2_1$ & 109252 & 21.1& 1.08$\times 10^{-5}$ & 5\\
     &$3_3 - 2_2$ & 129138 & 25.5& 2.25$\times 10^{-5}$ & 7\\
     &$6_5 - 5_5$ & 136634 & 50.7& 1.75$\times 10^{-6}$ &11 \\
     &$3_4 - 2_3$ & 138178 & 15.9& 3.17$\times 10^{-5}$& 9\\
     &$4_3 - 3_2$ & 158971 & 28.7& 4.23$\times 10^{-5}$& 7\\
     &$4_4 - 3_3$ & 172181 & 33.8& 5.83$\times 10^{-5}$& 9\\
     &$4_5 - 3_4$ & 178605 & 24.4& 7.02$\times 10^{-5}$& 11\\
     &$5_5 - 4_4$ & 215220 & 44.1& 1.19$\times 10^{-4}$& 11\\
     &$5_6 - 4_5$ & 219949 & 35.0& 1.34$\times 10^{-4}$& 13\\
     &$9_8 - 8_8$ & 254573 & 99.7& 4.24$\times 10^{-6}$& 17\\
     &$7_6 - 6_5$ & 296550& 64.9& 3.23$\times 10^{-4}$&13\\
     &$7_8 - 6_7$ & 304077 & 62.1& 3.61$\times 10^{-4}$ &17\\
     & $2_2 - 1_2$& 309502 & 19.3& 1.42$\times 10^{-5}$ &5\\
     &$3_3 - 2_3$ & 339341 & 25.5& 1.45$\times 10^{-5}$&7\\
     & $8_8 - 7_7$& 344310 & 87.5& 5.19$\times 10^{-4}$&17\\
    \hline
     $^{34}$SO\tablefootmark{d} &$2_2 - 1_1$ & 84410 & 19.2& 4.95$\times 10^{-6}$&5\\
     &$2_3 - 1_2$ & 97715& 9.1&1.07$\times 10^{-5}$ &7\\
     &$3_3 - 2_2$ & 126613 & 25.3& 2.12$\times 10^{-5}$ &7\\
     &$4_5 - 3_4$ & 175352 & 24.0 & 6.65$\times 10^{-5}$&11\\
     &$6_5 - 5_4$ & 246663 & 49.9&1.81$\times 10^{-4}$ &11\\
     &$6_7 - 5_6$ & 256877 & 46.7& 2.15$\times 10^{-4}$&15\\
     &$7_8 - 6_7$ & 298257 & 61.0 & 3.41$\times 10^{-4}$&17\\
     &$8_8 - 7_7$ & 337582 & 86.1 & 4.89$\times 10^{-4}$&17\\
    \hline
     SO$_2$\tablefootmark{a} &$3_{1,3} - 2_{0,2}$ &104029 &7.7 &1.01$\times 10^{-5}$ & 7\\
     &$10_{1,9} - 10_{0,10}$ &104239 & 54.7&1.12$\times 10^{-5}$ & 21\\
     &$10_{2,8} - 10_{2,8}$ &129514 & 60.9&2.50$\times 10^{-5}$ & 21\\
     &$12_{1,11} - 12_{0,12}$ &131014 & 76.4&1.86$\times 10^{-5}$ & 25\\
     &$14_{2,12} - 14_{1,13}$ &132744 & 108.1&2.93$\times 10^{-5}$ &29 \\
     &$6_{2,4} - 6_{1,5}$ &140306 & 29.2&2.53$\times 10^{-5}$ & 13\\
     &$3_{2,2} - 3_{1,3}$ &158199 & 15.3&2.53$\times 10^{-5}$ & 7\\
     &$2_{2,0} - 1_{1,1}$ &192651 & 12.6&6.56$\times 10^{-5}$ & 5\\
     &$24_{3,21} - 24_{2,22}$ &200287 & 302.3& 9.87$\times 10^{-5}$& 49\\
     &$18_{3,15} - 18_{2,16}$ &204246 & 180.6 &9.27$\times 10^{-5}$ &37 \\
     &$11_{2,10} - 11_{1,11}$ &205300 & 70.2&5.32$\times 10^{-5}$ & 23\\
     &$3_{2,2} - 2_{1,1}$ &208700 & 15.3&9.72$\times 10^{-5}$ & 7\\
     &$28_{3,25} - 28_{2,26}$ &234187 & 403.0&1.45$\times 10^{-4}$  & 57\\
     &$5_{2,4} - 4_{1,3}$ &241615 & 23.6&8.46$\times 10^{-5}$ & 11\\
     &$15_{2,14} - 15_{1,15}$ &248057 & 119.3&8.06$\times 10^{-5}$ & 31\\
     &$28_{4,24} - 28_{3,25}$ &267719 & 415.9&2.16$\times 10^{-4}$  &57 \\
     &$7_{2,6} - 6_{1,5}$ &271529 &35.5 &1.11$\times 10^{-4}$  & 15\\
     &$18_{1,17} - 17_{2,16}$ &288519 & 163.1&1.57$\times 10^{-4}$  & 37\\
     SO$_2$\tablefootmark{a}&$26_{2,24} - 26_{1,25}$ &296168 & 340.6&1.87$\times 10^{-4}$  & 53\\
     &$4_{3,1} - 3_{2,2}$ &332505 & 31.3&3.29$\times 10^{-4}$  & 9\\
     &$8_{2,6} - 7_{1,7}$ &334673 & 43.1&1.27$\times 10^{-4}$  & 17\\
     &$23_{3,21} - 23_{2,22}$ &336089 & 276.0&2.67$\times 10^{-4}$  & 47\\
     &$19_{1,19} - 18_{0,18}$ &346652 & 168.1& 5.22$\times 10^{-4}$ & 39\\
     &$5_{3,3} - 4_{2,2}$ &351257 & 35.9 & 3.36$\times 10^{-4}$ & 11\\
     &$12_{4,8} - 12_{3,9}$ &355045 &111.0 &3.40$\times 10^{-4}$  & 25\\
     &$13_{4,10} - 13_{3,11}$ &357165 &123.0 & 3.51$\times 10^{-4}$ & 27\\
     &$15_{4,12} - 15_{3,13}$ &357241 &149.7 &3.62$\times 10^{-4}$  & 31\\
     &$20_{0,20} - 19_{1,19}$ &358215 &185.3 & 5.83$\times 10^{-4}$ & 41\\
     &$19_{4,16} - 19_{3,17}$ &359770 &214.3 & 3.85$\times 10^{-4}$ & 39\\
     &$21_{4,18} - 21_{3,19}$ & 363159&252.1 &4.00$\times 10^{-4}$  &43 \\
     &$6_{3,3} - 5_{2,4}$ &371172 &41.4 &3.55$\times 10^{-4}$  &13 \\
    \hline
    \multicolumn{6}{l}{\tablefoottext{a}{The quantum numbers given are $J(K_a,K_b)$}}\\
    \multicolumn{6}{l}{\tablefoottext{b}{The quantum numbers given are $J$}}\\
    \multicolumn{6}{l}{\tablefoottext{c}{The quantum numbers given are $N(J)$}}\\
    \multicolumn{6}{l}{\tablefoottext{d}{The quantum numbers given are $J(K)$}}\\
\end{longtable}


\section{Spectra \& Gaussian fit results}\label{app:spectra}

Representative spectra for each species and region. The transitions shown are those used to perform the velocity-integrated maps in Figures~\ref{fig:maps_extended}, \ref{fig:maps_ocs_h2cs}, \ref{fig:maps_ccs_so2} and \ref{fig:maps_isotopologues}. SO$_2$ being detected only towards the inner regions GMC6, SSC5 and SSC2, no spectra outside these regions are shown for this species. The result for the Gaussian fit, the measured integrated intensity and the rms for each transition and region are presented in Table \ref{tab: fit_results}. 

\begin{figure*}
    \centering
    \includegraphics[width=0.9\textwidth]{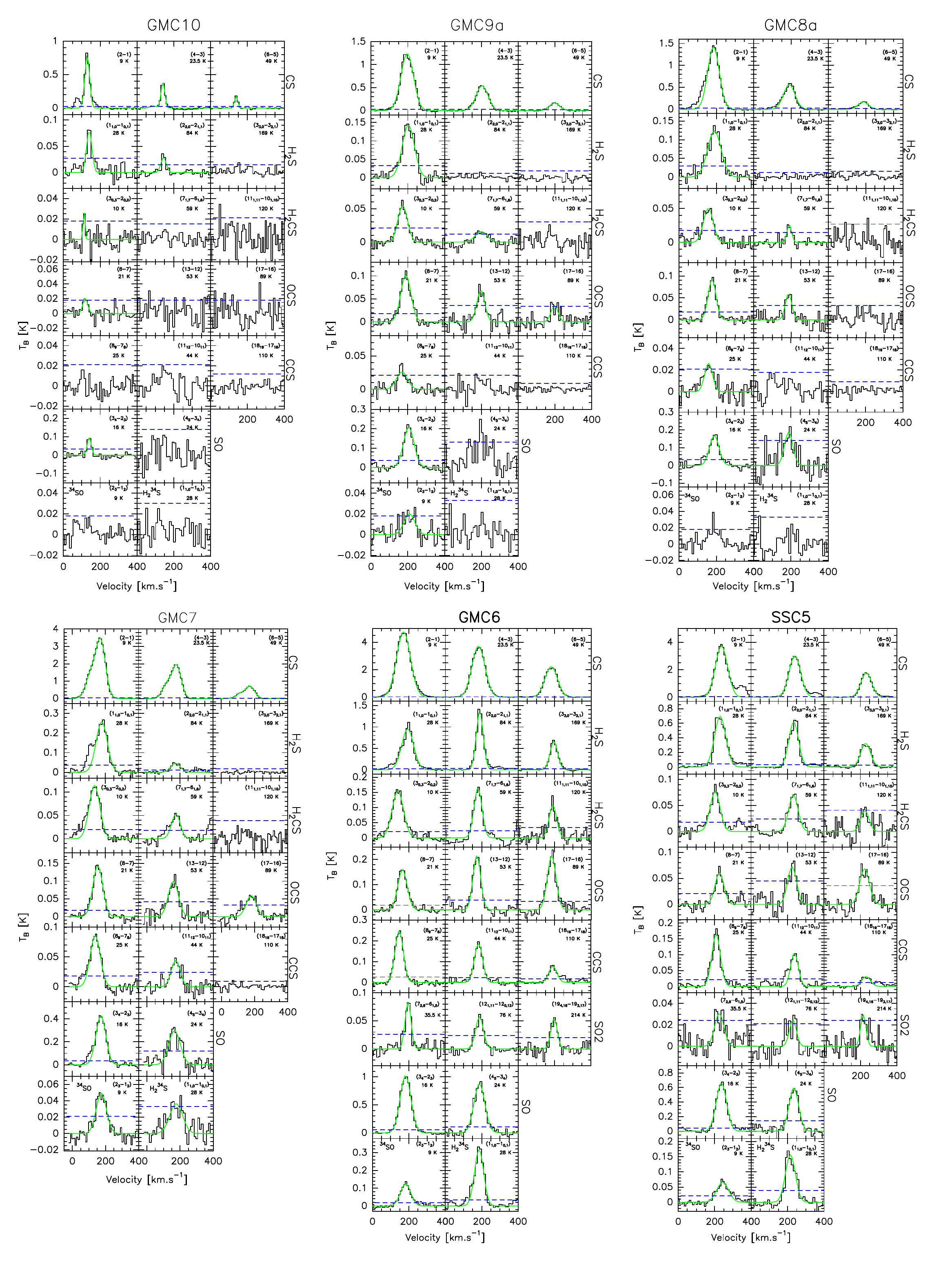}
    \caption{Representative spectra for each species towards GMC10, GMC9a, GMC8a, GMC7, GMC6, and SSC5. The full green line shows the Gaussian fit and the dashed blue line marks the 3$\sigma$ level.}
    \label{fig:spec_part1}
\end{figure*}

\begin{figure*}
    \centering
    \includegraphics[width=0.9\textwidth]{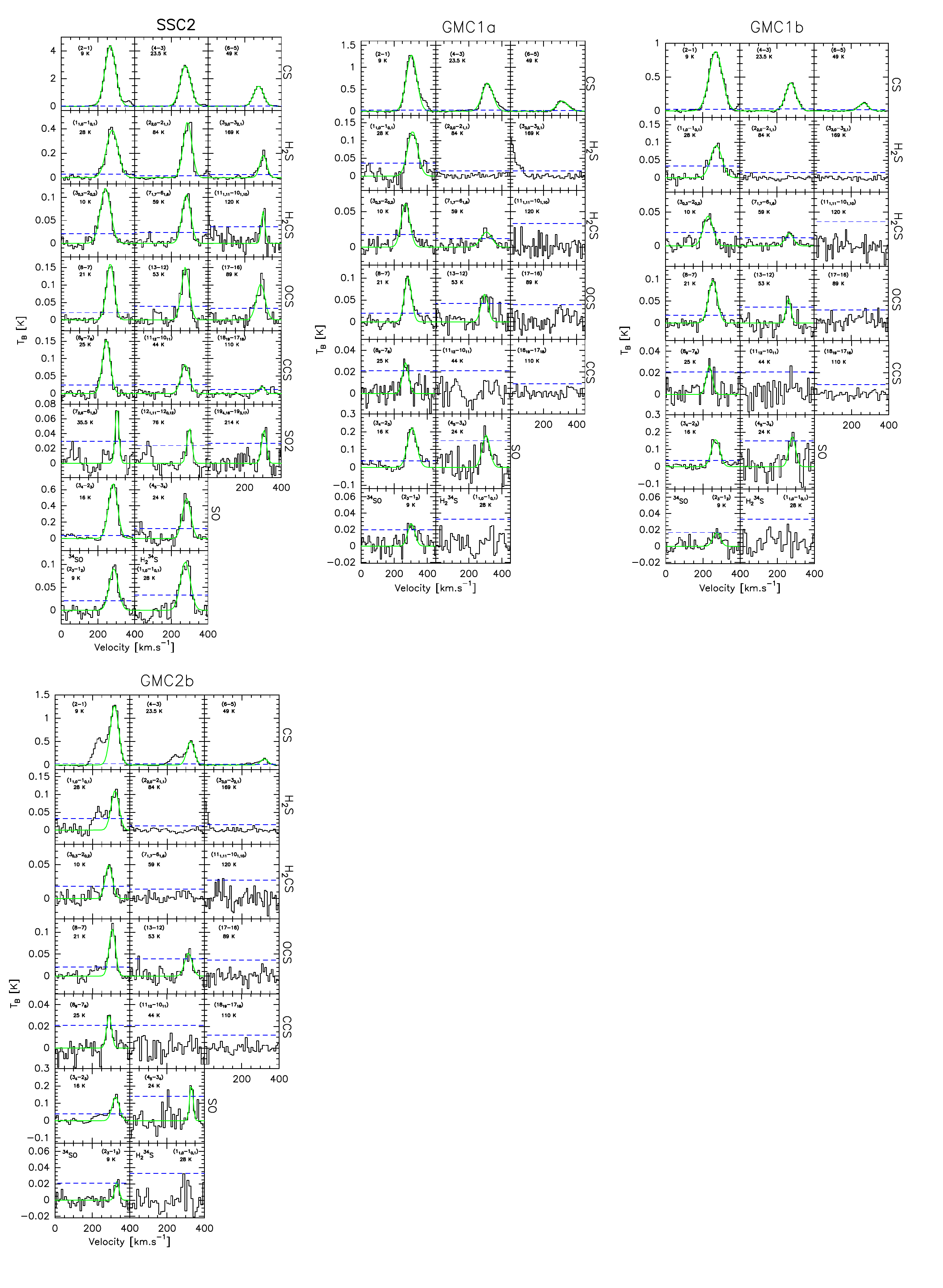}
    \caption{Representative spectra for each species towards SSC5, GMC1a, GMC1b, and GMC2b. The full green line shows the Gaussian fit and the dashed blue line marks the 3$\sigma$ level.}
    \label{fig:spec_part2}
\end{figure*}

\begin{landscape}
\tiny
    \begin{longtable}{cccccc|cccccc}
      \caption{Frequencies of the detected lines, their associated Gaussian fit result, integrated intensities and rms per region. The errors do not include the 15\% calibration uncertainty. The spectral resolution is 10 km.s$^{-1}$.}
    \label{tab: fit_results}\\
    \hline \hline
Region & Rest Frequency & $\int T_B dV$ & $V_{lsr}$ & $FWHM$ & rms&Region&Rest Frequency & $\int T_B dV$ & $V_{lsr}$ & $FWHM$ & rms\\
& (MHz) & (K.km.s$^{-1}$)& (km.s$^{-1}$)&(km.s$^{-1}$)&(mK)&&(MHz) & (K.km.s$^{-1}$)& (km.s$^{-1}$)&(km.s$^{-1}$)&(mK)\\
    \hline
    \multicolumn{6}{c|}{CS}&\multicolumn{6}{c}{OCS}\\
    \cline{2-12}
    \multirow{24}{*}{GMC10}&97980 &18.49 $\pm$ 1.45&128.0 $\pm$ 0.2 & 25.2 $\pm$ 0.9 &8.5&\multirow{15}{*}{GMC9a} &170267 &2.78 $\pm$ 0.56 &201.9 $\pm$ 7.8 &76.5 $\pm$ 17.4 &8.2 \\
    &146969 & 13.65 $\pm$ 0.43 &124.4 $\pm$ 0.1 & 23.4 $\pm$ 0.5 & 7.3 & &194586 &1.32 $\pm$ 0.21 &197.0 $\pm$ 4.6 &48.0 $\pm$ 9.9  &8.0 \\
    & 195954 &9.92 $\pm$ 0.05 & 140.6 $\pm$ 0.3 & 23.9 $\pm$ 0.5 & 8.2& &206745 &1.42 $\pm$ 0.54 &197.7 $\pm$ 5.0 &45.3 $\pm$ 10.9 &12.0 \\
    & 244935&6.72 $\pm$ 0.18 & 145.5 $\pm$ 0.3 & 24.3 $\pm$ 0.8 & 4.9& &218903&1.14 $\pm$ 0.12 &205.9 $\pm$ 3.3 &58.1 $\pm$ 6.7 &3.4\\
    & 293912& 3.82 $\pm$ 0.15 &138.5 $\pm$ 0.4 &19.7 $\pm$ 0.9 & 7.1& & 231060 &0.81 $\pm$ 0.17 &215.1 $\pm$ 4.4 &47.7 $\pm$ 10.2 &4.7\\
    \cline{8-12}
    &342883 &2.56 $\pm$ 0.17 &142.7 $\pm$ 0.6 &18.4 $\pm$ 1.5 &7.4 &\multicolumn{6}{c}{H$_2$CS} \\
    \cline{2-6}\cline{8-12}
    \multicolumn{6}{c|}{H$_2$S}&&101477&5.41 $\pm$ 0.34&172.9 $\pm$ 1.5 &61.0 $\pm$ 3.8 &8.0\\
    \cline{2-6}
    &168762 & 3.24 $\pm$ 0.30&140.2 $\pm$ 2.0 & 34.1 $\pm$ 6.1 &8.8 & &103040 &3.53 $\pm$ 0.28 &171.7 $\pm$ 1.8 &61.0 $\pm$ 4.4 &6.7 \\
    & 216710& 0.98 $\pm$ 0.17 &144.2 $\pm$ 1.9 & 34.0 $\pm$ 10.0 &4.8& &104617 &4.94 $\pm$ 0.26 &170.8 $\pm$ 1.5 &65.7 $\pm$ 3.3 &6.3 \\
    \cline{2-6}
    \multicolumn{6}{c|}{SO}&&137371&3.89 $\pm$ 0.56 &193.2 $\pm$ 4.0 &63.4 $\pm$ 11.6 &12.0\\
    \cline{2-6}
    &109252 &0.50 $\pm$ 0.15 & 142.2 $\pm$ 1.6 &14.2 $\pm$ 4.1 & 6.7& &139483 &3.42 $\pm$ 0.41 &187.9 $\pm$ 4.2 &64.2 $\pm$ 10.4 &13.0\\
    &138178 &2.51 $\pm$ 0.25 & 140.5 $\pm$ 1.3 &24.3 $\pm$ 2.7 & 10.5 & &169113 &2.54 $\pm$  0.35&206.6 $\pm$ 4.4 &60.3 $\pm$ 9.1 &11.0\\
    &215220 &0.53 $\pm$ 0.10 & 142.4 $\pm$ 2.6 &22.9 $\pm$ 5.3 &4.4& &236727 &1.09 $\pm$ 0.13 &198.4 $\pm$ 9.6 &75.6 $\pm$ 19.4 &4.2\\
    \cline{8-12}
    &219949 &0.80 $\pm$ 0.10 & 144.5 $\pm$ 1.4 &23.8 $\pm$ 4.8 &4.0&  \multicolumn{6}{c|}{CCS}\\
    \cline{8-12}
    &304077 &0.37 $\pm$ 0.15 &141.5 $\pm$ 2.6 &17.9 $\pm$ 6.1 &6.8&&93870 &3.25 $\pm$ 0.24 &165.2 $\pm$ 2.9 &70.1 $\pm$ 6.7 & 7.0  \\
    \cline{2-6}
    \multicolumn{6}{c|}{OCS}&&106347&1.78 $\pm$ 0.28 &167.8 $\pm$ 2.9 & 70.0 $\pm$ 10.0&6.9\\
    \cline{2-12}
    &85139 &1.03 $\pm$ 0.35 &110.0 $\pm$ 10.0 &20.0 $\pm$ 10.0 &8.1& \multicolumn{6}{c}{CS} \\
    \cline{8-12}
    &97301 &0.68 $\pm$ 0.15 &118.7 $\pm$ 3.1 &30.9 $\pm$ 6.8 &6.1 &\multirow{34}{*}{GMC8a} &97980 &123.97 $\pm$ 0.43 &180.6 $\pm$ 0.2 &83.8 $\pm$ 0.4&9.0\\
    &109463 &0.45 $\pm$ 0.21 &108.9 $\pm$ 2.3 &19.4 $\pm$ 5.4 &8.3 & &146969 &83.15 $\pm$ 0.43 &176.0 $\pm$ 0.4 &78.3 $\pm$ 1.1 &8.4\\
    &133785 &0.47 $\pm$ 0.16 &121.0 $\pm$ 2.2 &18.6 $\pm$ 5.8 &6.5 & &195954 &44.67 $\pm$ 0.36 &190.0 $\pm$ 0.4 &72.0 $\pm$ 0.9 &8.5\\
    \cline{2-6}
    \multicolumn{6}{c|}{H$_2$CS}&&244935&24.21 $\pm$ 0.18 &194.2 $\pm$ 0.3 &66.9 $\pm$ 0.8 &4.3\\
    \cline{2-6}  
    &101477 &0.37 $\pm$ 0.11 &112.2 $\pm$ 3.6 &10.9 $\pm$ 12.1 &7.8 & &293912 & 11.84 $\pm$ 0.26&196.8 $\pm$ 0.7 &63.7 $\pm$ 1.8 &6.3 \\
    &103040 &0.47 $\pm$ 0.13 &103.6 $\pm$ 1.6 &11.2 $\pm$ 7.3 &5.9& & 342883&7.10 $\pm$ 0.29 &193.4 $\pm$ 1.5 &60.4 $\pm$ 4.2 &8.3 \\
    \cline{8-12}
    &104617 &0.58 $\pm$ 0.22 &105.7 $\pm$ 1.8 &17.5 $\pm$ 4.1 &6.5 & \multicolumn{6}{c}{H$_2$S}\\
    \cline{1-6} \cline{8-12}
    \multicolumn{6}{c|}{CS}&&168762 &11.18 $\pm$ 0.43 &188.4 $\pm$ 2.2 &86.9 $\pm$ 5.5 & 9.8\\
    \cline{2-6}\cline{8-12}
    \multirow{27}{*}{GMC9a}&97980 &105.72 $\pm$ 0.43 &195.6 $\pm$ 0.1 &80.4 $\pm$ 0.3 &9.4 & \multicolumn{6}{c}{SO}\\
    \cline{8-12}
    &146969 &77.83 $\pm$ 0.41&187.0 $\pm$ 0.3 &80.5 $\pm$ 0.6 &8.1 & &109252 &2.19 $\pm$ 0.20 &193.8 $\pm$ 2.3 & 53.4 $\pm$ 5.4&6.6\\
    &195954 &43.53 $\pm$ 0.42&203.5 $\pm$ 0.3 &76.3 $\pm$ 0.8 &9.6 & &138178 &10.74 $\pm$ 0.51 &188.7 $\pm$ 1.3 &59.7 $\pm$ 3.6 &11.0\\
    &244935 &24.06 $\pm$ 0.19&205.1 $\pm$ 0.3 &69.8 $\pm$ 0.6 &4.7 & &158971 &2.61 $\pm$ 0.44 &194.5 $\pm$ 6.5 &68.1 $\pm$ 16.4 &11.0\\
    &293912 &11.06 $\pm$ 0.24&198.7 $\pm$ 0.7 &64.2 $\pm$ 1.6 &5.5 & &172181 &1.27 $\pm$ 0.20 &188.4$\pm$ 4.3 &36.4 $\pm$ 9.4 &6.6\\
    &342883 &5.60 $\pm$ 0.29&204.0 $\pm$ 1.8 &63.7 $\pm$ 4.4 &8.6 & &178605&10.51 $\pm$ 2.07 &187.7 $\pm$ 8.4 &53.0 $\pm$ 10.0 &48.0\\
    \cline{2-6}
    \multicolumn{6}{c|}{H$_2$S}&&215220 & 0.99 $\pm$ 0.14&185.2 $\pm$  6.0&59.5 $\pm$ 13.9&4.4 \\
    \cline{2-6} 
    &168762 &12.18 $\pm$ 0.51 &202.5 $\pm$ 1.5 &77.8 $\pm$ 3.6 &11.2 &&219949&3.12 $\pm$ 1.08 &198.1 $\pm$ 1.4 &52.3 $\pm$ 4.0 &3.4\\
     \cline{2-6}
    \multicolumn{6}{c|}{SO}&&304077&0.82 $\pm$ 0.14 &198.2 $\pm$ 3.1 &34.6 $\pm$ 7.5 &5.9\\
    \cline{2-6}\cline{8-12}
    &86093 &1.09 $\pm$ 0.20 &203.8 $\pm$ 8.4 &60.6 $\pm$ 20.8 &6.5 & \multicolumn{6}{c}{OCS}\\
    \cline{8-12}
    &109252 &2.27 $\pm$ 0.23 &203.7 $\pm$ 2.0 &50.2 $\pm$ 5.1 &6.6 & &85139 &5.82 $\pm$ 0.31 &157.0 $\pm$ 2.4 &55.7 $\pm$ 6.2 &8.0\\
    &129138 &2.28 $\pm$ 0.21 &198.2 $\pm$ 5.5 &64.1 $\pm$ 15.5 &6.4 & &97301 &4.78 $\pm$ 0.18 &176.1 $\pm$ 0.9 &50.2 $\pm$ 2.3 &6.0\\
    &138178 &14.84 $\pm$ 0.49 &205.0 $\pm$ 1.3 &67.4 $\pm$ 3.0 &12.0 & &109463 &4.35 $\pm$ 0.21 &157.9 $\pm$ 1.1 &42.0 $\pm$ 2.6 &8.3\\
    &158971 &3.36 $\pm$ 0.52 &198.7 $\pm$ 3.1 &43.4 $\pm$ 8.7 &43.0 & &133785 & 4.04 $\pm$ 0.26&175.7 $\pm$ 1.6 &51.0 $\pm$ 4.2 &8.0\\
    &178605 &8.85 $\pm$ 2.11 &199.6 $\pm$ 5.4 &43.4 $\pm$ 16.3 &43.0 & &145946 &3.02 $\pm$ 0.26 &190.7 $\pm$ 1.7 &46.7 $\pm$ 4.2 &7.2\\
    &215220 &1.26 $\pm$ 0.13 &211.6 $\pm$ 3.2 &52.2 $\pm$ 6.4 &3.9 & &158107 &2.28 $\pm$ 0.23 &186.2 $\pm$ 4,3 &41.5 $\pm$ 11.0 &11.0\\
    &219949 &3.85 $\pm$ 0.13 &209.9 $\pm$ 1.3 &55.0 $\pm$ 10.0 &2.2 & &170267 &2.00 $\pm$  0.27&191.6 $\pm$ 4.3 &43.0 $\pm$ 12.1 &9.2\\
    \cline{2-6}
    \multicolumn{6}{c|}{$^{34}$SO}& &194586 &1.35 $\pm$ 0.26 &189.7 $\pm$ 7.5 &63.2 $\pm$ 14.7 &9.1\\
    \cline{2-6} 
    &97715 &1.36 $\pm$ 0.20 &208.5 $\pm$ 5.5 &69.7 $\pm$ 11.0 &5.9 & &218903 &0.62 $\pm$ 0.10 &188.5 $\pm$ 3.4 &41.8 $\pm$ 8.3 &3.5\\
   \endfirsthead
    \caption{continued.}\\
    \hline \hline
    Region & Rest Frequency & $\int T_B dV$ & $V_{lsr}$ & $FWHM$ & rms&Region&Rest Frequency & $\int T_B dV$ & $V_{lsr}$ & $FWHM$ & rms\\
& (MHz) & (K.km.s$^{-1}$)& (km.s$^{-1}$)&(km.s$^{-1}$)&(mK)&&(MHz) & (K.km.s$^{-1}$)& (km.s$^{-1}$)&(km.s$^{-1}$)&(mK)\\
    \hline
    \endhead
    \endfoot
    \cline{2-6}\cline{8-12} 
    \multicolumn{6}{c|}{OCS}&\multicolumn{6}{c}{H$_2$CS}\\
    \cline{2-6}\cline{8-12}
    &85139 &8.14 $\pm$ 0.77 &167.1 $\pm$ 3.3 &73.2 $\pm$ 8.2& 7.7&&101477 &4.47 $\pm$ 0.29 &160.9 $\pm$ 1.4 & 54.1 $\pm$ 3.9& 7.7 \\
    &97301 &7.13 $\pm$ 0.24 &188.8 $\pm$ 1.0 & 65.7 $\pm$ 2.4&5.9 & &103040 &3.47 $\pm$ 0.24 &156.2 $\pm$ 2.1 &61.5 $\pm$ 5.0 &6.0\\
    &109463 &5.94 $\pm$ 0.25 &167.2 $\pm$ 1.7 &62.8 $\pm$ 4.0 &8.4 & &104617 & 3.43 $\pm$ 0.24 &158.0 $\pm$ 1.4 &56.9 $\pm$ 3.7 &6.0\\
    &133785 &5.33 $\pm$ 0.26 &188.1 $\pm$ 1.6 &67.2 $\pm$ 3.7&7.4 & &137371 &3.42 $\pm$ 0.51 &182.3 $\pm$ 2.7 &45.2 $\pm$ 9.6 &12.0\\
    &145946 &4.87 $\pm$ 0.24 &199.8 $\pm$ 2.0 &62.8 $\pm$ 5.6&6.3 & &139483 &2.41 $\pm$ 0.35 &176.1 $\pm$ 1.8 &35.6 $\pm$ 3.9 &11.0\\
    &158107 &3.79 $\pm$ 0.44 &196.9 $\pm$ 2.9 &51.0 $\pm$ 7.2&12.0 &  &169113 &3.30 $\pm$ 0.32 &190.7 $\pm$ 3.9 &62.4 $\pm$ 9.2 & 9.8 \\
    \hline 
    \multicolumn{6}{c|}{H$_2$CS} &\multicolumn{6}{c|}{H$_2$CS}\\
    \cline{2-6}\cline{8-12}
    \multirow{4}{*}{GMC8a}&236727&0.83 $\pm$ 0.13 &190.1 $\pm$ 2.4 &29.6 $\pm$ 5.4&5.0 & \multirow{12}{*}{GMC7} &240548 & 0.70 $\pm$ 0.14&170.8 $\pm$ 22.2 &63.1 $\pm$ 10.0 &4.5\\
    \cline{2-6}
   \multicolumn{6}{c|}{CCS}& &278886 &1.94 $\pm$ 0.22 &180.7 $\pm$ 3.3 &63.4 $\pm$ 9.1 &4.5\\
   \cline{2-6}\cline{8-12}
    &93870 &2.10 $\pm$ 0.20 &154.2 $\pm$ 2.1 &45.7 $\pm$ 5.1 &7.0 & \multicolumn{6}{c}{CCS}\\
    \cline{8-12}
    &106347 &1.58 $\pm$ 0.26 &158.0 $\pm$ 4.7 &50.0 $\pm$ 10.0 &7.0&&86181&2.62 $\pm$ 0.20&158.6 $\pm$ 3.6 &58.6 $\pm$ 11.0 &6.8\\
    \cline{1-6}
    \multicolumn{6}{c|}{CS} & &93870 &8.82 $\pm$ 0.29 & 133.1 $\pm$ 1.1& 72.3 $\pm$ 2.4&7.2\\
    \cline{2-6}
    \multirow{49}{*}{GMC7}&97980 &231.10 $\pm$ 4.44 &172.3 $\pm$ 0.5 &67.1 $\pm$ 0.5&10.0 & &99866 &2.03 $\pm$ 0.29 &144.9 $\pm$ 5.7 &63.5 $\pm$ 9.7 &8.7\\
    & 146969&176.25 $\pm$ 10.60 &167.7 $\pm$ 1.2 &63.3 $\pm$ 1.3&7.7 & &106347 &6.56 $\pm$ 0.26 &137.6 $\pm$ 3.4 &72.7 $\pm$ 7.8 &6.2\\
    &195954 &133.75 $\pm$ 1.84 &179.0 $\pm$ 0.4 &66.0 $\pm$ 0.6&11.4 & &129548 &1.23 $\pm$ 0.24 &177.2 $\pm$ 5.5 &39.7 $\pm$ 11.2 &7.7\\
    &244935 &76.09 $\pm$ 0.02 &184.6 $\pm$ 0.2 &62.5 $\pm$ 0.4& 3.3& &131551 &3.32 $\pm$ 0.26 &163.4 $\pm$ 3.3 &70.4 $\pm$ 7.8 &7.1\\
    &293912 &46.90 $\pm$ 2.77  &173.7 $\pm$ 1.6 &63.8 $\pm$ 2.1&6.6 & &140180 &1.61 $\pm$ 0.26 &175.3 $\pm$ 5.4 &72.9 $\pm$ 14.0 &6.4\\
    &342883 &25.22 $\pm$ 1.89 &178.5 $\pm$ 2.1 &63.3 $\pm$ 2.8&8.8 & &144244 & 2.93 $\pm$ 0.27&173.2 $\pm$ 3.7 &64.3 $\pm$ 8.2 &7.7\\
     \cline{2-6}
    \multicolumn{6}{c|}{H$_2$S}&&156981&1.62 $\pm$ 0.36 &174.2 $\pm$ 6.0 &54.5 $\pm$ 11.2 &1.2 \\
    \cline{2-6}\cline{7-12}
    &168762 &26.42 $\pm$ 0.54 &177.7 $\pm$ 1.1 &79.7 $\pm$ 2.8&11.8 & \multicolumn{6}{c}{H$_2$S}\\
    \cline{8-12}
    &216710 &3.30 $\pm$ 0.16 &173.7 $\pm$ 1.5 &66.2 $\pm$ 3.6&4.4 & \multirow{38}{*}{GMC6}&168762 & 78.13 $\pm$ 0.36& 192.7 $\pm$ 0.4&77.2 $\pm$ 1.2 &8.1\\
    \cline{2-6}
    \multicolumn{6}{c|}{H$_2^{34}$S}&&216710&67.79 $\pm$ 0.42 &188.5 $\pm$ 0.3 &45.0 $\pm$ 0.9 &10.9\\
    \cline{2-6}
    &167910 &3.41 $\pm$ 0.38 &174.0 $\pm$ 8.0 &95.6 $\pm$ 18.8 &8.9 & & 300505&35.49 $\pm$ 0.33 &195.4 $\pm$ 0.3 &49.3 $\pm$ 0.8 &8.9\\
    \cline{2-6}\cline{8-12}
    \multicolumn{6}{c|}{SO} &\multicolumn{6}{c}{H$_2^{34}$S}\\
    \cline{2-6}\cline{8-12}
    &86093 &4.75 $\pm$ 0.47 &174.8 $\pm$ 3.2 &79.7 $\pm$ 10.1&6.8 & & 167910&20.03 $\pm$ 0.41 &185.2 $\pm$ 0.8 &59.0 $\pm$ 2.1 &11.0\\
    \cline{8-12}
    &100029 &1.00 $\pm$ 0.16 &181.2 $\pm$ 4.0 &81.2 $\pm$ 7.0&8.7 & \multicolumn{6}{c}{CS}\\
    \cline{8-12}
    &109252 &8.52 $\pm$ 0.30 &171.7 $\pm$ 1.3 &77.8 $\pm$ 3.9&7.0 & & 97980&462.67 $\pm$ 0.49 &170.0 $\pm$ 0.1&89.5 $\pm$ 0.2&9.0\\
    &129138 &6.13 $\pm$ 0.30 &166.4 $\pm$ 2.6 &73.4 $\pm$ 6.2&7.7 & & 146969&413.18 $\pm$ 0.43&166.0 $\pm$ 0.7&85.7 $\pm$ 1.7&8.8\\
    &138178 &32.22 $\pm$ 1.04 &171.6 $\pm$ 1.1 &70.1 $\pm$ 2.8 &11.0&& 195954&332.55 $\pm$ 0.47 &181.4 $\pm$ 0.1 &84.0 $\pm$ 0.1 &9.7\\
    &158971 &8.75 $\pm$ 0.58 &170.5 $\pm$ 2.6 &74.9 $\pm$ 5.7& 14.0&& 244935&261.53 $\pm$ 0.27 &185.9 $\pm$ 0.1 &82.8 $\pm$ 0.2 &5.7\\
    &172181 &5.27 $\pm$ 0.63 &174.4 $\pm$ 5.1 &91.5 $\pm$ 13.6&30.0 & &293912&190.67 $\pm$ 0.34 &180.8 $\pm$ 0.8 &80.7 $\pm$ 1.5 &7.0\\
    &178605 &25.31 $\pm$ 1.22 &166.1 $\pm$ 3.0 &82.2 $\pm$ 6.4&8.9&&342883&150.23 $\pm$ 0.39&185.0 $\pm$ 0.3&80.5 $\pm$ 0.7&8.7\\
    \cline{8-12}
    &215220 &3.53 $\pm$ 0.21 &177.6 $\pm$ 2.3 &75.6 $\pm$ 5.2&5.7 & \multicolumn{6}{c}{SO}\\
    \cline{8-12}
    &219949&11.07 $\pm$ 0.26 &180.7 $\pm$ 0.8 &68.0 $\pm$ 2.0 &6.2&&86093&20.28 $\pm$ 0.43 & 179.2 $\pm$ 0.7&67.7 $\pm$ 1.8 &8.8\\
      &296550 &2.88 $\pm$ 0.19&170.4 $\pm$ 3.4 &78.3 $\pm$ 9.0&5.0 & &100029 &5.07 $\pm$ 0.31 &161.8 $\pm$ 2.9 &51.4 $\pm$ 6.4 &10.2\\
    &304077 &3.76 $\pm$ 0.42 &176.5 $\pm$ 4.3 &92.0 $\pm$ 13.6&6.5 & &109252 &30.84 $\pm$ 0.20 &184.6 $\pm$ 0.2 &57.1 $\pm$ 0.4 &7.2\\
    &309502 &1.00 $\pm$ 0.20 &181.7 $\pm$ 8.4 &75.0 $\pm$ 10.0&5.9 & &129138 &36.62 $\pm$ 0.31 &183.7 $\pm$ 0.7 &64.1 $\pm$ 1.8 &8.3\\
    \cline{2-6}
    \multicolumn{6}{c|}{$^{34}$SO}& &136634 &5.16 $\pm$ 0.87 &186.4 $\pm$ 4.0 &57.1 $\pm$ 10.4 &17.2\\
     \cline{2-6}
      &97715 &3.44 $\pm$ 0.20 &174.1 $\pm$ 2.5 &72.4 $\pm$ 6.6&5.8 & &138178 & 76.26 $\pm$ 0.63&181.3 $\pm$ 0.4 &69.7 $\pm$ 0.9 &13.8\\
      \cline{2-6}
   \multicolumn{6}{c|}{OCS} & & 158971&41.02 $\pm$ 0.79 &187.0 $\pm$ 0.5 &56.8 $\pm$ 1.3 &12.8\\
  \cline{2-6}
  &85139 &7.04 $\pm$ 1.66 &141.7 $\pm$ 4.5 &54.9 $\pm$ 9.6 &8.1& &172181 &33.51 $\pm$ 0.41 &187.4 $\pm$ 0.3 &59.0 $\pm$ 0.9 &15.9\\
  &97301 &10.75 $\pm$ 0.25 &154.1 $\pm$ 0.8 &70.2 $\pm$ 1.8&5.8 &&178605 &66.46 $\pm$ 0.93 &187.1 $\pm$ 0.9 & 70.7 $\pm$ 2.1&35.0\\
  &109463 &9.52 $\pm$ 0.30 &133.2 $\pm$ 1.0 &61.9 $\pm$ 2.2&10.0 & &215220 &34.02 $\pm$ 0.29 &190.2 $\pm$ 0.4 &56.2 $\pm$ 0.9 &7.8\\
  &133785 &11.37 $\pm$ 0.41 &156.4 $\pm$ 1.4 &80.7 $\pm$ 3.6&8.5 & &219949 &54.40 $\pm$ 2.10 &190.0 $\pm$ 1.3 &67.9 $\pm$ 3.1 &6.9\\
  &158107 &7.59 $\pm$ 0.54 &160.0 $\pm$ 2.7 &71.9 $\pm$ 6.7&14.0 & &254573 &2.90 $\pm$ 0.72 &197.9 $\pm$ 8.8 &76.0 $\pm$ 24.1 &7.4\\
  &170267 &4.78 $\pm$ 0.93 &173.6 $\pm$ 4.2 &57.0 $\pm$ 11.0&9.6 & &296550 &33.20 $\pm$ 0.54 &193.0 $\pm$ 0.5 &61.2 $\pm$ 1.2 &6.9\\
  &194586 &4.72 $\pm$ 0.36 &165.3 $\pm$ 3.1 &74.1 $\pm$ 7.7&8.7 & &304077 &35.90 $\pm$ 0.51 &194.0 $\pm$ 0.5 &65.6 $\pm$ 1.1 &6.7\\
  &206745 &4.48 $\pm$  0.48 &179.7 $\pm$ 4.2 &70.7 $\pm$ 10.2 &11.0&&309502 &3.33 $\pm$ 0.23 &191.4 $\pm$ 2.1 &77.7 $\pm$ 4.7 &5.9\\
  &218903 &3.26 $\pm$ 0.21 &175.5 $\pm$ 2.3 &75.4 $\pm$ 5.9&4.3 & &339341&3.55 $\pm$ 0.57 &182.0 $\pm$ 5.8 &69.5 $\pm$ 13.0 &9.6\\
  &231060 &2.38 $\pm$ 0.27 &174.8 $\pm$ 4.2 &63.0 $\pm$ 11.6&7.0&&344310 &23.90 $\pm$ 0.31 &192.7 $\pm$ 0.3 &53.6 $\pm$ 0.8 &7.8\\
  \cline{2-6} \cline{8-12}
  \multicolumn{6}{c|}{H$_2$CS} & \multicolumn{6}{c}{$^{34}$SO}\\
  \cline{2-6}\cline{8-12}
  &101477 &13.25 $\pm$ 0.40 &143.4 $\pm$ 1.0 &68.6 $\pm$ 2.6 &7.7&&84410 &1.59 $\pm$ 0.19 & 174.8 $\pm$ 4.3&53.5 $\pm$ 9.8 &9.3\\
  &103040 &9.35 $\pm$ 0.27 &134.3 $\pm$ 1.1 &76.8 $\pm$ 2.7&6.5 & &97715&8.95 $\pm$ 0.40 &181.7 $\pm$ 1.0 &66.7 $\pm$ 2.4 &6.0\\
  &104617 &11.02 $\pm$ 0.79 &139.9 $\pm$ 2.3 &65.9 $\pm$ 5.6 &5.9&&126613 & 2.65 $\pm$ 0.26&166.8 $\pm$ 2.4 &40.0 $\pm$ 5.5 &9.3\\
  &137371 &12.33 $\pm$ 0.78 &154.8 $\pm$ 2.6 &84.9 $\pm$ 6.6&14.0 & &175352&5.90 $\pm$ 0.90 &175.1 $\pm$ 3.7 &44.1 $\pm$ 9.2 &27.1\\
  &139483 &11.00 $\pm$ 0.45 &158.2 $\pm$ 1.8 &71.4 $\pm$ 4.4&11.9 & &246663 &3.65 $\pm$ 0.08 &197.6 $\pm$ 1.9 &43.1 $\pm$ 4.7 &2.5\\
  &169113 &8.76 $\pm$ 0.49 &166.3 $\pm$ 2.0 &81.4 $\pm$ 4.6&11.8& &256877 &7.65 $\pm$ 0.30 &197.8 $\pm$ 1.4 &58.9 $\pm$ 3.4 &8.1\\
  &205987 &2.84 $\pm$ 0.50 &176.3 $\pm$ 5.2 &70.7 $\pm$ 12.9&11.2 & &298257 &5.91 $\pm$ 0.15 &194.9 $\pm$ 1.2 &49.0 $\pm$ 2.9 &4.2\\
  &236727 &4.18 $\pm$ 0.23 &168.9 $\pm$ 2.8 &80.7 $\pm$ 6.4&5.6 &&337582 &3.86 $\pm$ 0.29 &198.3 $\pm$ 1.8 &51.0 $\pm$ 4.9 &11.0\\
   \hline
  \multicolumn{6}{c|}{OCS} &\multicolumn{6}{c}{CCS}\\
  \cline{2-6}\cline{8-12}
  \multirow{54}{*}{GMC6}&85139 &9.38 $\pm$ 1.47 &140.6 $\pm$ 4.0 &57.1 $\pm$ 10.0&9.3 &\multirow{37}{*}{GMC6} &271292 &1.96 $\pm$ 0.27 &193.1 $\pm$ 2.2 &36.9 $\pm$ 6.2 &8.0\\
  &97301&10.36 $\pm$ 0.23 &164.4 $\pm$ 0.7 &60.2 $\pm$ 1.8&6.0 & &272592 &2.59 $\pm$ 0.48 &193.5 $\pm$ 3.4 &50.0 $\pm$ 10.0 &9.6\\
  &109463 &9.94 $\pm$ 0.58 & 143.6 $\pm$ 1.6&59.5 $\pm$ 4.2&10.2 & &310837 &1.24 $\pm$ 0.25 &199.7 $\pm$ 5.5 &40.3 $\pm$ 15.5 &9.0\\
  &133785 & 14.15 $\pm$ 0.22&166.0 $\pm$ 2.3 &66.5 $\pm$ 6.1&5.7 & &336258 &1.32 $\pm$ 0.25 &201.3 $\pm$ 10.1 &45.3 $\pm$ 8.2 &9.1\\
  &145946 &13.31 $\pm$ 0.71 &180.1 $\pm$ 1.4 &57.0 $\pm$ 3.6&5.5 & &336722 &1.12 $\pm$ 0.26 &201.3 $\pm$ 3.7 &44.8 $\pm$ 8.5 &9.1\\
  \cline{8-12}
   &158107 &11.33 $\pm$ 0.46 &172.6 $\pm$ 1.5 &49.6 $\pm$ 3.6&12.8 & \multicolumn{6}{c}{SO$_2$}\\
   \cline{8-12}
   &170267&12.36 $\pm$ 0.80 &185.8 $\pm$ 1.8 &61.5 $\pm$ 4.7&10.6 & & 104029&3.41 $\pm$ 1.89 &181.3 $\pm$ 2.0 &40.1 $\pm$ 10.0 &7.1\\
  &194586&11.59 $\pm$ 0.36 &183.8 $\pm$ 0.8 &53.4 $\pm$ 2.0&10.4 & & 104239&3.89 $\pm$ 0.34 &189.4 $\pm$ 1.9 &50.1 $\pm$ 4.0 &7.1\\
  &206745 &13.07 $\pm$ 0.38 & 183.4 $\pm$ 0.8&56.6 $\pm$ 2.3&10.9 &&129514&4.44 $\pm$ 0.38 &187.3 $\pm$ 2.3 &53.9 $\pm$ 5.7 &8.5 \\
  &218903 &10.88 $\pm$ 1.86 &188.2 $\pm$ 4.5 &56.9 $\pm$ 10.7&7.8 & & 131014&3.05 $\pm$ 0.25 &187.5 $\pm$ 1.6 & 50.1 $\pm$ 3.7&8.1 \\
  &231060 &10.60 $\pm$ 0.17 &191.2 $\pm$ 1.1 &56.0 $\pm$ 2.8&5.1 & &132744 &3.66 $\pm$ 0.23 &191.6 $\pm$ 2.3 &49.2 $\pm$ 5.1 &7.2\\
  &279685 &7.84 $\pm$ 0.30 &190.6 $\pm$ 4.0 &50.0 $\pm$ 9.9 &6.0&&140306&3.77 $\pm$ 0.42 &190.6 $\pm$ 2.2 & 47.5 $\pm$ 6.6&13.7\\
  &291839 &7.99 $\pm$ 0.85 &193.2 $\pm$ 2.3 &52.6 $\pm$ 6.6&5.6 & &158199 &1.68 $\pm$ 0.45 &187.5 $\pm$ 5.7 &42.3 $\pm$ 13.7 &15.7\\
   & 316146&9.31 $\pm$ 0.22 &197.6 $\pm$ 1.8 &67.9 $\pm$ 4.9&5.7 & &192651 &3.15 $\pm$ 0.41 &184.3 $\pm$ 4.6 &47.9 $\pm$ 11.7 &11.5\\
  & 352599&5.58 $\pm$ 0.35 &196.0 $\pm$ 1.8 &55.5 $\pm$ 3.8&9.9 & &200287 &1.99 $\pm$ 0.84 &191.7 $\pm$ 8.8 &43.5 $\pm$ 21.6 &29.3\\
  \cline{2-6}
  \multicolumn{6}{c|}{H$_2$CS} & &204246 &2.26 $\pm$ 0.41 &184.9 $\pm$ 1.5 & 37.1 $\pm$ 3.1&13.3\\
  \cline{2-6}
  &101477 &17.70 $\pm$ 0.37 &146.9 $\pm$ 0.6 &62.3 $\pm$ 1.6&8.0 & &205300 &2.63 $\pm$ 0.45 &195.3 $\pm$ 2.6 &41.3 $\pm$ 5.8 &13.3\\
  &103040&12.36 $\pm$ 0.31 &138.6 $\pm$ 1.5 &73.9 $\pm$ 4.0& 6.9& & 208700&2.96 $\pm$ 0.44 &187.4 $\pm$ 3.5 &53.9 $\pm$ 7.2 &13.5\\
   &104617 &15.60 $\pm$ 0.27 &142.5 $\pm$ 1.1 &65.8 $\pm$ 2.7&6.0 & &234187 &1.66 $\pm$ 0.10 &186.4 $\pm$ 2.5 &47.4 $\pm$ 5.6 &3.2\\
  &137371&14.43 $\pm$ 1.70 &160.8 $\pm$ 2.7 &72.5 $\pm$ 7.5&17.2 & &241615 &5.46 $\pm$ 0.19 & 192.3 $\pm$ 1.2&64.9 $\pm$ 3.3 &5.8\\
  &139483 &15.55 $\pm$ 0.52 &162.5 $\pm$ 1.1 &56.8 $\pm$ 2.8 &14.7 &&248057&1.92 $\pm$ 0.10 &193.1 $\pm$ 1.1 &37.0 $\pm$ 2.5 &3.5\\
   &169113 &15.95 $\pm$ 0.34 &178.0 $\pm$ 0.7 &66.4 $\pm$ 1.7& 8.1& &271529 &3.38 $\pm$ 0.26 &195.2 $\pm$ 1.3 &38.5 $\pm$ 3.6 &8.8\\
   &205987 &8.24 $\pm$ 1.10 &169.8 $\pm$ 4.0 &75.8 $\pm$ 13.2&10.9 & &288519 &2.34 $\pm$ 0.26 &195.9 $\pm$ 2.2 &43.2 $\pm$ 5.3 &8.5\\
  &236727 &8.85 $\pm$ 0.25 &178.7 $\pm$ 1.4 & 51.8 $\pm$ 3.2&7.8&&296168&2.23 $\pm$ 0.17&197.9 $\pm$ 2.5&57.2 $\pm$ 6.3&5.8\\
  &240548 & 2.35 $\pm$ 0.24&181.9 $\pm$ 12.8 &56.4 $\pm$ 7.8&12.2 &  & 332505& 4.21 $\pm$ 0.57&194.5 $\pm$ 2.7 &44.4 $\pm$ 7.3 &6.9\\
  &278886 &8.80 $\pm$ 0.35 &183.3 $\pm$ 1.3 &67.1 $\pm$ 3.1&6.0 & &334673 &2.61 $\pm$ 0.22 &199.9 $\pm$ 1.1 & 38.6 $\pm$ 4.2&5.7\\
  &313716 &5.87 $\pm$ 1.07 &190.4 $\pm$ 3.7 &48.5 $\pm$ 6.7&4.6 & &336089&1.96 $\pm$ 0.42 &203.2 $\pm$ 4.2 &50.5 $\pm$ 10.0 &8.5\\
 &348534 &7.94 $\pm$ 0.36 &200.2 $\pm$ 1.6 &58.7 $\pm$ 3.6&8.2 & &346652 &5.38 $\pm$ 0.52 & 199.4 $\pm$ 2.7&40.0 $\pm$ 10.0 &6.9\\
  &371847 &6.17 $\pm$ 0.40 &189.8 $\pm$ 2.1 &49.0 $\pm$ 6.0&11.4 & &351257&3.47 $\pm$ 0.31&195.5 $\pm$ 1.1&40.4 $\pm$ 2.8&9.9\\
  \cline{2-6}
  \multicolumn{6}{c|}{CCS}& &355045 & 2.16 $\pm$ 0.32& 197.6 $\pm$ 3.8&34.8 $\pm$ 7.7 &10.0\\
  \cline{2-6}
  &86181&7.18 $\pm$ 0.25 &166.8 $\pm$ 1.1 &49.9 $\pm$ 2.7&8.8 & &357165 &2.87 $\pm$ 0.70 &196.0 $\pm$ 10.0 & 50.0 $\pm$ 10.0&8.8\\
  &93870 &19.38 $\pm$ 0.27 &140.7 $\pm$ 0.8 &57.8 $\pm$ 2.0&7.0 & &357241 &3.19 $\pm$ 0.71 &198.0 $\pm$ 10.0 & 50.0 $\pm$ 10.0&8.8\\
  &99866 &7.48 $\pm$ 0.40 &148.8 $\pm$ 1.2 &56.1 $\pm$ 3.3&10.2 & &358215 &5.72 $\pm$ 0.43 &194.1 $\pm$ 1.7 &41.9 $\pm$ 4.4 &12.1\\
  &103640 &6.67 $\pm$ 0.33 &151.7 $\pm$ 1.6 &54.2 $\pm$ 3.7& 10.1& &359770 &2.33 $\pm$ 0.23 &197.1 $\pm$ 1.7 &41.5 $\pm$ 3.5 &6.8\\
   &106347 &14.23 $\pm$ 0.39 &146.2 $\pm$ 0.8 &54.0 $\pm$ 1.2&8.7 & &363159 &2.31 $\pm$ 0.37 &193.0 $\pm$ 3.3 &38.7 $\pm$ 6.0 &9.7\\
   &113410& 6.82 $\pm$ 0.32&158.5 $\pm$ 1.5 &62.5 $\pm$ 3.6&11.2 & &371172 &3.62 $\pm$ 0.67 &202.1 $\pm$ 3.9 &48.7 $\pm$ 11.3 &13.0\\
   \cline{7-12}
  &129548&8.12 $\pm$ 0.56 &174.1 $\pm$ 1.9 &60.1 $\pm$ 5.3&8.3 &\multicolumn{6}{c}{CS}\\
  \cline{8-12}
  &131551 &12.00 $\pm$ 0.33 &172.2 $\pm$ 0.9 &55.5 $\pm$ 2.2 &8.3 &\multirow{15}{*}{SSC5}&97980 &318.19 $\pm$ 0.46 & 238.0 $\pm$ 0.1& 79.2 $\pm$ 0.1&10.0\\
  &140180 &6.25 $\pm$ 0.28 &176.6 $\pm$ 1.1 &51.9 $\pm$ 2.6&8.4 & &146969 &291.58 $\pm$ 0.30 &227.2 $\pm$ 1.0 &76.8 $\pm$ 2.4 &7.1\\
   &144244 &10.71 $\pm$ 0.29 &181.8 $\pm$ 0.6 &52.7 $\pm$ 1.6&8.0 & &195954 &244.05 $\pm$ 0.44 &240.3 $\pm$ 0.1 &75.0 $\pm$ 0.2 &10.5\\
  &156981 & 8.26 $\pm$ 0.45&184.3 $\pm$ 4.2 &52.0 $\pm$ 10.8&12.6 & &244935 &190.87 $\pm$ 0.24 &241.9 $\pm$ 0.1 &73.9 $\pm$ 0.2 &5.4\\
  &166662 & 7.18 $\pm$ 0.43&190.5 $\pm$ 3.0 &65.9 $\pm$ 7.3&11.0 & &293912 &132.10 $\pm$ 0.36 &232.3 $\pm$ 0.3 &68.8 $\pm$ 0.8 &8.1\\
  &168406 &5.55 $\pm$ 0.24 &190.5 $\pm$ 2.1 &45.8 $\pm$ 5.3&8.1 & &342883&100.35 $\pm$ 0.36 &232.8 $\pm$ 0.3 &69.4 $\pm$ 0.8 &9.1\\
  \cline{8-12}
  &169753 & 8.99 $\pm$ 9.28&174.2 $\pm$ 1.9 &70.7 $\pm$ 5.3&8.1 & \multicolumn{6}{c}{H$_2$S}\\
  \cline{8-12}
   &192961 &4.30 $\pm$ 0.80 &185.0 $\pm$ 3.4 &42.7 $\pm$ 17.5&11.1 & &168762 &53.22 $\pm$ 0.55 &230.6 $\pm$ 0.6 &72.4 $\pm$ 1.4 &13.8\\
  &195375&6.78 $\pm$ 0.35 &183.9 $\pm$ 2.2 &53.0 $\pm$ 6.2& 10.4& &216710 &40.32 $\pm$ 0.36 &234.0 $\pm$ 0.3 &62.7 $\pm$ 0.6 &8.9\\
  &207260&3.72 $\pm$ 0.38 &189.9 $\pm$ 3.4 &55.6 $\pm$ 9.1&10.9 & &300505 &20.77 $\pm$ 0.32 &228.9 $\pm$ 0.4 &62.7 $\pm$ 0.9 &7.6\\
  \cline{8-12}
  &221071&3.56 $\pm$ 0.23 &185.4 $\pm$ 1.1 &37.7 $\pm$ 3.2&6.9 & \multicolumn{6}{c}{H$_2^{34}$S}\\
  \cline{8-12}
  &232201&2.70 $\pm$ 0.30 &191.1 $\pm$ 2.1 &47.0 $\pm$ 5.7&8.2 & &167910 &11.48 $\pm$ 0.49 &215.3 $\pm$ 1.8 &63.7 $\pm$ 4.0 &13.0\\
  \cline{8-12}
  &233159&2.86 $\pm$ 0.24 &186.4 $\pm$ 2.7 &41.6 $\pm$ 4.2&5.5 & \multicolumn{6}{c}{SO}\\
  \cline{8-12}
  &233938&4.14 $\pm$ 0.47 &190.2 $\pm$ 2.6 &49.0 $\pm$ 6.1&5.5 & & 86093&12.04 $\pm$ 0.75 &228.4 $\pm$ 1.8 &63.8 $\pm$ 4.9 &7.7\\
  &245244&2.48 $\pm$ 0.15 &196.4 $\pm$ 4.3 &46.7 $\pm$ 11.1&5.8 & & 109252&19.85 $\pm$ 0.30 &237.6 $\pm$ 0.4 &59.1 $\pm$ 1.0 &7.4\\
 \hline 
  \multicolumn{6}{c|}{SO}& \multicolumn{6}{c}{CCS}\\
  \cline{2-6}\cline{8-12}
  \multirow{50}{*}{SSC5}& 129138&20.42 $\pm$ 0.30 &229.3 $\pm$ 1.2 &70.5 $\pm$ 2.7&8.7 &\multirow{25}{*}{SSC5} &166662 &2.99 $\pm$ 0.41 &238.6 $\pm$ 4.8 &62.5 $\pm$ 10.9 &13.0\\
  & 138178&45.71 $\pm$ 0.48 &236.7 $\pm$ 0.7 &66.5 $\pm$ 1.7&13.7 & &169753 &3.61 $\pm$ 0.42 &217.0 $\pm$ 4.1 & 67.8 $\pm$ 8.3&13.8\\
  &158971 &27.93 $\pm$ 0.52 &235.5 $\pm$ 0.6 &63.4 $\pm$ 1.4&14.9 & &195375 &4.04 $\pm$ 0.30 &220.0 $\pm$ 10.0 &60.0 $\pm$ 10.0 &9.5\\
  &178605&37.75 $\pm$ 1.63 &234.1 $\pm$ 1.6 &61.7 $\pm$ 3.7&46.6 & &207260 &2.34 $\pm$ 0.50 &219.1 $\pm$ 5.1 &49.4 $\pm$ 13.0 &12.0\\
  &172181 &19.77 $\pm$ 0.38 & 227.9 $\pm$ 1.5 &70.4 $\pm$ 3.6&10.6 & &232201 & 1.57 $\pm$ 0.14&219.8 $\pm$ 4.4 & 54.2 $\pm$ 9.2&4.9\\
  &215220 &19.61 $\pm$ 0.25 &226.2 $\pm$ 0.4 &65.3 $\pm$ 0.8& 6.4& &233159 &1.71 $\pm$ 0.15 &208.9 $\pm$ 4.5&71.9 $\pm$ 9.7 &4.0\\
  &219949 &33.25 $\pm$ 1.11 &235.4 $\pm$ 1.1 &68.1 $\pm$ 2.6&3.7 & &233938 &1.85 $\pm$ 0.24 & 230.7 $\pm$ 3.8& 56.0 $\pm$ 8.2&4.9\\
   &296550 &18.36 $\pm$ 0.25 & 224.2 $\pm$ 0.6&64.1 $\pm$ 1.3& 7.2&  &271292 &1.47 $\pm$ 0.29 &225.8 $\pm$ 5.6 &48.2 $\pm$ 10.6 &9.7\\
   &304077&27.19 $\pm$ 2.29 &226.1 $\pm$ 2.9 &80.0 $\pm$ 8.9&7.9 & &272592 &1.24 $\pm$ 0.38 &203.5 $\pm$ 4.4 &57.0 $\pm$ 12.3 &12.2\\
   \cline{8-12}
  &309502&2.07 $\pm$ 0.23 &236.3 $\pm$ 4.7 &56.0 $\pm$ 10.1&6.5 & \multicolumn{6}{c}{SO$_2$}\\
  \cline{8-12}
    &344310 &12.80 $\pm$ 0.35 &218.6 $\pm$ 0.6 &59.1 $\pm$ 1.3&10.3 & &129514 &1.46 $\pm$ 0.40 &233.0 $\pm$ 7.6 & 53.1 $\pm$ 14.2&8.1\\
\cline{2-6}
\multicolumn{6}{c|}{$^{34}$SO}&&131014 & 1.18 $\pm$ 0.23&229.2 $\pm$ 5.6 &45.2 $\pm$ 9.8 &7.2\\
\cline{2-6}
&97715&5.40 $\pm$ 0.35 &243.4 $\pm$ 2.4 &77.3 $\pm$ 6.1&6.5 & &132744 &1.55 $\pm$ 0.26 &233.9 $\pm$ 10.0 & 75.4 $\pm$ 18.7&7.5\\
  &175352 &3.30 $\pm$ 0.52 &213.8 $\pm$ 4.2 &52.2 $\pm$ 8.8&27.1 & &140306& 1.16 $\pm$ 0.33&225.1 $\pm$ 8.3 & 40.4 $\pm$ 15.8&12.3\\
  &246663 &1.80 $\pm$ 0.50 &216.5 $\pm$ 6.8 &47.5 $\pm$ 15.2&14.9 & &158199 & 1.12 $\pm$ 0.42& 228.5 $\pm$ 10.3&45.4 $\pm$ 16.9 &14.7\\
   &256877 &4.26 $\pm$ 0.27 & 231.0 $\pm$ 4.9&69.5 $\pm$ 9.2&7.6 & &205300 &0.98 $\pm$ 0.36 & 235.0 $\pm$ 10.0& 45.0 $\pm$ 10.0&11.5\\
  &298257&3.97 $\pm$ 0.20 &219.4 $\pm$ 4.2 &65.2 $\pm$ 9.8&6.0 & & 271529 &1.71 $\pm$ 0.22 &225.2 $\pm$ 4.9&59.8 $\pm$ 0.1 &7.8\\
  &337582&2.29 $\pm$ 0.27 & 218.4 $\pm$ 3.3&55.4 $\pm$ 7.9&7.6 & & 296168 &0.52 $\pm$ 0.16 & 235.9 $\pm$ 3.9& 31.2 $\pm$ 8.1&6.7\\
  \cline{2-6}
  \multicolumn{6}{c|}{OCS} & &334673&1.07 $\pm$ 0.31 &246.1 $\pm$ 4.8 & 30.3 $\pm$ 10.3&5.6\\
  \cline{2-6}
  &97301&3.70 $\pm$ 0.23&227.0 $\pm$ 1.6 & 57.1 $\pm$ 4.5&6.5 & &351257 &1.67 $\pm$ 0.20 &213.7 $\pm$ 4.9 &53.3 $\pm$ 10.9&6.0\\
  &109463&4.58 $\pm$ 0.33 & 227.0 $\pm$ 10.0&57.0 $\pm$ 10.0&8.9 & &355045 &1.16 $\pm$ 0.20 & 223.6 $\pm$ 5.5&49.0 $\pm$ 9.2 &6.3\\
  &133785&5.36 $\pm$ 0.75 &228.5 $\pm$ 8.2 &57.0 $\pm$ 10.0&7.7 & &358215 &2.99 $\pm$ 0.35 &223.7 $\pm$ 4.9 & 75.6 $\pm$ 17.1&9.3\\
  &158107&4.17 $\pm$ 0.52 &225.3 $\pm$ 3.3 &58.3 $\pm$ 7.2&14.9 & &359770 &1.03 $\pm$ 0.20 &215.0 $\pm$ 3.4 &35.1 $\pm$ 6.6 &8.0\\
  &194586&5.45 $\pm$ 0.39 &229.0 $\pm$ 3.9 &79.8 $\pm$ 9.0&9.5 & &363159 &1.34 $\pm$ 0.24 &239.6 $\pm$ 4.8 &49.5 $\pm$ 10.5 &8.4\\
  &206745&4.21 $\pm$ 0.42 &222.7 $\pm$ 2.8 &61.7 $\pm$ 7.9&12.0 & &371172 &2.95 $\pm$ 0.43 &243.5 $\pm$ 3.5 &53.0 $\pm$ 9.1 &13.1\\
  \cline{7-12}
  & 231060&4.87 $\pm$ 0.21 &230.4 $\pm$ 2.4 &72.1 $\pm$ 5.2 &5.3& \multicolumn{6}{c}{CS}\\
  \cline{8-12}
  &279685 &3.30 $\pm$ 0.21 &221.2 $\pm$ 3.6 &66.7 $\pm$ 7.7& 5.7&\multirow{29}{*}{SSC2} &97980 &322.20 $\pm$ 0.32 &267.3 $\pm$ 0.1 & 69.3 $\pm$ 0.1&9.1\\
   &316146 &4.00 $\pm$ 0.29 &222.7 $\pm$ 2.8 &72.4 $\pm$ 6.0&8.3 & &146969 & 269.30 $\pm$ 0.20&260.8 $\pm$ 0.9 &67.6 $\pm$ 2.2 &5.0\\
  &352599 &1.58 $\pm$ 0.38 &225.0 $\pm$ 10.0 &70.0 $\pm$ 10.0&10.5 & &195954 &207.80 $\pm$ 0.38 &276.9 $\pm$ 0.1 &67.4 $\pm$ 0.1 &9.4\\
  \cline{2-6}
  \multicolumn{6}{c|}{H$_2$CS}& &244935 &151.23 $\pm$ 0.25 &280.9 $\pm$ 0.1 &65.4 $\pm$ 0.2 &6.4\\
  \cline{2-6}
 &101477 &8.32 $\pm$ 0.36 &211.6 $\pm$ 1.3 &61.6 $\pm$ 3.3&8.9&&293912&96.33 $\pm$ 0.25 &276.0 $\pm$ 0.5 &62.1 $\pm$ 1.1 &6.3\\
 &103040&5.15 $\pm$ 0.24 &202.8 $\pm$ 1.7 &64.2 $\pm$ 4.6&6.5& &342883 &64.97 $\pm$ 0.33 &281.1 $\pm$ 0.5 &63.5 $\pm$ 1.2 &8.3\\
 \cline{8-12}
  &104617&7.41 $\pm$ 0.22 &206.9 $\pm$ 1.0 &57.0 $\pm$ 2.6&5.9 & \multicolumn{6}{c}{H$_2$S}\\
  \cline{8-12}
  &139483&8.37 $\pm$ 0.39 &224.8 $\pm$ 2.0 &58.3 $\pm$ 5.0&12.8 & &168762 &31.22 $\pm$ 0.47 &277.4 $\pm$ 0.6 &74.8 $\pm$ 1.5 &10.6\\
  &169113&5.88 $\pm$ 0.44 &231.9 $\pm$ 1.9 &54.4 $\pm$ 4.9&13.8 & &216710 &26.35 $\pm$ 0.18 &288.4 $\pm$ 0.3 &55.5 $\pm$ 0.6 &5.0\\
  &236727&4.11 $\pm$ 0.26 &232.2 $\pm$ 1.5 &61.4 $\pm$ 3.8&8.2 & &300505 &9.21 $\pm$ 0.28 &289.7 $\pm$ 0.6 & 44.3 $\pm$ 11.9&8.0\\
  \cline{8-12}
  &240548&0.86 $\pm$ 0.20 &216.2 $\pm$ 9.6 &57.8 $\pm$ 17.1&6.3 &  \multicolumn{6}{c}{H$_2^{34}$S}\\
  \cline{8-12}
  &278886 &3.83 $\pm$ 0.21 &226.2 $\pm$ 2.7 &65.5 $\pm$ 5.5&5.7 & &167910 &8.54 $\pm$ 0.47 &277.2 $\pm$ 2.8 &63.1 $\pm$ 6.1 &11.3\\
  \cline{8-12}
   &348534 &3.59 $\pm$ 0.40 &231.6 $\pm$ 3.1 &57.0 $\pm$ 6.0& 10.0& \multicolumn{6}{c}{SO}\\
   \cline{8-12}
  &371847 &2.27 $\pm$ 0.50 &224.8 $\pm$ 6.4 &53.3 $\pm$ 13.5&13.9 & &86093 &9.88 $\pm$ 0.29 &280.9 $\pm$ 0.9 &54.5 $\pm$ 2.2 &9.3\\
  \cline{2-6}
  \multicolumn{6}{c|}{CCS}& &100029 &2.42 $\pm$ 0.20 &272.1 $\pm$ 1.5 &36.3 $\pm$ 3.7 &8.7\\
  \cline{2-6}
   &86181&5.75 $\pm$ 0.85 &231.9 $\pm$ 3.1 &55.1 $\pm$ 11.2&7.7 & &109252 &17.20 $\pm$ 0.22 &284.2 $\pm$ 0.7 &55.1 $\pm$ 1.6 &6.0\\
  &93870&13.37 $\pm$ 0.32 &205.7 $\pm$ 0.7 &57.7 $\pm$ 2.0&8.0 & &129138 &14.85 $\pm$ 0.66 &287.6 $\pm$ 1.1 &50.9 $\pm$ 2.7 &7.8\\
  &99866& 4.60 $\pm$ 0.35&206.2 $\pm$ 2.1 &55.4 $\pm$ 4.9&9.7 & &136634 &2.76 $\pm$ 0.46 &288.6 $\pm$ 4.9 &54.8 $\pm$ 9.9 &14.6\\
  &103640&5.55 $\pm$ 0.35 &208.4 $\pm$ 1.7 &53.2 $\pm$ 4.2&9.6 & &138178 &41.00 $\pm$ 0.42 &282.3 $\pm$ 0.3 &57.2 $\pm$ 0.7 &11.7\\
  &106347&9.94 $\pm$ 0.29 &207.7 $\pm$ 0.7 &55.5 $\pm$ 1.9&7.0 & &158971 &21.97 $\pm$ 0.50 &285.5 $\pm$ 0.4 &56.6 $\pm$ 1.0 &13.4\\
  &113410 &4.16 $\pm$ 0.30 &218.3 $\pm$ 1.6 &44.9 $\pm$ 4.0& 10.9& &172181 &12.34 $\pm$ 0.41 &290.0 $\pm$ 2.3 &53.1 $\pm$ 6.0 &11.8\\
  &129548 &5.00 $\pm$ 0.34 &231.0 $\pm$ 2.2 &66.8 $\pm$ 5.6& 7.6& & 178605&27.76 $\pm$ 1.29 &283.6 $\pm$ 1.6 &53.5 $\pm$ 3.7 &40.9\\
   &131551 &5.58 $\pm$ 0.27 &226.5 $\pm$ 1.5 &56.1 $\pm$ 3.4&6.8 & &215220 &11.33 $\pm$ 0.13 &295.4 $\pm$ 0.6 &46.1 $\pm$ 1.7 &4.2\\
  &140180 &3.90 $\pm$ 0.23 &227.5 $\pm$ 2.4 &58.5 $\pm$ 5.5&7.2 & &219949 &22.76 $\pm$ 0.18 &290.8 $\pm$ 1.0 &61.2 $\pm$ 2.4 &5.6\\
  &144244 &5.51 $\pm$ 0.28 &237.6 $\pm$ 1.1 &48.9 $\pm$ 3.0&7.5 & &296550 &10.89 $\pm$ 0.20 &296.4 $\pm$ 0.6 &55.9 $\pm$ 1.8 &6.0\\
 &156981 &5.47 $\pm$ 0.47 &233.4 $\pm$ 5.5 &75.3 $\pm$ 20.8 &13.1&&304077&9.85 $\pm$ 0.56 &299.6 $\pm$ 1.2 &43.7 $\pm$ 3.1 &8.1\\
 \hline 
 \multicolumn{6}{c|}{SO} &\multicolumn{6}{c}{CCS} \\
 \hline
  \multirow{52}{*}{SSC2}&309502&1.46 $\pm$ 0.19 &291.5 $\pm$ 5.0 &42.9 $\pm$ 10.0&7.8 &\multirow{33}{*}{SSC2} & 169753&3.62 $\pm$ 0.36 &277.4 $\pm$ 5.1 &58.4 $\pm$ 14.2 &11.0\\
  &339341 &1.47 $\pm$ 0.28 &295.6 $\pm$ 4.0 &43.6 $\pm$ 10.1 &9.7&&195375 &1.77 $\pm$ 0.36 &276.1 $\pm$ 5.2 &68.9 $\pm$ 14.0 &10.5\\
  &344310 &5.13 $\pm$ 0.29 & 300.4 $\pm$ 0.7&30.4 $\pm$ 1.8& 11.7& &221071 &1.08 $\pm$ 0.22 &277.8 $\pm$ 6.1 &52.0 $\pm$ 11.2 &5.6\\
  \cline{2-6}
  \multicolumn{6}{c|}{$^{34}$SO}& &233938 &0.82 $\pm$ 0.11 &295.1 $\pm$ 6.6 &37.0 $\pm$ 17.4 &4.1\\
  \cline{2-6}\cline{8-12}
  &97715 &5.93 $\pm$ 0.30 & 286.5 $\pm$ 1.7&58.1 $\pm$ 4.6&7.3&\multicolumn{6}{c}{SO$_2$}\\
  \cline{8-12}
  &126613 &1.36 $\pm$ 0.20 &274.2 $\pm$ 5.9 &39.2 $\pm$ 13.5&8.3 & &129514 &2.20 $\pm$ 0.18 &297.4 $\pm$ 1.6 &30.7 $\pm$ 0.1 &8.2\\
   &175352 &2.30 $\pm$ 0.55 &278.7 $\pm$ 2.4 &22.9 $\pm$ 6.4&25.1 & &131014 &1.69 $\pm$ 0.20 &298.9 $\pm$ 2.1 &30.8 $\pm$ 5.3 &8.5\\
  & 246663&1.52 $\pm$ 0.19 &305.6 $\pm$ 2.5 &23.8 $\pm$ 6.0&8.3 & &132744 &2.17 $\pm$ 0.17  &298.2 $\pm$ 1.6 &24.4 $\pm$ 4.1 &7.0\\
  & 256877&2.46 $\pm$ 0.23 & 306.7 $\pm$ 1.5&28.7 $\pm$ 4.6&8.9 & &134004 &1.53 $\pm$ 0.30 &283.3 $\pm$ 1.6 &22.6 $\pm$ 4.3 &7.0\\
  &298257 &1.91 $\pm$ 0.17 &303.9 $\pm$ 1.7 &30.0 $\pm$ 3.9&7.1 & &140306 &2.09 $\pm$ 0.34 &291.3 $\pm$ 3.3 &32.8 $\pm$ 8.1 &14.3\\
  &339858 &1.08 $\pm$ 0.43 &302.4 $\pm$ 3.9 &20.1 $\pm$ 9.3&13.2& &158199&0.83 $\pm$ 0.35 &296.6 $\pm$ 5.8 &32.8 $\pm$ 8.1 &13.2\\
  \cline{2-6}
   \multicolumn{6}{c|}{OCS}& &192651 &1.26 $\pm$ 0.29 &300.0 $\pm$ 10.0 &25.0 $\pm$ 10.0 &13.0\\
   \cline{2-6}
  &85139&9.16 $\pm$ 1.27 &247.0 $\pm$ 3.6 &52.3 $\pm$ 8.6&9.9 & &204246 &1.02 $\pm$ 0.19 &293.5 $\pm$ 1.9 &20.9 $\pm$ 4.4 &11.4\\
  &97301 &8.31 $\pm$ 0.29 &267.1 $\pm$ 0.7 &50.0 $\pm$ 1.6 &7.3&&205300&1.27 $\pm$ 0.27 &299.0 $\pm$ 1.4 &17.8 $\pm$ 3.3 &11.4\\
  &109463 &8.83 $\pm$ 0.29 &245.8 $\pm$ 1.2 &51.7 $\pm$ 2.7&19.2 & &208700 &1.38 $\pm$ 0.30 &297.4 $\pm$ 1.6 &21.9 $\pm$ 4.1 &11.4\\
   &133785 &9.81 $\pm$ 0.24 &270.8 $\pm$ 1.3 &57.5 $\pm$ 3.2&7.7 & &234187 &0.96 $\pm$ 0.11 &300.2 $\pm$ 2.6 &26.8 $\pm$ 5.5 &4.3\\
  &145946 &8.66 $\pm$ 0.73 &284.1 $\pm$ 0.8 &51.0 $\pm$ 1.9&20.5 & &241615 &2.21 $\pm$ 0.26 &295.2 $\pm$ 1.8 &32.5 $\pm$ 4.9 &4.4\\
  &158107 & 8.24 $\pm$ 0.43&276.6 $\pm$ 1.8 &50.6 $\pm$ 3.9&13.4 & &248057 &1.09 $\pm$ 0.21 &305.5 $\pm$ 1.5 &24.6 $\pm$ 2.9 &8.7\\
  &170267 & 7.70 $\pm$ 0.31&286.6 $\pm$ 1.5 &57.0 $\pm$ 3.6&9.4 & &255553 &1.10 $\pm$ 0.52 &305.3 $\pm$ 5.6 &24.5 $\pm$ 14.5 &6.7\\
  &194586 & 6.69 $\pm$ 0.41&290.4 $\pm$ 1.3 &50.8 $\pm$ 3.2&10.6 & &267719 &0.92 $\pm$ 0.21 &305.5 $\pm$ 2.9 &23.5 $\pm$ 5.0 &6.0\\
   &206745 &6.07 $\pm$ 0.38 &287.2 $\pm$ 2.1 &55.3 $\pm$ 5.2&11.2 & &271529 &2.01 $\pm$ 0.25 &304.5 $\pm$ 1.0 &23.9 $\pm$ 2.7 &9.6\\
  &218903&4.66 $\pm$ 0.25 &295.3 $\pm$ 4.4 &44.9 $\pm$ 12.2&8.2 & &296168 &0.82 $\pm$ 0.10 &301.6 $\pm$ 2.1 &24.7 $\pm$ 7.3 &5.9\\
  &231060 &4.95 $\pm$ 0.22 &294.7 $\pm$ 1.5 &49.6 $\pm$ 4.1&7.2 & &332505 &1.65 $\pm$ 0.25 &300.8 $\pm$ 2.0 &26.3 $\pm$ 4.8 &8.4\\
   &279685 &3.20 $\pm$ 0.17 &300.8 $\pm$ 1.1 &30.9 $\pm$ 4.0&5.1 & &334673 & 1.28 $\pm$ 0.14&302.3 $\pm$ 1.3&22.7 $\pm$ 3.0 &6.4\\
  &291839 &3.22 $\pm$ 0.14 &304.0 $\pm$ 2.1 &36.7 $\pm$ 7.8&5.2 & &336089 &0.87 $\pm$ 0.21 &311.9 $\pm$ 3.0 &27.3 $\pm$ 6.1 &8.4\\
  &316146 &2.70 $\pm$ 0.61 &298.8 $\pm$ 3.8 &35.1 $\pm$ 9.9& 5.6& &351257 & 1.58 $\pm$ 0.26&303.0 $\pm$ 5.4 &31.0 $\pm$ 6.1 &7.7\\
  &352599 &1.73 $\pm$ 0.26 &304.0 $\pm$ 2.6 &32.0 $\pm$ 6.1&10.3 & &355045 &1.50 $\pm$ 0.11 &305.1 $\pm$ 1.5 &23.8 $\pm$ 3.3 &4.3\\
  \cline{2-6}
   \multicolumn{6}{c|}{H$_2$CS}& &357165  &1.78 $\pm$ 0.17 &307.6 $\pm$ 3.0 &27.6 $\pm$ 7.1 &6.6\\
   \cline{2-6}
   &101477 &12.08 $\pm$ 0.29 &248.5 $\pm$ 0.7 &51.6 $\pm$ 1.7&7.6 & &357241  &1.81 $\pm$ 0.17 &306.5 $\pm$ 4.5 &34.9 $\pm$ 12.8 &6.6\\
   &103040 &8.33 $\pm$ 0.28 & 240.7 $\pm$ 1.0&66.6 $\pm$ 2.3 &7.3&&358215&1.97 $\pm$ 0.24 &306.0 $\pm$ 1.6 &24.9 $\pm$ 3.2 &9.6\\
  &104617 & 10.89 $\pm$ 0.27& 244.8 $\pm$ 1.2&55.2 $\pm$ 3.0&6.7 & &359770 &1.41 $\pm$ 0.24 &305.9 $\pm$ 2.5 &31.3 $\pm$ 5.4 &9.1\\
   &137371 &12.68 $\pm$ 1.22 &265.4 $\pm$ 3.1 &64.3 $\pm$ 7.3& 14.6& &363159 &1.69 $\pm$ 0.26 &299.6 $\pm$ 1.6 &28.6 $\pm$ 3.5 &7.9\\
  &139483 &11.08 $\pm$ 0.42 &265.0 $\pm$ 1.3 &56.6 $\pm$ 2.9&11.6 & &371172 &2.01 $\pm$ 0.34 &307.7 $\pm$ 3.0 &28.9 $\pm$ 8.0 &11.7\\
  \cline{7-12}
  &169113 &8.88 $\pm$ 0.35 & 277.8 $\pm$ 1.5&57.9 $\pm$ 3.6& 10.6&\multicolumn{6}{c}{CS}\\
  \cline{8-12}
  &205987 &3.89 $\pm$ 1.23 &274.7 $\pm$ 9.2 &60.1 $\pm$ 10.4&11.2 &\multirow{19}{*}{GMC1a} &97980 &109.18 $\pm$ 0.42 &304.2 $\pm$ 0.2 &80.3 $\pm$ 0.4 &9.5\\
  &236727 &6.15 $\pm$ 0.25 & 282.2 $\pm$ 1.2&55.5 $\pm$ 3.0&7.6 & &146969 &86.71 $\pm$ 0.37 &298.4 $\pm$ 0.9 &81.3 $\pm$ 2.3 &8.7\\
   &240548 &0.88 $\pm$ 0.15 &280.9 $\pm$ 3.3 &42.4 $\pm$ 6.6&5.3 & &195954 & 55.15 $\pm$ 0.40&315.3 $\pm$ 0.4 &81.7 $\pm$ 1.0 &9.1\\
  &278886&3.96 $\pm$ 0.42 &282.5 $\pm$ 4.1 &76.1 $\pm$ 9.2& 5.1& &244935 &37.83 $\pm$ 0.21 &319.8 $\pm$ 0.4 &79.8 $\pm$ 1.0 &5.1\\
  &348534 &2.90 $\pm$ 0.29 &304.9 $\pm$ 3.3 &52.1 $\pm$ 6.6&8.4 & &293912 & 18.89 $\pm$ 0.27&314.0 $\pm$ 0.7 &78.3 $\pm$ 1.6 &7.0\\
   &371847 &1.79 $\pm$ 0.32 &300.1 $\pm$ 3.0 &23.4 $\pm$ 5.9& 12.2& &342883 & 12.24 $\pm$ 0.31&318.3 $\pm$ 1.0 &85.5 $\pm$ 2.5 &7.4\\
   \cline{2-6}\cline{8-12}
 \multicolumn{6}{c|}{CCS} & \multicolumn{6}{c}{H$_2$S}\\
 \cline{2-6}\cline{8-12}
  &86181&4.38 $\pm$ 0.31  & 270.2 $\pm$ 1.5&50.2 $\pm$ 3.2&9.3 & &168762 & 8.68 $\pm$ 0.45& 311.0 $\pm$ 2.4& 72.8 $\pm$ 5.8&12.0\\
  \cline{8-12}
  &93870& 11.86 $\pm$ 0.28 &243.9 $\pm$ 0.6 &53.8 $\pm$ 1.5 & 7.5&\multicolumn{6}{c}{SO}\\
  \cline{8-12}
   &99866&3.86 $\pm$ 0.29  & 247.7 $\pm$ 2.7&48.9 $\pm$ 5.9&8.7 & &109252&2.24 $\pm$ 0.24 &297.5 $\pm$ 2.2 & 43.2 $\pm$ 7.0&6.9\\
   &103640&3.03 $\pm$ 0.40&246.6 $\pm$ 3.9&59.0 $\pm$ 7.5&9.6&&129138 &2.19 $\pm$ 0.20 &299.2 $\pm$ 3.2 &40.0 $\pm$ 10.0 &8.8\\
   & 106347&8.32 $\pm$ 0.31 &246.0 $\pm$ 0.8 &50.9 $\pm$ 2.0& 7.7&  &138178 &15.17 $\pm$ 0.45 &307.9 $\pm$ 1.1 &65.6 $\pm$ 2.4 &11.5\\
   & 113410&2.92 $\pm$ 0.38 &258.8 $\pm$ 2.9 &49.5 $\pm$ 5.9 &11.2 &&158971&3.58 $\pm$ 0.47 &308.9 $\pm$ 5.0 &61.8 $\pm$ 12.5 &13.7\\
   &129548 &3.56 $\pm$ 0.23 &277.9 $\pm$ 3.9 &64.3 $\pm$ 10.7&7.8 & &178605 &10.01 $\pm$ 1.58 & 303.5 $\pm$ 6.3& 54.8 $\pm$ 19.4&49.5\\
   &131551 & 5.37 $\pm$ 0.26&271.5 $\pm$ 1.4 &50.4 $\pm$ 3.3&8.1 & &215220 &0.67 $\pm$ 0.12 &298.7 $\pm$ 3.7 &37.7 $\pm$ 8.5 &4.2\\
  &140180 & 2.72 $\pm$ 0.24&275.4 $\pm$ 2.1 &53.9 $\pm$ 4.4&7.5 & &219949 & 3.64 $\pm$ 0.25& 306.5 $\pm$ 1.9&57.1 $\pm$ 4.7 &4.1\\
  \cline{8-12}
  &144244 &4.59 $\pm$ 0.28 &276.2 $\pm$ 1.5 &54.4 $\pm$ 3.4&8.5 & \multicolumn{6}{c}{$^{34}$SO}\\
  \cline{8-12}
  &156981 & 3.95 $\pm$ 0.55&284.4 $\pm$ 4.7 &65.7 $\pm$ 10.6&15.9 & &97715&1.27 $\pm$ 0.18 &303.4 $\pm$ 3.7 &50.6 $\pm$ 8.5 & 6.5\\
  \hline 
  \multicolumn{6}{c|}{OCS}&\multicolumn{6}{c}{H$_2$CS}\\
  \cline{2-6}\cline{8-12}
   \multirow{21}{*}{GMC1a}&85139&7.60 $\pm$ 0.30 &266.9 $\pm$ 1.3 & 58.3 $\pm$ 4.1& 8.3&\multirow{8}{*}{GMC1b}& 104617&3.55 $\pm$ 0.23 &235.8 $\pm$ 1.6 &59.2 $\pm$ 3.9 &6.2\\
   &97301 &6.15 $\pm$ 0.26 &282.6 $\pm$ 1.1 &50.3 $\pm$ 2.5&6.5 & &137371 &2.50 $\pm$ 0.32 &249.4 $\pm$ 5.4 &65.0 $\pm$ 15.0 &10.5\\
   &109463 &4.47 $\pm$ 0.35 &281.0 $\pm$ 10.0 &50.0 $\pm$ 10.0 &8.6&&139483&2.40 $\pm$ 0.27 &252.2 $\pm$ 2.6 &38.3 $\pm$ 6.0 &10.9\\
   &133785 &5.19 $\pm$ 0.37 &283.4 $\pm$ 1.7 &53.9 $\pm$ 5.0&7.5 & &169113 &2.10 $\pm$ 0.36 &266.8 $\pm$ 6.3 &52.7 $\pm$ 16.9 &10.5\\
   &145946 &4.20 $\pm$ 0.38 &297.1 $\pm$ 2.1 &50.6 $\pm$ 5.6& 7.4& &236727 &0.92 $\pm$ 0.12 &265.3 $\pm$ 3.3 &52.7 $\pm$ 10.7 &4.2\\
   \cline{8-12}
  &158107 &3.35 $\pm$ 0.52 &294.6 $\pm$ 4.1 &49.9 $\pm$ 8.3&13.7 &\multicolumn{6}{c}{CCS}\\
  \cline{8-12}
  &170267 &2.13 $\pm$ 0.28 &300.1 $\pm$ 3.9 &51.0 $\pm$ 14.3&9.1 &&93870&1.65 $\pm$ 0.22&235.7 $\pm$ 3.3&47.0 $\pm$ 8.6&7.5\\
  &194586 &1.91 $\pm$ 0.30 &311.2 $\pm$ 4.3 &53.6 $\pm$ 9.4&9.7 & & 106347&1.09 $\pm$ 0.21 &234.2 $\pm$ 3.0 &38.0 $\pm$ 7.3 &6.9\\
  \cline{7-12}
  &231060 &0.54 $\pm$ 0.17 &303.1 $\pm$ 4.6 &30.3 $\pm$ 12.2&4.7 & \multicolumn{6}{c}{CS}\\
  \cline{8-12}
   &218903 &0.69 $\pm$ 0.10 &295.3 $\pm$ 2.2 &29.8 $\pm$ 5.1&4.0 &\multirow{43}{*}{GMC2b} &97980 &72.20 $\pm$ 0.64 &317.7 $\pm$ 0.2 &53.1 $\pm$ 0.4 &9.0\\
   \cline{2-6}
   \multicolumn{6}{c|}{H$_2$CS}&& 146969&49.63 $\pm$ 0.57 &309.6 $\pm$ 0.3 &52.1 $\pm$ 0.5 &7.3\\
   \cline{2-6}
   &101477 &5.19 $\pm$ 0.30 &275.3 $\pm$ 1.6 &57.9 $\pm$ 3.8& 8.1& &195954 &23.82 $\pm$ 0.63 &326.6 $\pm$ 0.4 &45.6 $\pm$ 1.0 &8.4\\
   &103040&3.80 $\pm$ 0.21&268.0 $\pm$ 1.7&60.7 $\pm$ 4.6&6.1 &&244935&14.50 $\pm$ 0.54 &328.7 $\pm$ 0.5 &46.4 $\pm$ 1.7 &5.8\\
   &104617 & 3.97 $\pm$ 0.25&268.5 $\pm$ 2.3 &55.6 $\pm$ 5.2&6.6 & &293912 &6.32 $\pm$ 0.46 & 320.0 $\pm$ 10.0&45.0 $\pm$ 10.0 &5.8\\
   &137371 &3.80 $\pm$ 0.47 &289.2 $\pm$ 3.0 &51.7 $\pm$ 8.1&12.5 & &342883 &2.49 $\pm$ 0.23 &320.0 $\pm$ 10.0 &43.0 $\pm$ 10.0 &8.2\\
   \cline{8-12}
  &139483 &4.46 $\pm$ 0.36 &289.3 $\pm$ 3.1 &64.5 $\pm$ 6.7&10.7 &\multicolumn{6}{c}{H$_2$S}\\
  \cline{8-12}
  &169113 &3.44 $\pm$ 0.43 &298.8 $\pm$ 3.8 &59.1 $\pm$ 8.0&12.3 & &168762 &5.63 $\pm$ 0.30 &320.7 $\pm$ 1.6 &50.0 $\pm$ 3.6 &11.0\\
  \cline{8-12}
  &236727 &1.52 $\pm$ 0.27 & 302.6 $\pm$ 4.9&62.7 $\pm$ 15.4&3.8 & \multicolumn{6}{c}{SO}\\
  \cline{2-6}\cline{8-12}
   \multicolumn{6}{c|}{CCS} & &86093 &1.30 $\pm$ 0.23 & 323.9 $\pm$ 5.5& 51.9 $\pm$ 11.7&7.2\\
  \cline{2-6}
   &93870 &2.77 $\pm$ 0.27 &264.4 $\pm$ 2.0 &46.6 $\pm$ 5.7&7.6 & & 109252&1.46 $\pm$ 0.22 &326.6 $\pm$ 2.1 &40.8 $\pm$ 4.9 &7.2\\
     &106347 &1.30 $\pm$ 0.22 &268.2 $\pm$ 2.8 &43.4 $\pm$ 6.1&6.6 & &138178 & 6.69 $\pm$ 0.37& 325.0 $\pm$ 1.5&45.4 $\pm$ 4.1 &12.8\\
     \cline{1-6}
   \multicolumn{6}{c|}{CS} & &178605 & 4.92 $\pm$ 1.10&330.2 $\pm$ 1.8 &20.7 $\pm$ 4.7 &47.8\\
     \cline{2-6}
  \multirow{31}{*}{GMC1b}&97980&76.95 $\pm$ 0.35 &269.8 $\pm$ 0.2 &83.7 $\pm$ 0.5& 8.4& &219949 &1.35 $\pm$ 0.13 &328.3 $\pm$ 1.3 &33.9 $\pm$ 3.5 &3.1\\
  \cline{8-12}
   & 146969&52.40 $\pm$ 0.32 &262.0 $\pm$ 0.3 &77.0 $\pm$ 0.6 &7.2&\multicolumn{6}{c}{$^{34}$SO}\\
   \cline{8-12}
     &195954 &29.42 $\pm$ 0.33 &274.7 $\pm$ 0.3 &66.5 $\pm$ 0.8&9.0 & &97715 & 0.64 $\pm$ 0.15&331.9 $\pm$ 3.6 &26.0 $\pm$ 6.4 &6.6\\
     \cline{8-12}
   &244935 &15.55 $\pm$ 0.21 & 275.1 $\pm$ 0.3&55.9 $\pm$ 0.8&5.5 & \multicolumn{6}{c}{OCS}\\
   \cline{8-12}
  &283912 &7.34 $\pm$ 0.25 & 265.3 $\pm$ 0.9&56.7 $\pm$ 2.2&5.8 & &85139 &5.88 $\pm$ 0.24 &291.4 $\pm$ 1.2 &38.4 $\pm$ 3.3 &7.9\\
  &342883 &3.47 $\pm$ 0.22 &267.3 $\pm$ 1.8 &52.8 $\pm$ 4.2&7.0 &&97301 &4.64 $\pm$ 0.20 &305.9 $\pm$ 0.9 &40.4 $\pm$ 2.4 &6.6 \\
  \cline{2-6}
  \multicolumn{6}{c|}{H$_2$S} & &109463 &4.38 $\pm$ 0.28 &286.9 $\pm$ 1.3 &46.8 $\pm$ 3.8 &9.3\\
  \cline{2-6}
  &168762 & 6.96 $\pm$ 0.39&272.9 $\pm$ 2.6 &71.8 $\pm$ 8.0& 10.5& &133785 &3.68 $\pm$ 0.22 &305.0 $\pm$ 10.0 &50.0 $\pm$ 10.0 &7.4\\
  \cline{2-6}
   \multicolumn{6}{c|}{SO} & &145946 &2.79 $\pm$ 0.24 &319.8 $\pm$ 1.3 &38.4 $\pm$ 3.2 &6.8\\
   \cline{2-6} 
   &86093&0.92 $\pm$ 0.24&255.5 $\pm$ 4.6&40.2 $\pm$ 8.5&7.9&&158107& 2.45 $\pm$ 0.39&313.6 $\pm$ 3.8 & 47.4 $\pm$ 9.1&12.7\\
      &109252 &1.63 $\pm$ 0.25 &269.1 $\pm$ 2.8 &44.9 $\pm$ 7.2&7.2 & &170267 &1.01 $\pm$ 0.17 &327.5 $\pm$ 1.7 &16.5 $\pm$ 5.2 &9.8\\
   &129138 &1.72 $\pm$ 0.27 &265.2 $\pm$ 8.3 &63.6 $\pm$ 23.8& 7.7& &194586 &0.91 $\pm$ 0.25 &330.0 $\pm$ 2.8 &24.1 $\pm$ 7.1 &9.3\\
   & 138178&9.15 $\pm$ 0.41 &268.4 $\pm$ 1.2 &52.6 $\pm$ 2.7 &11.9 &&218903 &0.47 $\pm$ 0.10 &327.8 $\pm$ 2.5 &19.8 $\pm$ 9.5&3.7\\
   &158971 &1.48 $\pm$ 0.36 &277.3 $\pm$ 4.7 &38.2 $\pm$ 8.2&12.3 & &231060 &0.17 $\pm$ 0.10  &323.7 $\pm$ 2.9 &15.2 $\pm$ 5.1 &3.9\\
   \cline{8-12}
  &178605&7.47 $\pm$ 1.28 &280.8 $\pm$ 5.1 &40.8 $\pm$ 11.2& 42.8& \multicolumn{6}{c}{H$_2$CS}\\
  \cline{8-12}
  &219949 &2.10 $\pm$ 0.21 &274.0 $\pm$ 2.4 &52.8 $\pm$ 6.6&4.2 & &101477 &4.08 $\pm$ 0.27 &289.6 $\pm$ 1.7 &46.3 $\pm$ 5.1 &7.4\\
  \cline{2-6}
  \multicolumn{6}{c|}{$^{34}$SO}&&103040&2.27 $\pm$ 0.19 &287.3 $\pm$ 2.1 &45.5 $\pm$ 4.8 &6.0\\
   \cline{2-6}
    &97715 & 0.85 $\pm$ 0.24&274.2 $\pm$ 5.9 &46.0 $\pm$ 16.1&6.4 & &104617 &3.06 $\pm$ 0.22 &287.5 $\pm$ 1.8 &41.3 $\pm$ 4.8 &6.6\\
    \cline{2-6}
  \multicolumn{6}{c|}{OCS}&&137371&3.07 $\pm$ 0.73 & 310.0 $\pm$ 4.7&44.5 $\pm$ 17.4 &12.9\\
   \cline{2-6}
   &85139 &6.68 $\pm$ 0.51 &236.2 $\pm$ 2.1 &57.3 $\pm$ 5.3&8.5 & &139483 &2.58 $\pm$ 0.37 &307.5 $\pm$ 4.0 &48.4 $\pm$ 8.2 &12.5\\
   &97301 &5.86 $\pm$ 0.24 &254.4 $\pm$ 1.1 &56.1 $\pm$ 2.8&6.4 & &169113 &2.44 $\pm$ 0.33 &317.0 $\pm$ 5.8 &61.8 $\pm$ 16.3&11.1\\
   \cline{8-12}
   &109463&4.55 $\pm$ 0.32 &232.7 $\pm$ 1.6 &51.8 $\pm$ 3.4&8.6 &\multicolumn{6}{c}{CCS}\\
   \cline{8-12}
   &133785 &4.37 $\pm$ 0.22 &259.5 $\pm$ 1.9 &53.3 $\pm$ 5.0&7.3 & &93870 &2.15 $\pm$ 0.23 &285.6 $\pm$ 2.3 & 40.2 $\pm$ 5.9&7.0\\
   &145946 &2.94 $\pm$ 0.25 &269.6 $\pm$ 2.6 &53.6 $\pm$ 5.7&7.1 & &106347 &1.06 $\pm$ 0.20 & 289.7 $\pm$ 2.5&31.4 $\pm$ 6.4 &6.8\\
   \cline{7-12}
  &158107 &1.92 $\pm$ 0.34 &261.7 $\pm$ 3.2 &55.0 $\pm$ 10.0 &12.3&&& & & &\\
   &170267 &1.62 $\pm$ 0.26 &261.7 $\pm$ 5.4 &60.0 $\pm$ 13.1&8.8 & & & & & &\\
  &194586 &1.02 $\pm$ 0.21 &266.3 $\pm$ 4.8 &49.7 $\pm$ 13.3 &8.8&&& & & &\\
  \cline{2-6}
   \multicolumn{6}{c|}{H$_2$CS} & \\
   \cline{2-6}
   &101477 &3.99 $\pm$ 0.25 &241.3 $\pm$ 1.6 &57.3 $\pm$ 4.2& 6.4& & & & & &\\
   &103040 &2.68 $\pm$ 0.22 &228.9 $\pm$ 2.4 &60.8 $\pm$ 5.4& 6.5& & & & & &\\
   \cline{1-6}
\end{longtable}
\end{landscape}

\section{LTE analysis: Rotation diagrams}\label{app:RDs}

We present here the LTE results for CS, H$_2$S, SO, $^{34}$SO, and H$_2$CS. At least two rotation temperature components were needed in order to fit the observations. This is the case of CS in all the regions, and for most of the species outwards the inner CMZ. For CS,  $^{34}$SO, and H$_2$CS, the break in rotation temperature seen seem to correspond to the where the shift in systemic velocity occurs (see Sec.~\ref{subsub:velocities}) which could favour the presence of multiple components with different physical conditions rather than opacity/non-LTE effects. SO presents a particular case: The transitions at frequencies 86093 MHz, 100029 MHz, 136634 MHz, and 254573 MHz, detected towards the inner CMZ, have the lowest Einstein coefficients ($A_{\text{ij}} \sim [1-5]\times 10^{-6}$ s$^{-1}$). Except for the 86093 MHz transition, the other transitions show a big deviation from the other transitions in the rotation diagrams, highly impacting the fit. As these transitions are not likely to be contaminated by another species (see also \citealt{holdship_energizing_2022}), non-LTE effects are probably the culprit. For the LTE analysis, we thus excluded these transitions from the fit for GMC6, GMC7, and SSC2. For SSC2 and SSC5 some of these transitions are either too blended or undetected i.e. with an integrated intensity below our threshold of 3$\sigma$, and were thus naturally excluded from the analysis. For some regions and species, only one transition is detected. In these cases, we could not construct the RD. \\

\begin{figure}
    \centering
    \includegraphics[width=\textwidth]{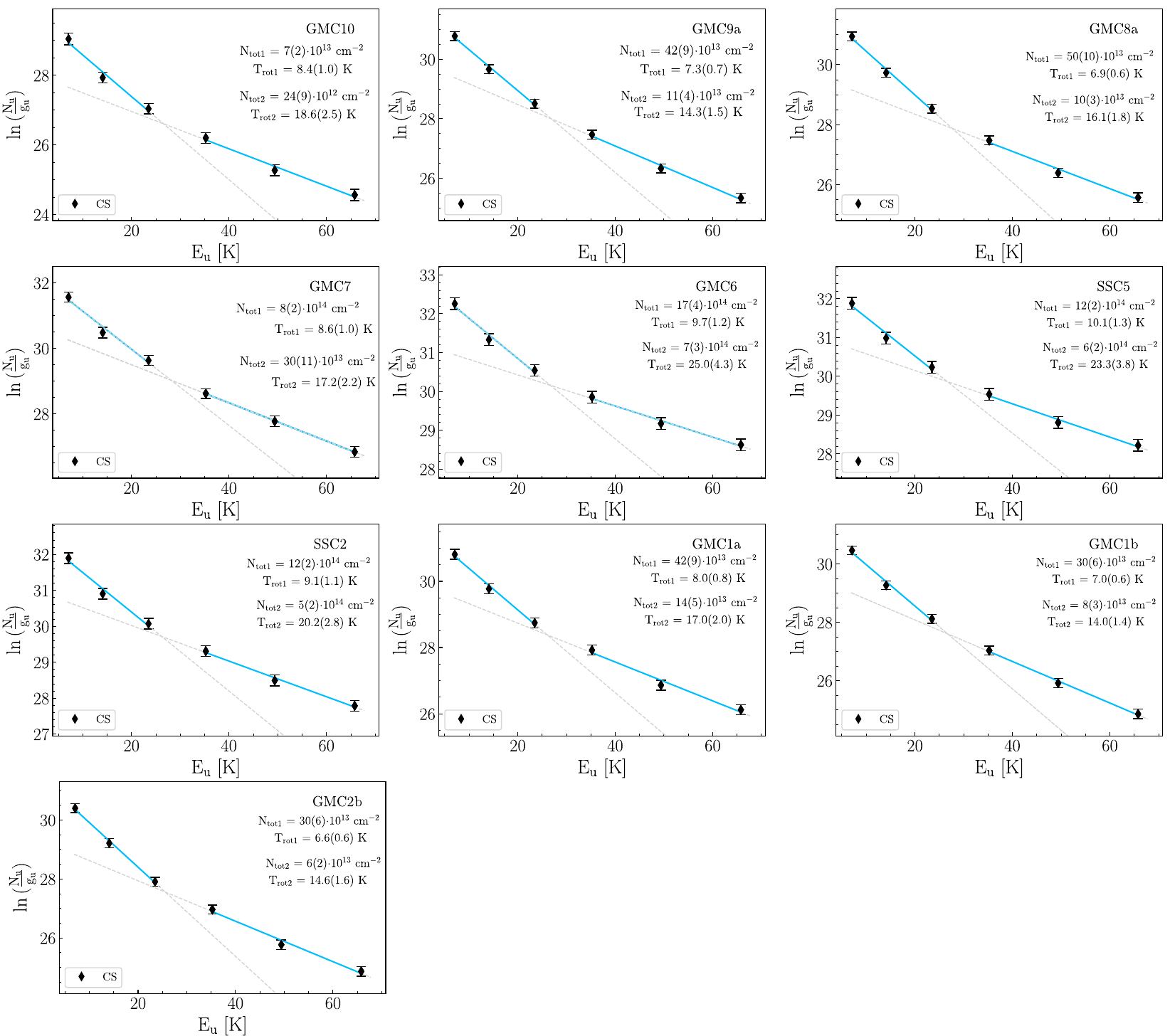}
    \caption{Rotation diagrams of CS. }
    \label{fig:RD_CS}
\end{figure}

In all the RDs, the parameters $N_{\text{u}}, g_{\text{u}},$ and $E_{\text{u}}$ are the column density, degeneracy and level energy (with respect to the ground state level) of the upper level, respectively. Each fit performed is shown by a blue line and the dashed grey lines are the extrapolations of these fit for the full range of $E_{\text{u}}$. Finally, we do not differentiate between the ortho- and para- transitions of H$_2$S and H$_2$CS in the RDs, as we fitted both types of transitions together.The rotation diagrams are presented in Fig. ~\ref{fig:RD_CS} to \ref{fig:RD_SO2}, and the derived beam-averaged column densities, $N_{\text{beam}}$, and rotation temperatures, $T_{\text{rot}}$, are shown in Table \ref{tab: LTE_results}.\\

In the outer CMZ, we obtained relatively cold ($T_{\text{rot}} \leq 50$ K) rotation temperatures for all the species. The highest rotational temperatures are derived for H$_2$S towards GMC10 whilst for the other species GMC10 is rather the region for which the lowest rotation temperature is derived. The higher beam-average column densities derived were found for CS, SO and OCS with column densities up to $(3-6)\times10^{14}$\pcms. We thus can only give an upper limit for the column density for this species. In the inner CMZ, for most of the species we could fit two rotation temperatures. H$_2$S always has one component as we have only three transitions available. The first component of SO, $^{34}$SO and H$_2$CS, CCS have the coldest derived rotation temperatures with $T_{\text{rot}} < 20$ K whilst the highest ones are derived for the second component of OCS and SO$_2$ with $T_{\text{rot}} > 100$ K (see also Sec.~\ref{subsec:LTE_main}. All the other components and species have intermediate rotation temperatures. For the beam-averaged column densities, the highest values are found for CS and H$_2$S with $N_{\text{tot}}\sim 10^{14}-10^{15}$ \pcms.

\begin{figure}
    \centering
    \includegraphics[width=\textwidth]{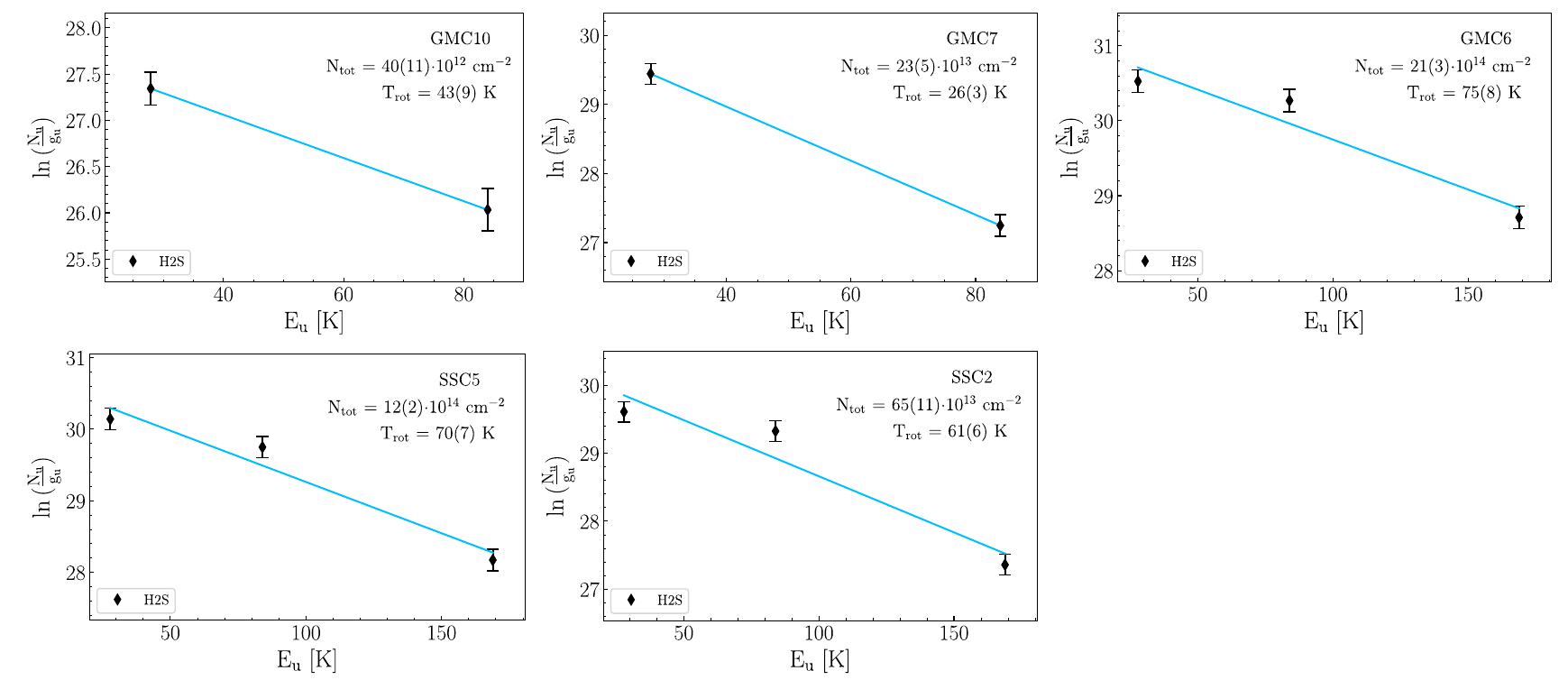}
    \caption{Rotation diagrams of H$_2$S.}
    \label{fig:RD_H2S}
\end{figure}

\begin{figure}
    \centering
    \includegraphics[width=\textwidth]{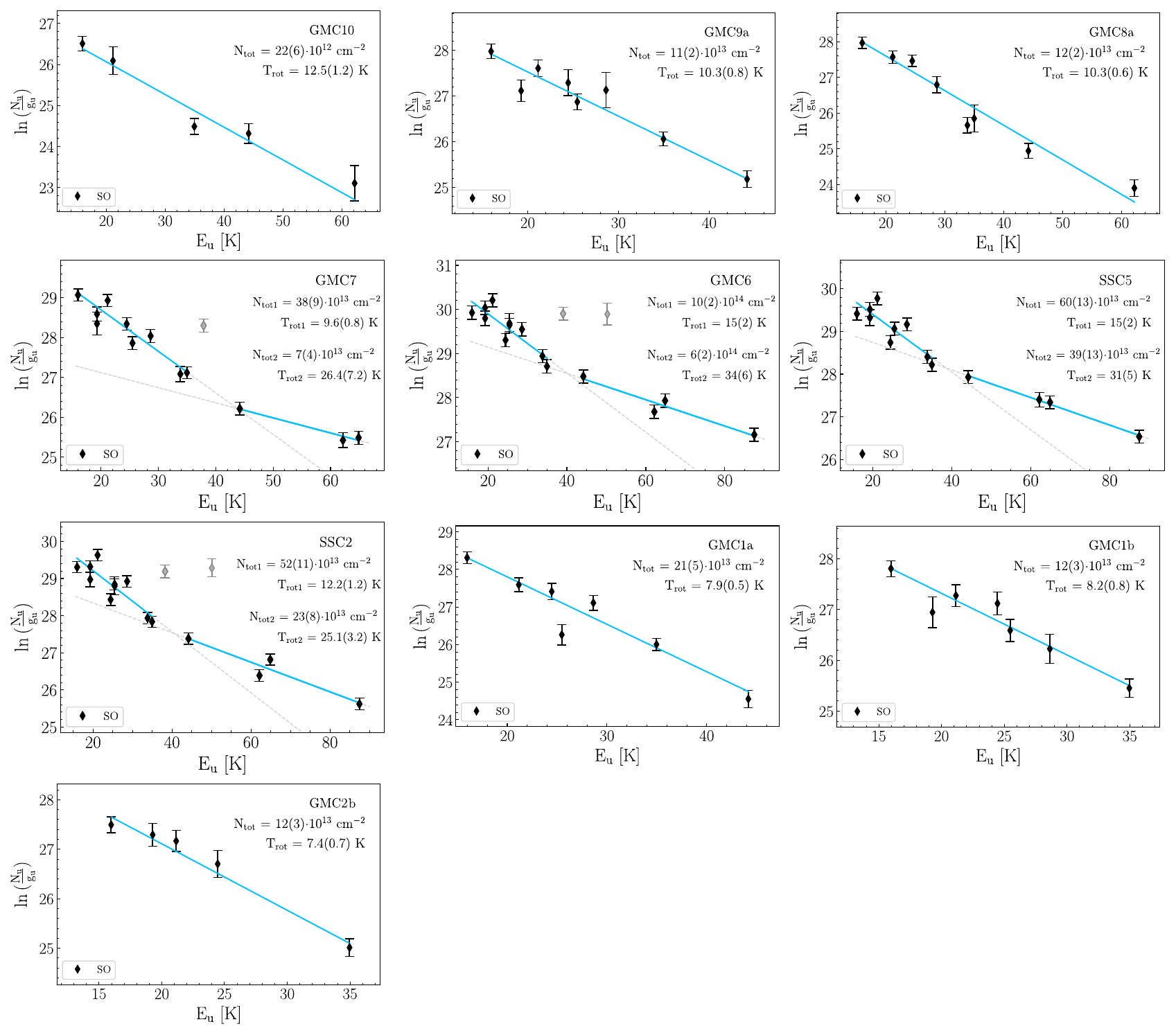}
    \caption{Rotation diagrams of SO. In GMC7, GMC6, and SSC2, the grey symbols represent the lines at 100029 MHz and 136634 MHz, which were excluded from the fit (see Sec.~\ref{app:RDs}). The line at 254573 MHz has $E_{\text{u}}=100$ K and is not shown in the RDs.}
    \label{fig:RD_SO}
\end{figure}

\begin{figure}
    \centering
    \includegraphics[width=\textwidth]{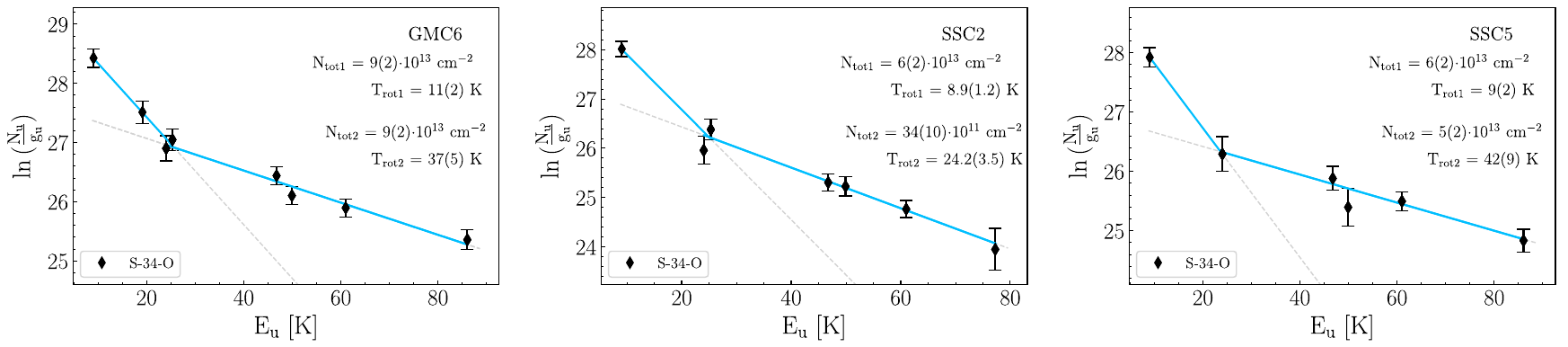}
    \caption{Rotation diagrams of $^{34}$SO.}
    \label{fig:RD_S34O}
\end{figure}

\begin{figure}
    \centering
    \includegraphics[width=\textwidth]{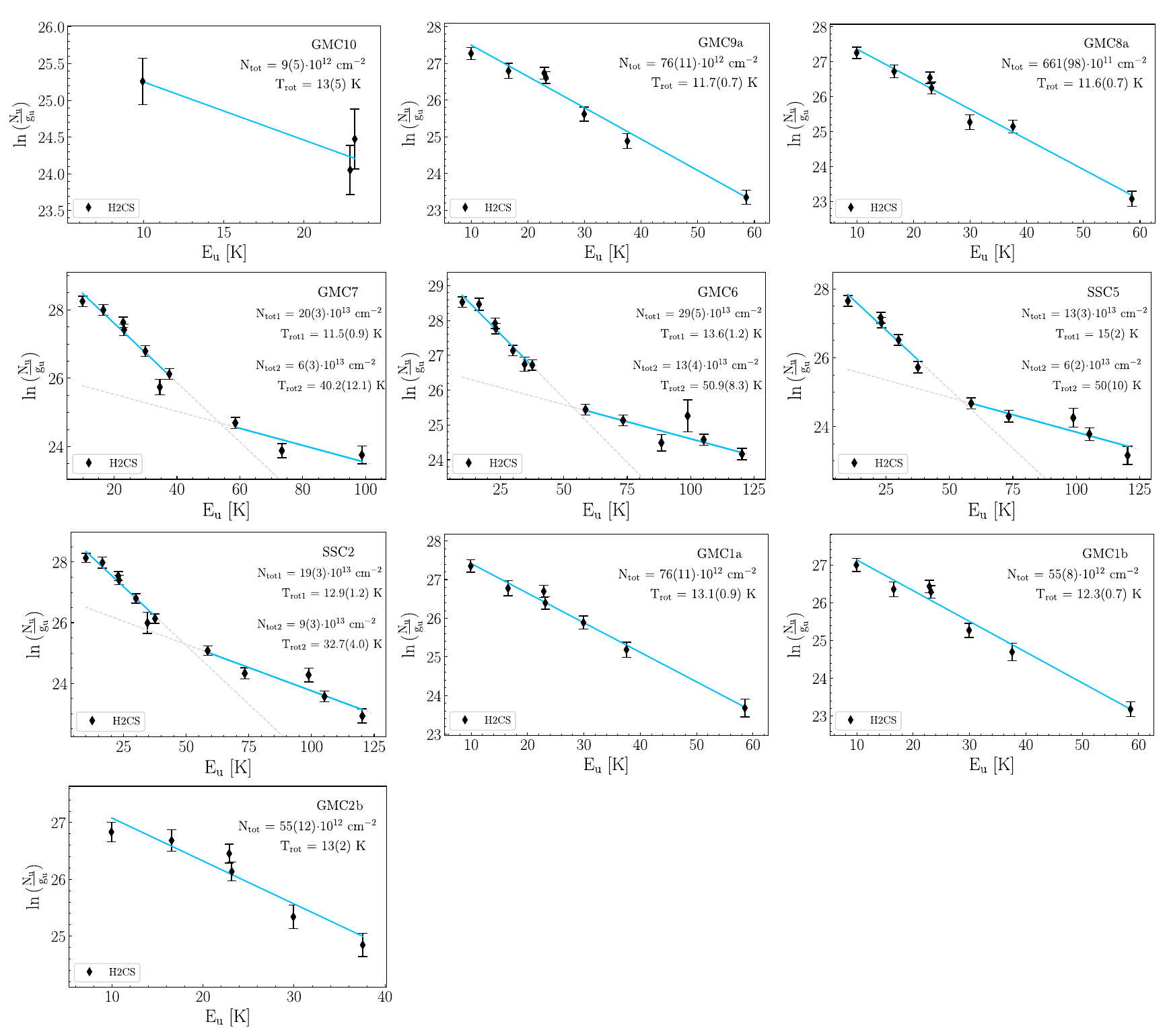}
    \caption{Rotation diagrams of H$_2$CS.}
    \label{fig:RD_H2CS}
\end{figure}

\begin{table*}
\centering
    \caption{Results from the Rotation Diagrams.}
    \label{tab: LTE_results}
    \begin{tabular}{c|cccc|c|cccc}
    \hline \hline
         \multirow{2}{*}{Region}&\multirow{2}{*}{Species}& \multirow{2}{*}{Component} & $T_{\mathrm{rot}}$ & $N_{\mathrm{beam}}$&\multirow{2}{*}{Region}&\multirow{2}{*}{Species}& \multirow{2}{*}{Component} & $T_{\mathrm{rot}}$ & $N_{\mathrm{beam}}$    \\
         \multirow{9}{*}{GMC10}&& & (K) & ($\times10^{14}$\pcms)&\multirow{9}{*}{GMC9a}&& &(K)  &($\times10^{14}$\pcms) \\
         \hline
         &\multirow{2}{*}{CS} & 1 & 8 (1) &0.7 (0.2) & & \multirow{2}{*}{CS} & 1 &7 (1) &4 (1)  \\
         && 2 &18.6 (2.5) &0.2 (0.1) & & & 2 &14.3 (1.5) &1.1 (0.4) \\
         \cline{2-5} \cline{7-10} 
          &H$_2$S & 1 & 43 (9) &0.4 (0.1) &&H$_2$CS & 1 &12 (1)& 0.8 (0.1) \\
         \cline{2-5}\cline{7-10} 
          &H$_2$CS& 1 & 13 (5) &0.09 (0.05)&&OCS & 1 &22 (1) & 5 (1)\\
         \cline{2-5}\cline{7-10} 
         &OCS & 1 &14 (4) &2(1)&&SO & 1 &10 (1) &1.1 (0.2) \\
         \cline{2-5}\cline{7-10} 
          &SO & 1 &12.5 (1.2) &0.2 (0.1)&&CCS & 1 &6 (2) & 1 (1) \\
         \cline{1-5}\cline{6-10} 
           \multirow{6}{*}{GMC8a}&\multirow{2}{*}{CS} & 1 &7 (1) &5 (1) & \multirow{10}{*}{GMC7}& \multirow{2}{*}{CS} & 1 &9 (1) & 8 (2)  \\
         && 2 &16 (2) &1.0 (0.3) & & & 2 &17 (2) &3 (1) \\
         \cline{2-5} 
         \cline{7-10} 
          &H$_2$CS & 1 &12 (1) &0.7 (0.1)&&H$_2$S & 1 &26 (3) &2.3 (0.5) \\
         \cline{2-5}\cline{7-10} 
          &OCS & 1 & 21 (1)&3.9 (0.5)&&\multirow{2}{*}{H$_2$CS} & 1 & 11.5 (0.9)&2.0 (0.3) \\
          \cline{2-5}
         &SO & 1 &10 (1) &1.2 (0.2)&&& 2 &40 (12) &0.6 (0.3) \\
         \cline{2-5}\cline{7-10} 
         & CCS & 1 &7 (3) &0.4 (0.5)&&OCS & 1 &28.9 (1.5) & 8 (1) \\
         \cline{1-5}\cline{7-10} 
          \multirow{16}{*}{GMC6}&\multirow{2}{*}{CS} & 1 &10 (1) &17 (4) &&\multirow{2}{*}{SO} & 1 & 10 (1) &4 (1) \\
          && 2 &25 (4) &7 (3) &&& 2 &26 (7) &0.7 (0.4) \\
         \cline{2-5}\cline{7-10} 
         &H$_2$S & 1 & 75 (8)&21 (3)  && \multirow{2}{*}{CCS} & 1 &7.0 (1.5) & 2 (1) \\
         \cline{2-5}
        &\multirow{2}{*}{H$_2$CS} & 1 & 14 (1)&2.9 (0.5)&& & 2 &18 (4) &0.4 (0.2) \\
         \cline{6-10}
         && 2 &51 (8) &1.3 (0.4)  & \multirow{15}{*}{SSC5}& \multirow{2}{*}{CS} & 1 &10 (1) &12 (2)  \\
         \cline{2-5}
         &\multirow{2}{*}{ OCS} & 1 &53 (5) &10 (1) & & & 2 &23 (4) &6 (2) \\
          \cline{7-10} 
          && 2 &250 (62) &7 (2) &&H$_2$S & 1 &70 (7) &12 (4) \\
         \cline{2-5}\cline{7-10} 
         &\multirow{2}{*}{SO} & 1 &15 (2) &10 (2)&&\multirow{2}{*}{H$_2$CS} & 1 &15 (2) &1.3 (0.3) \\
        && 2 &34 (6) &6 (2)&&& 2 &50 (10) &0.6 (0.2) \\
         \cline{2-5}\cline{7-10} 
         &\multirow{2}{*}{$^{34}$SO} & 1 & 11 (2)&0.9 (0.2)&&\multirow{2}{*}{ OCS} & 1 &58 (7) & 4.1 (0.6)\\
         && 2 & 37 (5)&0.9 (0.2)&&& 2 &169 (45) &3 (1) \\
         \cline{2-5}\cline{7-10} 
           & \multirow{3}{*}{CCS} & 1 &12 (2) &1.7 (0.6)&&\multirow{2}{*}{SO} & 1 &15 (2) &6 (1) \\
         & & 2 &30 (2) &1.1 (0.1)&&& 2 &31 (5) &4 (1) \\
         \cline{7-10} 
          & & 3 & 65 (7)& 0.4 (0.1) &&\multirow{2}{*}{$^{34}$SO} & 1 &9 (2) &0.6 (0.2) \\
          \cline{2-5}
         &\multirow{2}{*}{SO$_2$} & 1 &62 (5) &3.3 (0.3)&&& 2 &42 (9) &0.5 (0.2) \\
         \cline{7-10} 
        && 2 &250 (44) &5 (1)&& \multirow{3}{*}{CCS} & 1 &11 (2) &1.4 (0.5) \\
        \cline{1-5}
       \multirow{15}{*}{SSC2}&\multirow{2}{*}{CS} & 1 &9 (1) &12 (2) && & 2 &23 (4) & 0.7 (0.2)\\
      && 2 &20 (3) &5 (2) & & &3 & 38 (5)& 0.5 (0.2)\\
         \cline{2-5}\cline{7-10} 
            &H$_2$S & 1 &61 (6) &6.5 (1.1)&&\multirow{2}{*}{SO$_2$} & 1 &65 (10) & 1.4 (0.2)\\
            \cline{2-5}
          &\multirow{2}{*}{H$_2$CS} & 1  &13 (1)&1.9 (0.3)& && 2 &144 (29) &1.9 (0.5) \\
         \cline{6-10}
         && 2 &33 (4) &0.9 (0.3) & \multirow{6}{*}{GMC1a}& \multirow{2}{*}{CS} & 1 &8 (1) &4 (1)  \\
         \cline{2-5}
         &\multirow{2}{*}{ OCS} & 1 & 38 (2)&8 (1)& & & 2 &17 (2) &1.4 (0.5) \\
         \cline{7-10} 
            && 2 & 191 (44)&2.4 (0.6) &&H$_2$CS & 1 &13 (1) &0.8 (0.1) \\
         \cline{2-5}\cline{7-10} 
         &\multirow{2}{*}{SO} & 1 &12 (1) &5 (1)&&OCS & 1 &21 (1) & 5 (1) \\
         \cline{7-10}
         && 2 &25 (3) &2 (1)&&SO & 1 &7.9 (0.5) &2.1 (0.5) \\
         \cline{2-5}\cline{7-10} 
        &\multirow{2}{*}{$^{34}$SO} & 1 & 8.9 (1.2)&0.6 (0.2)&& CCS & 1 &5.2 (1.5) &1 (2) \\
        \cline{6-10}
        && 2 & 24.2 (3.5)&0.3 (0.1)&\multirow{6}{*}{GMC2b}& \multirow{2}{*}{CS} & 1 &7 (1) & 3 (1) \\
         \cline{2-5}
          & \multirow{2}{*}{CCS} & 1 &9 (1) &2 (1)&&& 2 &15 (2) &0.6 (0.2) \\
          \cline{7-10}
         & & 2 &25 (2) &0.5 (0.1)&&H$_2$CS & 1 & 13 (2)&0.6 (0.1) \\
         \cline{2-5}\cline{7-10} 
         &\multirow{2}{*}{SO$_2$} & 1 &74 (5) &1.9 (0.2)&&OCS & 1 &18 (1) & 4.1 (0.5) \\
         \cline{7-10}
          && 2 &251 (47) &3 (1)&&SO & 1 & 7 (1)& 1.2 (0.3) \\
         \cline{1-5}\cline{7-10} 
          \multirow{6}{*}{GMC1b}&\multirow{2}{*}{CS} & 1 &7 (1) &3 (1) && CCS & 1 &5 (2) &1 (1) \\
          \cline{6-10}
        && 2 &14(1) &0.8(0.3) \\
         \cline{2-5}
        &H$_2$CS & 1 & 12 (1)&0.6 (0.1)\\
        \cline{2-5}
        &OCS & 1 & 17 (1)&5 (1) \\
         \cline{2-5} 
          &SO & 1 &8 (1) &1.2 (0.3) \\
          \cline{2-5}
         & CCS & 1 & 8 (4) &0.3 (0.4) \\
         \cline{1-5}
    \end{tabular}
\end{table*}


\section{LVG modeling and results}\label{app:LVG}

We give in this section detailed information about the LVG modeling of each species. The list of collision coefficients used to perform the analysis are presented in Table~\ref{tab:coeffs} with their corresponding references. The results of the non-LTE analysis for the regions GMC10, GMC8a, GMC7, SSC5, GMC1a, and GMC1b are displayed in Figures \ref{fig:LVG_others} and in Table~\ref{Tab:LVG_results_others}.

\subsection{CS}
From the LTE analysis, two components were found, the first one including the transitions J=2--1 to J=4--3 and the second one including the transitions J=5--4 to J=7--6. For each component, only three lines were available. Hence, we fixed the size of the emission to be able to constraint the other parameters ($N_{\text{tot}}, T_{\text{gas}}, n_{\text{gas}}$). As we do not have information on where the CS could be emitted from we chose to fix the size of the emission to the beam size of 1.6$\arcsec$, which corresponds in having a beam filling factor of 0.5. The resulting reduced $\chi^2$ were good enough ($\chi^2_{\text{red}}\sim 0.1-1.9$) to be considered as possible solutions. Note however that other solutions with emission size more compact could be possible.

\subsection{H$_2$S + H$_2^{34}$S}
In the case of H$_2$, we also included the line of H$_2^{34}$S at 167 GHz that we detect in several regions. We assumed a value of 10 for the $^{32}$S/$^{34}$S ratio, as derived in Martin et al. 2021. We could perform the LVG analysis for the inner CMZ and GMC10 where we detect at least two transitions. Except for GMC10 where only two transitions were available, we left the size as a free parameter. For GMC10, we took the size found for GMC7 (0.34$\arcsec$), assuming the emission comes for the same component in the two regions. Finally, as we detected both ortho- and para- transitions of H$_2$S, we had to assume an ortho-to-para ratio (OPR). The value usually taken in the literature is the equilibrium value of 3. However, as some studies found that the OPR could be lower in the case of H$_2$S, and in particular in Orion where a value of 1.7$\pm$0.8 has been measured in Orion (Crockett et al. 2014), we ran two types of models, one with an OPR of 3 and the other with an OPR of 1. We found that the best OPR reproducing the observations in all the regions is 1. 

\subsection{SO+$^{34}$SO}\label{app:LVG_SO}
From the rotation diagram analysis, we found one component in the inner CMZ, and two components in the outer CMZ for SO and $^{34}$SO. We thus modeled the two components separately. As for H$_2$S, we used the $^{32}$S/$^{34}$S ratio of 10 derived in Martin et al. 2021 to include the $^{34}$SO lines in the model. This ratio was reproducing well the observations for the first component in the inner CMZ but for the second component, we found that a lower ratio of 8 was necessary to reproduce the observations. In the outer CMZ, only the 99 GHz line of $^{34}$SO is detected. We found that including this line was driving the fit to very low temperatures. We do not have the main isotopologue of this line and its $E_{\text{u}}$ of 9K is lower than that of the other detected transitions. Hence, this transition could trace a different component and we thus chose to not include it in the LVG modeling.

For most of the regions we had enough lines to leave the size as a free parameter, except for the second component of GMC7 where we had only three lines. In this case, we fixed the size to 0.2$\arcsec$ assuming the emission comes from the same component as for the other regions in the inner CMZ (for which we derived sizes between 0.15 and 0.3 $\arcsec$). On the other hand, for some components of regions, the best fit provided by the LVG analysis provided the product $\theta_s \times N_{\text{tot}}$ because of a degeneracy between the size and column density. This happened because the all the lines of SO are optically thin ($\tau << 1$). We used the method described in Sec.~\ref{subsec:method_LVG} to constrained the column density. Finally, we could not include the SO lines at 100, 136 and 254 GHz with $A_{ij} \sim 10^{-6}$ s$^{-1}$, as for the LTE analysis, as these lines are not well reproduced by the LVG model when including them with the other lines. These lines are probably tracing a third component that we could not model due to the poor number of lines. 

\subsection{H$_2$CS}
From the LTE analysis, up to two components were found in the inner CMZ. We thus modelled the two components separately in these regions. As for H$_2$S, H$_2$CS is detected in both ortho- and para- forms. We assumed an OPR ratio of 3, which reproduces well the observations. For most regions, enough lines were available to leave the size as a free parameter but the column density was not always constrained. In these cases, we used the method described in Sec.\ref{subsec:method_LVG} to better constrain the column density. For the region GMC10, and the second component in GMC7, there are only three lines available, preventing us from leaving the size as a free parameter. In the case of GMC10, we fixed the size to both extended emission and 0.2$\arcsec$ as these two values were found as best-fit values in the other regions. We adopted the solution with the best $\chi^2_{\text{red}}$, which is with a size of 0.2$\arcsec$. For GMC7, we fixed it to 0.12$\arcsec$, which is the average size found in the best-fit of the other regions for this component.

\subsection{OCS}
From the rotation diagram analysis, we found up to two components in the inner CMZ.  In all the regions but SSC5 and SSC2, we left the source as a free parameter. For SSC5 not enough lines were available whilst for SSC2, no $\chi^2$ minimum was found. For these sources, we then fixed the size to 0.17$\arcsec$, which corresponds to the average best-fit found in the other sources.

\subsection{Results}
\begin{figure*}
    \centering
    \includegraphics[width=0.9\textwidth]{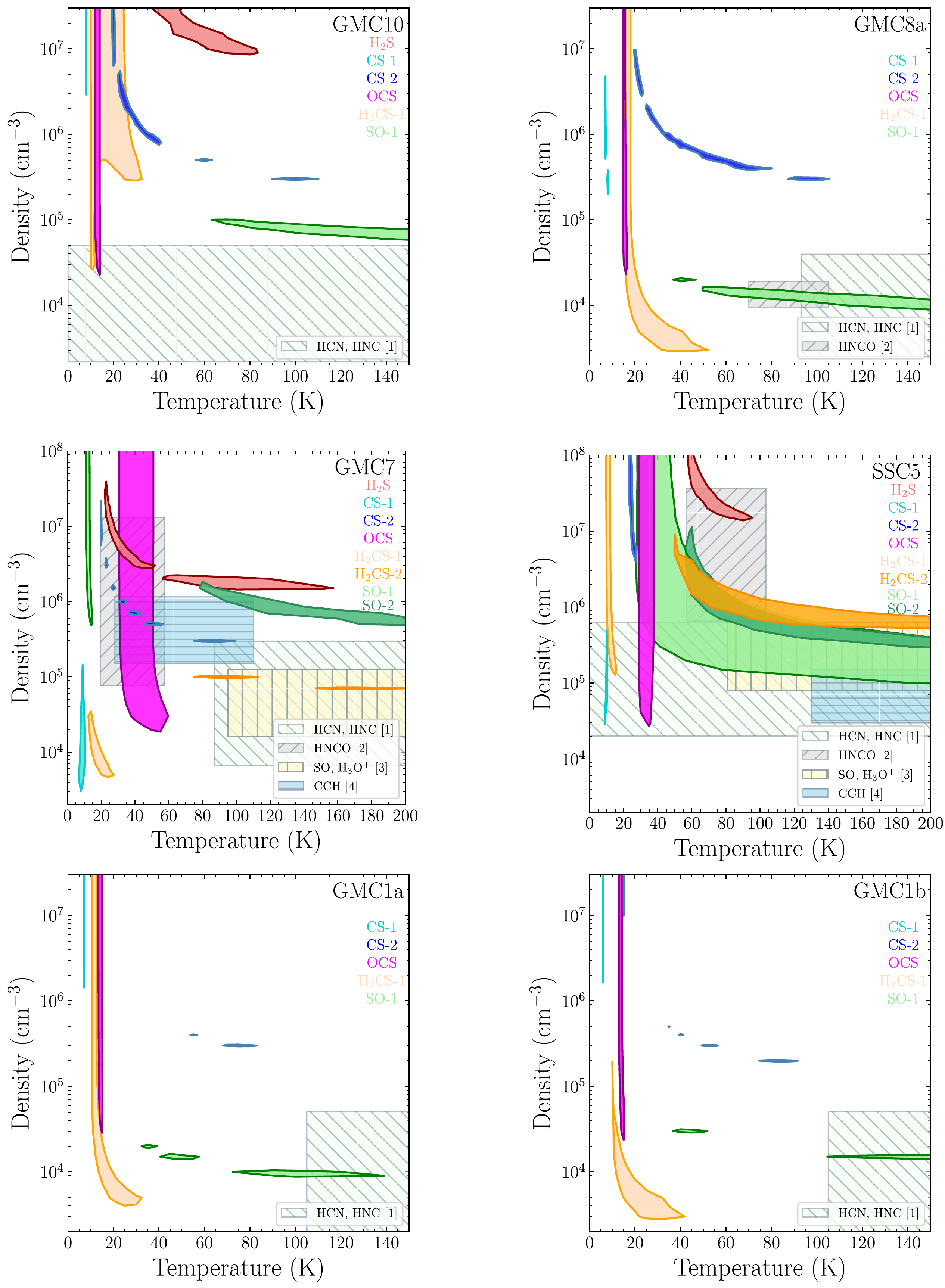}
    \caption{Density and temperature contour plots for GMC10, GMC8a, GMC7, SSC5, GMC1a, and GMC1b. The contours show the $1\sigma$ solutions obtained for the minimum value of the $\chi^2$ in the column density parameter derived for each species (and components) and region. Derived gas density and parameters for other species (HCN, HNC; HNCO; SO, H$_3$O$^+$; CCH) studied in previous ALCHEMI studies were plotted for comparison: [1] Behrens et al. 2022, [2] Huang et al. 2023, [3] Holdship et al. 2022, [4] Holdship et al. 2021.}
    \label{fig:LVG_others}
\end{figure*}

The physical parameters derived from the non-LTE LVG modelling for the regions GMC10, GMC8a, GMC7, SSC2, GMC1a, and GMC1b are displayed in Table~\ref{Tab:LVG_results_others} and their temperature-density contour plots are shown in Figure~\ref{fig:LVG_others}. The abundances derived for these regions are also indicated in Table ~\ref{Tab:LVG_results_others}.

\begin{table*}[ht]
\tiny
    \centering
    \caption{Best-fit results and $1\sigma$ confidence level range from the non-LTE LVG analysis for GMC10, GMC7, SSC5, GMC1a, and GMC1b, respectively (from top to bottom).}
    \label{Tab:LVG_results_others}
    \begin{tabular}{cc|cc|cc|cc|cc|c}
    \hline \hline
        \multirow{2}{*}{Species}& \multirow{2}{*}{Component} & $N_{\mathrm{tot}}$ & Range $N_{\mathrm{tot}}$ & $T_{\mathrm{kin}}$ & Range $T_{\mathrm{kin}}$ &$n_{\mathrm{H2}}$ & Range $n_{\mathrm{H2}}$ & Size  $(\theta_s)$&Range $\theta_s$&\multirow{2}{*}{Range $\tau_L$\tablefootmark{b}}\\
         & & (\pcms) &(\pcms) &(K) & (K)&(\pcmc) &(\pcmc) &($\arcsec$) &($\arcsec$)& \\
\hline
\multicolumn{11}{c}{GMC10}\\
\hline
\multirow{2}{*}{CS}  & 1 &$1.7\times10^{14}$ &$(1.5-3)\times10^{14}$ &8 &$7-9$ &... &$\geq3\times10^6$ &1.60\tablefootmark{a} &$\geq 1.00$ &$\leq2.5$ \\
 & 2 &$5\times10^{13}$ &$(0.2-15)\times10^{14}$ &20 &$18-110$ &... &$\geq 3\times10^5$ &1.60\tablefootmark{a} &$\geq 0.30$ & $\leq 0.3$\\
\hline
H$_2$S  & 1 &$1.2\times10^{15}$ &$(0.6-3)\times10^{15}$ &40 &$30-82$ &... &$\geq9\times10^6$ &0.34\tablefootmark{a} & $0.20-0.50$&$0.01-0.1$ \\
\hline
H$_2$CS & 1 &$8\times10^{14}$ &$(5.3-40)\times10^{14}$ &10 &$10-33$ &... &$\geq2.5\times10^4$ &0.20\tablefootmark{a}&$0.13-0.60$ &$0.02-0.8$\\
\hline
OCS & 1 &$1.5\times10^{16}$ & $(0.3-3)\times10^{16}$&13 &$12-14$ &... &$\geq 2.2\times10^4$ &0.13 & $0.11-0.25$&$0.3-1.4$ \\
\hline
SO & 1 &$4\times10^{14}$ &$(0.3-15)\times10^{14}$ &200 &$62-300\tablefootmark{c}$ &$6\times10^4$ & $(5-10)\times10^4$&... &$\geq 0.25$ &$\leq 0.1$ \\
\hline
\multicolumn{11}{c}{GMC8a}\\
\hline
\multirow{2}{*}{CS} & 1 &$1.5\times10^{15}$ &$(0.5-5)\times^{15}$ &7 &$6-8$ &$1\times10^6$ &$(0.2-6)\times10^6$ &$1.60\tablefootmark{a}$ & $\geq 1.00$&$\leq$ 2.4  \\
& 2 &$1.5\times10^{14}$ &$(1.2-6)\times10^{14}$ &100 &$20-105$ &$3\times10^5$ &$(0.3-10)\times10^6$ &$1.60\tablefootmark{a}$ &$\geq0.60$ &$\leq 0.3$ \\
\hline
H$_2$CS & 1 &$4\times10^{16}$ &$(0.2-20)\times10^{16}$ &19 &$16-53$ &... &$\geq 3\times10^3$ &0.12 & $0.11-0.30$&$0.2-4.0$ \\
\hline
OCS & 1 &$1\times10^{17}$ &$(0.2-1.5)\times10^{17}$ &15 &$14-16$ &... &$\geq 2\times10^4$ &0.16 & $0.14-0.30$&$0.9-1.9$ \\
\hline
SO & 1 &$1.5\times10^{16}$ &$(0.5-4)\times10^{16}$ &250 &$40-300\tablefootmark{c}$ &$9\times10^3$ &$(6-20)\times10^3$ &0.26  &$0.20-0.42$ &$0.1-1.6$ \\
\hline
\multicolumn{11}{c}{GMC7}\\
\hline
\multirow{2}{*}{CS}  & 1 & $9\times10^{15}$&$(4-15)\times10^{15}$ &9 &$6-10$ &$4\times10^4$ &$(0.3-15)\times10^4$ &1.60\tablefootmark{a} &$\geq 1.50$ &$\leq 4.3$ \\
 & 2 &$8\times10^{14}$ &$(3-15)\times10^{14}$ &90 &$20-100$ &... &$\geq 3\times10^5$ &1.60\tablefootmark{a}  &$\geq 1.00$ &$\leq 0.4$\\
\hline
H$_2$S  & 1 &$1\times10^{16}$ &$(0.8-1.4)\times10^{16}$ &25 & $22-159$&$1.2\times10^7$ &$(0.2-4)\times10^7$ &0.34 &$0.28-0.38$ &$0.05-0.4$ \\
\hline
\multirow{2}{*}{H$_2$CS} & 1 &$4\times10^{16}$ &$(2-9.3)\times10^{16}$ &15 &$13-27$ &$1.5\times10^4$ &$(0.5-3.5)\times10^4$ &0.20 &$0.19-0.26$ &$0.6-5.4$ \\
 & 2 & $6.7\times10^{15}$&$(4-9.3)\times10^{15}$ &90 &$75-200$ &$1\times10^5$ &$(0.7-1)\times10^5$ &0.20\tablefootmark{a} &$0.15-0.25$ &$0.03-0.3$\\
\hline
OCS & 1 &$5\times10^{16}$ &$(0.3-15)\times10^{16}$ &40 &$31-60$ &... & $\geq1.9\times10^4$&0.20 &$0.13-0.25$ &$0.1-1.0$ \\
\hline
\multirow{2}{*}{SO} & 1 &$5\times10^{14}$ &$(3-15)\times10^{14}$ &12 &$11-15$ &... &$\geq5\times10^5$ &...  &$\geq1.50$ &$\leq 0.1$\\
 & 2 &$5\times10^{15}$ &$(0.4-10)\times10^{15}$ &140 & $78-300$& $7\times10^5$&$(5-20)\times10^5$ &0.20\tablefootmark{a} &$0.15-0.80$ &$0.01-0.2$\\
\hline
\multicolumn{11}{c}{SSC5}\\
\hline
\multirow{2}{*}{CS}  & 1 &$1.5\times10^{16}$ &$\geq2\times10^{15}$ &10 &$9-11$ &$1\times10^5$ &$(0.3-5)\times10^5$ &1.60\tablefootmark{a}&$\geq1.60$ &$\leq 7.0$ \\
 & 2 & $1.3\times10^{15}$& $(0.5-3)\times10^{15}$&25 &$22-27$ &... &$\geq5\times10^6$ &1.60\tablefootmark{a} &$\geq1.00$ &$\leq0.7$\\
\hline
H$_2$S  & 1 & $2\times10^{17}$&$(1.4-3)\times10^{17}$ &70 &$57-96$ &...& $\geq1.5\times10^7$&0.20 &$0.17-0.23$ &$0.1-1.6$ \\
\hline
\multirow{2}{*}{H$_2$CS} & 1 &$2\times10^{14}$ &$(1.3-80)\times10^{14}$ &12 &$10-15$ &... &$\geq1.3\times10^5$ &... &$\geq0.25$ &$\leq0.9$ \\
 & 2 &$6.7\times10^{15}$ &$(0.7-20)\times10^{15}$ &90 & $50-300\tablefootmark{c}$&$1\times10^6$ &$(0.5-10)\times10^6$ &0.16 &$0.10-0.50$ &$0.02-0.5$\\
\hline
OCS & 1 &$3\times10^{16}$ &$(2.5-3.5)\times10^{16}$ &35 &$29-38$ &... &$\geq2.5\times10^4$ &0.17\tablefootmark{a}&$0.16-0.18$ &$0.1-0.2$ \\
\hline
\multirow{2}{*}{SO} & 1 &$7\times10^{14}$ &$(5-20)\times10^{14}$ &200 &$27-300\tablefootmark{c}$ &... &$\geq8\times10^4$ &... &$\geq1.00$ &$\leq0.1$ \\
 & 2 &$8\times10^{16}$ &$(6-10)\times10^{16}$ &80 &$55-260$ &$1\times10^6$ &$(0.3-10)\times10^6$ &0.15 &$0.13-0.17$ &$0.1-1.0$\\
\hline
\multicolumn{11}{c}{GMC1a}\\
\hline
\multirow{2}{*}{CS} & 1 &$1.5\times10^{15}$ &$(0.4-4)\times10^{15}$ &7 &$6-8$ &... &$\geq1.5\times10^6$ &1.60\tablefootmark{a}&$\geq1.00$ &$\leq3.0$  \\
 & 2 &$3\times10^{14}$ &$(1-3.5)\times10^{14}$ &75 &$50-85$ &... &$\geq3\times10^5$ &1.60\tablefootmark{a} &$\geq 1.50$ &$\leq 0.1$ \\
\hline
H$_2$CS & 1 &$2\times10^{16}$ &$(0.5-5.3)\times10^{16}$ &12 & $11-33$&... &$\geq4\times10^3$ & 0.16&$0.15-0.20$ &$0.7-2.2$  \\
\hline
OCS & 1 &$1.5\times10^{17}$ &$(0.3-1.5)\times10^{17}$ &14 &$13-16$ &... &$\geq3\times10^4$ &0.17& $0.15-0.25$&$1.0-3.0$ \\
\hline
SO & 1 &$1.5\times10^{16}$&$(0.8-3)\times10^{16}$  &47 &$30-300\tablefootmark{c}$ & $1.5\times10^4$& $(0.5-2)\times10^4$& 0.30& $0.24-0.40$&$0.1-1.6$ \\
\hline
\multicolumn{11}{c}{GMC1b}\\
\hline
\multirow{2}{*}{CS}  & 1 &$1.3\times10^{15}$ &$(0.3-2)\times10^{15}$ &6 &$5-8$ &... & $\geq1.7\times10^6$& 1.60\tablefootmark{a}& $\geq1.00$&$\leq1.3$ \\
 & 2 &$1.5\times10^{14}$ &$(0.5-6)\times10^{14}$ &85 &$35-92$ &$2\times10^5$ &$(2-5)\times10^5$ &1.60\tablefootmark{a}&$\geq0.60$ &$\leq0.15$\\
\hline
H$_2$CS & 1 &$1.3\times10^{16}$ &$(0.5-5.3)\times10^{16}$ &16 &$15-43$ &$7\times10^3$ &$(0.3-20)\times10^{4}$ & 0.16&$0.13-0.20$ &$0.6-2.4$ \\
\hline
OCS & 1 &$1\times10^{17}$ &$(0.2-50)\times10^{16}$ &14 &$13-15$ &... &$\geq2.2\times10^5$ &0.17 &$0.15-0.25$ &$0.2-2.4$ \\
\hline
SO & 1 &$1\times10^{16}$ &$(0.6-5)\times10^{16}$ &170 &$35-300\tablefootmark{c}$ &$1.5\times10^4$ &$(0.6-3)\times10^4$ & 0.28&$0.24-0.36$ & $0.2-1.2$\\
\hline
    \end{tabular}
        \tablefoot{
\tablefoottext{a}{Initial fixed size assumed before exploring the range of similar solution (see text Sec.\ref{subsec:method_LVG})}
\tablefoottext{b}{Average value for the optical depths of all the transitions over the range of size $\theta_s$.}
\tablefoottext{c}{Maximum limit due to the maximum temperature for which the collisional rates are available.}
}
\end{table*}

\section{Molecular abundance ratios}

The abundances ratios used in Fig.~\ref{fig:Comp_abundances} for NGC 253 are shown in Table~\ref{tab:abundance_ratios}.

\begin{landscape}

\begin{table}
    \centering
     \caption{Abundance ratios derived in each region. The underscored numbers 1 and 2 are the low- and high-$E_\text{u}$ component, respectively.}
    \label{tab:abundance_ratios}
    \resizebox{\linewidth}{!}{\begin{tabular}{lccccccccccc}
    \hline \hline
       Region & [H$_2$S]/[OCS\_1] & [H$_2$S]/[SO\_2]& [H$_2$S]/[H$_2$CS\_2]& [OCS\_1]/[H$_2$CS\_1]& [OCS\_1]/[SO\_1]& [H$_2$CS\_1]/[SO\_2] & [CS\_2]/[CCS]& [SO\_1]/[SO$_2$]& [CS\_1]/[SO\_1]& [OCS\_2]/[SO$_2$]&  [OCS\_2/CS\_2] \\
        \hline
        GMC10 & $0.01-1.0$ &... & ... &$0.75-55.6$ &$2.0-1000.0$& ... & ...&...&...&...&...\\
        GMC9a & ... &... &... &$3.7-1875.0$& $0.2-50.0$& ... & $\geq0.4$&...&...&...&...\\
        GMC8a & ... &... &... &$0.1-75.0$& $0.5-30$& ... &$\geq1.3$&...&...&...&... \\
        GMC7 & $0.05-4.7$ & $0.8-35.0$ & $0.9-3.5$ &$0.03-7.5$& ...&$2.0-232.5$&$1.0-75.0$ &...& $2.7-50.0$...&...\\
        GMC6 &$0.93 - 10.0$ & $1.4-15.0$ & 0.2$-100.0$ &$0.15-75.0$& ...&$0.02-10.0$ &$0.5-63.6$&$1.3-16.7$ &$\geq1.0$&$0.8-3.0$&$0.07-1.5$ \\
        SSC5 &$3.9 - 10.0$ & $1.4-5.0$ &$0.1-7.0$&$3.1-276.9$& ...&$0.1\times10^{-2}-0.1$&$2.6-100.0$&$2.1-16.7$&$\geq1.0$ &$0.8-2.2$ &$0.06-0.5$ \\
        SSC2 &$0.7-6.4$ & $1.4-533.3$ & $66.7 - 5.0$&$1.25-37.5$& ...&$0.06-66.7$ & $2.3-34.1$ &$1.2-11.8$&$0.1-0.5$&$0.5-1.8$& $0.1-0.6$\\
        GMC1a & ... & ... &... &$0.6-30.0$& $1.0-18.75$& ... &$\geq0.3$&...&...&...&...\\
        GMC1b &  ... &... &... &$0.04-100.0$& $0.04-83.3$& ... &$\geq0.7$&...&...&...&...\\
        GMC2b&  ... &... &... &$11.1-2000.0$& $1.0-50.0$& ... &$\geq0.7$&...&...&...&...\\
        \hline
    \end{tabular}}
\end{table}

\end{landscape}

%
%

%

\end{document}